\pdfoutput=1
\documentclass[12pt,a4paper,fleqn,english,intoc,bibliography=totoc,index=totoc,BCOR10mm,captions=tableheading,titlepage,openany]{scrbook}
\usepackage{lmodern}

\usepackage[T1]{fontenc}
\usepackage[latin9]{inputenc}
\usepackage{fancyhdr}
\usepackage{babel}
\usepackage{mathtools}
\usepackage{amsmath}
\usepackage{amssymb}
\usepackage{graphicx}
\usepackage{nomencl}
\providecommand{\printnomenclature}{\printglossary}
\providecommand{\makenomenclature}{\makeglossary}
\makenomenclature
\usepackage[unicode=true,pdfusetitle,
 bookmarks=true,bookmarksnumbered=true,bookmarksopen=true,bookmarksopenlevel=1,
 breaklinks=false,pdfborder={0 0 0},pdfborderstyle={},backref=false,colorlinks=false]
 {hyperref}
\hypersetup{linktoc=all,
 pdfpagelayout=OneColumn, pdfnewwindow=true, pdfstartview=XYZ, plainpages=false}

\makeatletter

\pdfpageheight\paperheight
\pdfpagewidth\paperwidth

\usepackage{tikz}
\usetikzlibrary{calc}

\@ifundefined{date}{}{\date{}}

\usepackage[figure]{hypcap}

\let\myTOC\tableofcontents
\renewcommand\tableofcontents{%
  \frontmatter
  \pdfbookmark[1]{\contentsname}{}
  \myTOC
  \mainmatter }

\setkomafont{captionlabel}{\bfseries}
\setcapindent{1em}

\usepackage{calc}

\let\mySection\section\renewcommand{\section}{\suppressfloats[t]\mySection}

\makeatother

\begin{document}
\subject{MSc Thesis}
\title{Finite volume corrections of non-diagonal form factors}
\author{István Vona}
\date{Budapest, 2019}
\publishers{\includegraphics[scale=0.24]{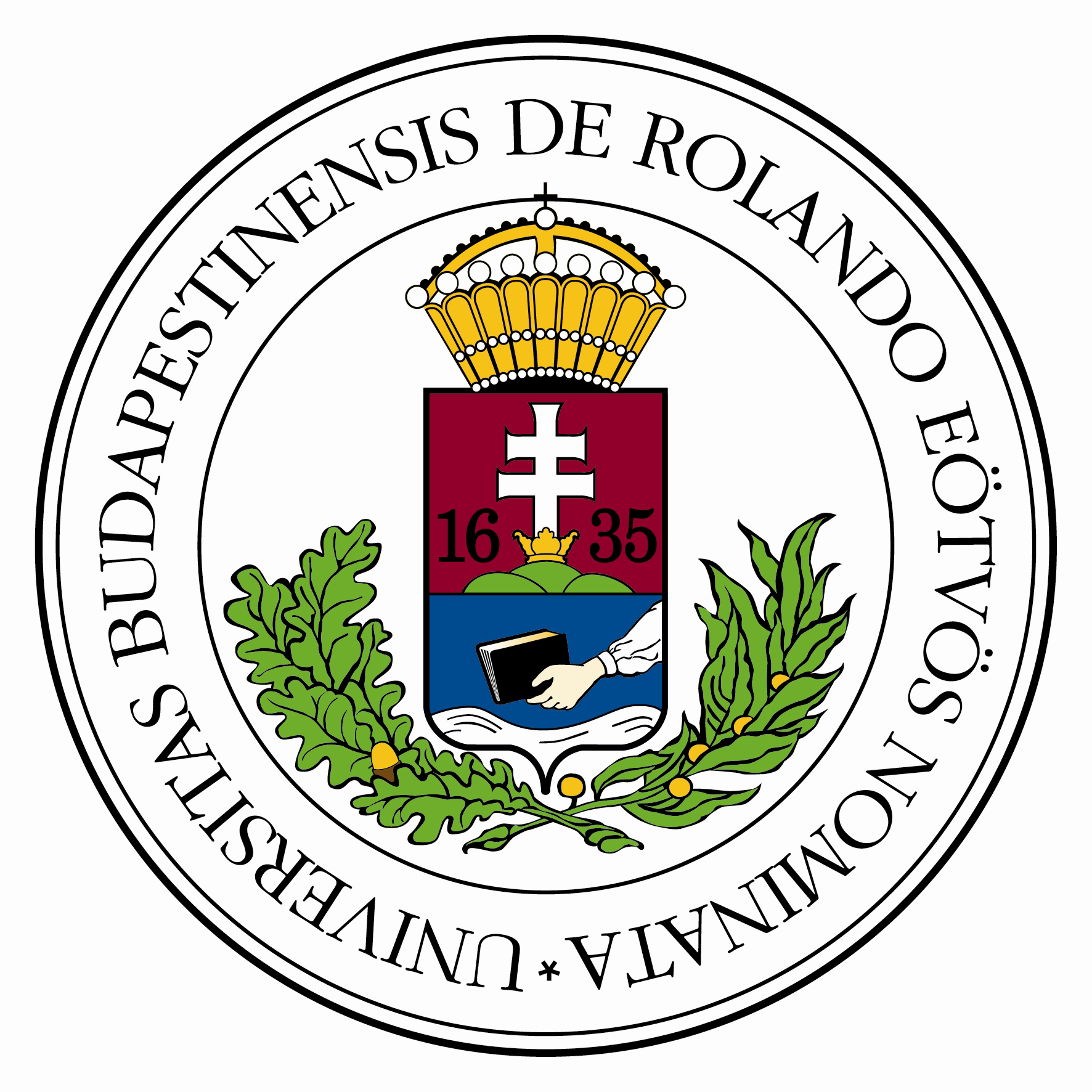}\vspace{\baselineskip}\\
Eötvös Loránd University\\
\vspace{1cm}
Supervisor:\\
Dr. Zoltán Bajnok\\
Wigner Research Centre for Physics\\
MTA Momentum Holographic Quantum Field Theory Research Group\vspace{-3cm}
}

\maketitle
\cleardoublepage{}

\lhead{\rightmark}

\rhead[\leftmark]{}

\lfoot[\thepage]{}

\cfoot{}

\rfoot[]{\thepage}

\tableofcontents{}

\cleardoublepage{}

\pagestyle{plain}

\chapter*{Abstract}

\addcontentsline{toc}{chapter}{Abstract} 

This thesis presents Lüscher's $\mu$- and $F$-term corrections to
volume dependence of non-diagonal finite volume form factors in the
scaling Lee-Yang  model. An explicit calculation proves the suspected
relation that the $\mu$-terms known previously from bound state quantization
can be obtained from the $F$-term integrals by modifying the contour
of integration such that it picks up residues of appropriate poles
in the integrand. The fact that these two different approaches for
getting the $\mu$-terms give the same result underpins the formal
derivation of the $F$-term in arXiv:1904.00492 which was not known until recently. In
the meantime, the notions of integrable quantum field theories and
those related to their treatment in finite volume are introduced to
help understand the topic for readers not familiar with it.

\cleardoublepage{}

\pagestyle{fancy}

\lhead[\chaptername~\thechapter]{\rightmark}

\lhead[\chaptername~\thechapter]{\rightmark}

\rhead[\leftmark]{}

\lfoot[\thepage]{}

\cfoot{}

\rfoot[]{\thepage}

\chapter{Introduction}

One of the mainstream tools in modern physics is quantum field theory.
QFT-s describe nature well in circumstances prevailing in experimental
setups of high energy physics where observations are made at different
length (energy) scales, they provide effective theories for low energies
where bound states of elementary particles interact, etc. From the
theoretical viewpoint, they combine special relativity and quantum
theory such that measurements at space-like separated events do not
affect each other; on the other hand, their statistical physics equivalents
are useful to study critical phenomena in condensed matter.

Outside ``real world'' examples of $3+1$ space-time dimensions,
in theoretical physics the framework of QFT finds its applications
in a wide range of models. Among the $1+1$ dimensional theories there
are some, the so-called integrable QFT-s, which - in contrast to theories
only defined perturbatively, or with lattice regularization - are
exactly solvable and analytic results for quantities are available.
These can serve as toy models to be studied for understanding questions
arising in physically relevant theories and they can also describe
low-dimensional physical systems themselves in some cases. 

Many applications of QFT requires us to work in a finite volume, e.g.
the case of lattice field theories. There, in the continuum limit,
the growth of the correlation length requires increasing the lattice
size, and so an extrapolation of quantities to infinite volume. Therefore,
the understanding of the effects caused by defining a quantum field
theory in finite volume  becomes important. Integrable models are
hoped to remain exactly solvable if defined at a finite (one-dimensional)
size, and this way they could give insight on how quantities (e.g.
the spectrum of masses) depend on the volume. A direct application
to $1+1$ dimensional field theories in finite volume is string theory,
where fields are defined on a finite world sheet. 

In the first part of this thesis I intend to give a (not too rigorous)
introduction to integrable quantum field theories building on well-known
field theoretical concepts, and motivating the steps along the path
that leads to an axiomatic formulation of the theory in infinite volume.
In this framework a complete solution to a 1+1 dimensional QFT \nomenclature{IQFT}{Integrable Quantum Field Theory}
is possible - meaning one can calculate the mass spectrum of the particles,
the scattering matrix and correlation functions of local operators
- not just perturbatively, but exactly, and also for models and ranges
of parameters where perturbation theory does not exist or cannot be
applied. 

Chapter \ref{chap:Integrable-quantum-field} covers this topic. In
the classical field theoretical picture the main ingredients - the
higher spin charges - are introduced and the effects of integrability
- like the absence of particle production - get explained in the realm
of the sine-Gordon model. Then - in quantum theory - the factorization
of scatterings into two-particle interactions due to the existence
of the mentioned charges allows one to describe them with the S-matrix
for a $2\to2$ process alone. Using the general analytic properties
of the latter results in restrictions on its functional form, and
also in consistency equations which determine both the mass spectrum
of the model and the exact scattering matrix elements between these
particles. This leads to the so-called bootstrap method: apart from
some minimal external information on the model, the solution to these
equations is based on the requirement that the system of the fusions
among the particles - being bound states of each other - has to close.
This method then gets generalized to the matrix elements of local
operators between momentum eigenstates - the so-called form factors
- whose exact expressions provide the expansion of the correlation
functions over multiparticle states in the spectral representation.
The form factors themselves have to satisfy a set of axioms whose
importance for us now - together with the general constraints on the
S-matrix - lies in being able to carry out model-independent calculations.

Then, in chapter \ref{chap:Finite-volume-effects} we explore the
consequences of compactifying the spatial dimension. As usual, the
momenta get quantized and the energy spectrum becomes discrete, however,
the dependence of the energy levels on the finite size of space is
unknown in a general interacting QFT. With integrability at hand,
we are able to get exact results here too. Taking the interactions
of physical particles on the space-time cylinder into account is based
on the infinite volume scattering data and gives polynomial corrections
in the inverse of the spatial extent at large volumes for both energy
levels and form factors - compared to a free theory. These are not
enough to describe the complete volume dependence since there exist
processes where virtual particles propagate around the finite world
resulting in corrections which become dominant for smaller volumes
only. With the help of a method originating from statistical physics,
it is possible to derive an integral equation whose solution contains
the information on all these corrections and gives exact results for
the energy spectrum. 

The complete solution to an IQFT in finite volume is however not available
yet. There exist results for special cases of form factors - expectation
values of operators in finite volume multiparticle states - which
sum up the volume corrections due to the mentioned virtual processes.
For the general case of non-diagonal form factors\footnote{Which means the matrix element of the operator is taken between two
different states. } only the leading order is known, and recent developments presented
in chapter \ref{chap:L=0000FCscher-corrections-of-non-diago} give
the next-to-leading order\footnote{Actually, for theories where fusion among the particles is absent
this latter becomes the leading correction.}. There is a mathematical relation connecting these two corrections
which was shown by my supervisor, Zoltán Bajnok for the vacuum-one-particle
form factor to hold. In my thesis, I demonstrate this property with
calculations for the generalized multiparticle case. A possible application
to this may be establishing similar relations for higher orders which
could help us on the way obtaining the exact volume dependence of
general form factors. The rather cumbersome parts of the derivations
are relegated to the appendix for those who are interested. In chapter
\ref{chap:Conclusion} we interpret and discuss the results and point
towards possible directions of development in this topic.

\lhead[\chaptername~\thechapter]{\rightmark}

\rhead[\leftmark]{}

\lfoot[\thepage]{}

\cfoot{}

\rfoot[]{\thepage}

\global\long\def\Res{\operatornamewithlimits{Res}}%

\global\long\def\Tr{\operatorname{Tr}}%

\global\long\def\diag{\operatorname{diag}}%

\global\long\def\re{\operatorname{Re}}%

\global\long\def\im{\operatorname{Im}}%

\global\long\def\Ordo{\operatorname{\mathcal{O}}}%

\chapter{Integrable quantum field theory\label{chap:Integrable-quantum-field}}

\section{Integrability in classical field theory}

In classical mechanical systems, the notion of integrability arises
when the continuous symmetry of the system is big enough to have as
many conserved quantities by Noether's theorem as degrees of freedom
in the system. Then the equations of motion can be solved by integrating
them.

In field theory - where the number of d.o.f is infinite - a system
has to contain infinitely (countably) many functionally independent
Noether charges to show this property. In this section we show how
to construct them, and illustrate what are their implications on the
solutions to the classical equations of motion of the system.

\subsection{Higher spin charges\label{subsec:Higher-spin-charges}}

Concerning a relativistic theory - beyond those coming from Poincaré
symmetry - local conserved currents can be constructed from the fields
and their derivatives. These transform generally as higher rank tensors.

For reasons which will be obvious later (see \ref{subsec:Effect-of-integrability}),
we consider 1+1 (space-time) dimensional theories only, and the case
of a real scalar field for simplicity
\begin{equation}
\mathcal{L}=\frac{1}{2}\partial_{\mu}\phi\partial^{\mu}\phi-V(\phi),\quad x^{\mu}=\begin{pmatrix}t\\
x
\end{pmatrix},\quad\eta_{\mu\nu}=\diag(1,-1),\quad\mu=0,1\label{eq:lagrange}
\end{equation}
and search for a potential which allows conservation of additional
(hopefully infinitely many) charges.

Special to 1+1 dimension, the Lorentz-transformation can be diagonalized
by a transition to the so-called light-cone coordinates defined by
$x^{\pm}=\frac{1}{2}(t\pm x)$ (we keep upper and lower indices $x^{\pm},x_{\pm}$):
\begin{align*}
x^{\mu} & \to\begin{pmatrix}x^{+}\\
x^{-}
\end{pmatrix} & \eta_{\mu\nu} & \to\begin{pmatrix}0 & 2\\
2 & 0
\end{pmatrix} & \Lambda_{\;\nu}^{\mu}=\begin{pmatrix}\cosh\theta & \sinh\theta\\
\sinh\theta & \cosh\theta
\end{pmatrix} & \to\begin{pmatrix}e^{\theta} & 0\\
0 & e^{-\theta}
\end{pmatrix}
\end{align*}
where the rapidity variable $\theta$ parametrizes the boost:
\[
\Lambda_{\;\nu}^{\mu}x^{\nu}\to e^{\pm\theta}x^{\pm}.
\]
Tensor components in these coordinates transform under boost simply
by scaling:
\begin{align*}
\Lambda_{\ \nu_{1}}^{\mu_{1}}\ldots\Lambda_{\ \nu_{n}}^{\mu_{n}}\Lambda_{\mu'_{1}}^{\;\nu'_{1}}\ldots\Lambda_{\mu'_{m}}^{\;\nu'_{m}}T_{\qquad\nu'_{1}\ldots\nu'_{m}}^{\nu_{1}\ldots\nu_{n}} & \to e^{s\theta}T_{\qquad s'_{1}\ldots s'_{m}}^{s_{1}\ldots s_{n}}\\
s_{i},s'_{i}=\pm, & s=\sum_{i=1}^{n}s_{i}-\sum_{i=1}^{m}s'_{i},
\end{align*}
therefore we can associate the integer $s$ - called ``spin'' -
to each of them.

The e.o.m. looks like
\begin{equation}
(\partial_{t}^{2}-\partial_{x}^{2})\phi=-V'(\phi)\to\partial_{+}\partial_{-}\phi=-V'(\phi),\quad\partial_{\pm}=\partial_{t}\pm\partial_{x};\label{eq:eom}
\end{equation}
note that the derivatives change the spin by one according to $\Lambda_{\mu}^{\;\nu}\partial_{\nu}\phi\to e^{\mp s\theta}\partial_{\pm}\phi$.

The continuity equation for the Noether current (omitting every other
tensorial indices)
\begin{equation}
\partial_{\mu}J^{\mu(\mu_{1}\ldots\mu_{n})}=0\to\partial_{+}J^{+}+\partial_{-}J^{-}=0\label{eq:conservation}
\end{equation}
is integrated over space to define a charge:
\[
Q_{s}[\phi]=\int_{-\infty}^{\infty}dxJ^{0}=\int_{-\infty}^{\infty}dx\left(T^{(s+1)}+\Theta^{(s-1)}\right)
\]
where we denoted the components of the current in light-cone coordinates
as
\[
\begin{pmatrix}J^{+}\\
J^{-}
\end{pmatrix}\equiv\begin{pmatrix}T^{(s+1)}\\
\Theta^{(s-1)}
\end{pmatrix},\quad\partial_{+}T^{(s+1)}+\partial_{-}\Theta^{(s-1)}=0
\]
indicating their spin. One calls $Q_{s}$ - transforming as $Q_{s}[\phi']=e^{s\theta}Q_{s}[\phi]$,
where $\phi'$ is the transformed field - a higher spin charge for
$\left|s\right|>1$.

For $V(\phi)=0$, the case of a free massless boson one can satisfy
\eqref{eq:conservation} by choosing
\[
J_{-}=\left(\partial_{-}^{j_{1}}\phi\right)^{k_{1}}\ldots\left(\partial_{-}^{j_{n}}\phi\right)^{k_{n}},\quad J_{+}=0\quad\Rightarrow\quad\partial_{+}J_{-}+\partial_{-}J_{+}=0
\]
for any powers $j_{i},k_{i}>0$, since $\partial_{+}J_{-}=0$ by using
the e.o.m. $\partial_{+}\partial_{-}\phi=0$. (Of course we can pick
$J_{-}$ to be zero and $J_{+}$ as a product of $\partial_{+}$-derivatives
as well.)

A non-zero potential will generally spoil the conservation of these
currents, however one can try some combinations of such $J_{-}$-s
along the ideas in \cite{how}.

Let us try the most simple choice
\begin{align*}
J_{-} & =-\partial_{-}\phi,\\
\partial_{+}J_{-} & =-\partial_{+}\partial_{-}\phi=-\partial_{-}\left(\partial_{+}\phi\right)\overset{!}{=}-\partial_{-}J_{+}\\
\Rightarrow J_{+} & =\partial_{+}\phi
\end{align*}
In the original coordinates this looks as $J^{\mu}=\epsilon^{\mu\nu}\partial_{\nu}\phi$
(notice that its divergence vanishes by Schwartz's theorem even without
requiring the e.o.m. to hold), and the corresponding charge
\[
Q_{0}=\int_{-\infty}^{\infty}dx\partial_{x}\phi=\phi(t,x=\infty)-\phi(t,x=-\infty)
\]
is non-zero for a general case: if we set the lower bound of the potential
to be zero ($V(\phi)\geq0$), to have finite energy field configurations
\begin{equation}
E[\phi]=\int_{-\infty}^{\infty}dx\{\frac{1}{2}(\partial_{t}\phi)^{2}+\frac{1}{2}(\partial_{x}\phi)^{2}+V(\phi)\}<\infty\label{eq:finiteEcond}
\end{equation}
the field should pick its asymptotic values from the minima of $V$,
since
\[
\lim_{x\to\pm\infty}V(\phi(t,x))=0
\]
(and also $\partial_{t}\phi,\partial_{x}\phi\overset{x\to\pm\infty}{\longrightarrow}0$)
is necessary to keep the integrand \eqref{eq:finiteEcond} non-vanishing
over a finite domain. If the set of minima is discrete (and $V$ has
at least two minima), the field can take different values at the infinities,
and the continuous time evolution cannot change this discrete asymptotic
property of the field-configuration; therefore $Q_{0}$ is a constant
- the so-called topological charge.

For $s=1$, we need a current-component with spin $s+1=2$, which
has overall two $\partial_{-}$ derivatives:
\begin{align*}
J_{-} & =\left(\partial_{-}\phi\right)^{2},\\
\partial_{+}J_{-} & =2\left(\partial_{-}\phi\right)\overbrace{\partial_{+}\partial_{-}\phi}^{-V'(\phi)}=-2\partial_{-}V(\phi)\overset{!}{=}-\partial_{-}J_{+}\\
\Rightarrow J_{+} & =2V(\phi)
\end{align*}
This works for any potential, and the corresponding charge is just
the ``$+$'' component of the energy-momentum 2-vector $(E,p)^{T}$
in light-cone coordinates:
\[
T^{(2)}+\Theta^{(0)}=\frac{1}{2}\left(\partial_{-}\phi\right)^{2}+V(\phi)\;\Rightarrow\;Q_{+1}=E+p,
\]
similarly there is a $Q_{-1}=E-p$. The other choice could have been
$J_{-}=\partial_{-}^{2}\phi$, but since it is a total derivative
(of $\partial_{-}\phi$, which is already zero at the infinities,
unlike $\phi$ itself), one can show that the corresponding charge
will be zero. We generally omit these total derivatives, the only
exceptional case being $Q_{0}$.

The only remaining element for $s=2$
\begin{align*}
J_{-} & =(\partial_{-}\phi)^{3},\\
\partial_{+}J_{-} & =-3(\partial_{-}\phi)^{2}V'(\phi)\neq-\partial_{-}J_{+}
\end{align*}
cannot be written as a $\partial_{-}$-derivative.

The type of currents with 4 derivatives, which could appear in a $s=3$
charge are
\begin{align*}
J_{-} & =\frac{1}{4}(\partial_{-}\phi)^{4}, & \partial_{+}J_{-} & =-(\partial_{-}\phi)^{3}V'(\phi)=-(\partial_{-}\phi)^{2}\partial_{-}V(\phi)\neq-\partial_{-}J_{+}\\
J_{-} & =(\partial_{-}^{2}\phi)^{2}, & \partial_{+}J_{-} & =-2(\partial_{-}^{2}\phi)(\partial_{-}V'(\phi))=-\partial_{-}\left[(\partial_{-}\phi)^{2}\right]V''(\phi)\neq-\partial_{-}J_{+}
\end{align*}
and we cannot find a $J_{+}$ for them for a general potential: however
if we impose a condition on it (for some constants $\beta,c$)
\begin{equation}
V''(\phi)=-\beta^{2}V(\phi)+c\label{eq:SGconstraint}
\end{equation}
the following pair of current components will satisfy the continuity
equation:
\[
T^{(4)}=-\beta^{2}\frac{1}{4}(\partial_{-}\phi)^{4}+(\partial_{-}^{2}\phi)^{2},\quad\Theta^{(2)}=(\partial_{-}\phi)^{2}\left[-\beta^{2}V(\phi)+c\right].
\]
Particular solutions to the constraint \eqref{eq:SGconstraint} 
\begin{align}
V(\phi) & =\frac{m^{2}}{\beta^{2}}(1-\cos(\beta\phi))\nonumber \\
V(\phi) & =\frac{m^{2}}{b^{2}}(\cosh(b\phi)-1), & (\beta^{2}<0\; & \Rightarrow\;\beta\to ib)\label{eq:preSG}
\end{align}
will be important to us later (they are tuned by parameter $c$ such
that their global minima are at $V(0)=0$ and $V(2\pi n)=0,\;n\in\mathbb{Z}$
respectively). These are the potentials for the sine- and sinh-Gordon
models, the names referring to their e.o.m., at first order in $\beta$
being the Klein-Gordon equation for free fields:
\[
\partial_{+}\partial_{-}\phi=-\frac{m^{2}}{\beta}\sin(\beta\phi)\;\Rightarrow\partial_{+}\partial_{-}\phi=-m^{2}\phi.
\]
In this way we obtained a $Q_{3}$ higher spin charge, whose existence
depended on the form of the potential but we did not prove that there
are infinitely many of these $Q_{s}$-s. There is a method specific
to the potentials like \eqref{eq:preSG} (see Bäcklund-transformation
in \cite[p. 521]{mussardo}) capable of generating these charges;
instead of this, we present a more universal method for proving integrability
of a model in the next subsection.

\subsection{Lax-integrability}

A sufficient condition for integrability of a 1+1 dimensional model
can be formulated as described in the following. Let us have a $N$-component
function $\psi(t,x,\lambda)$ which depends on a $\lambda\in\mathbb{C}$
parameter, and a non-abelian vector field $A_{\mu}(t,x,\lambda)$
acting as a $N\times N$ matrix on the internal space of $\psi$.
We define the covariant derivative as $D_{\mu}=\partial_{\mu}-A_{\mu}(t,x,\lambda)$,
and demand that
\begin{equation}
D_{\mu}\psi=0\label{eq:covariantderivative}
\end{equation}
which in components looks like (the components $A_{t},A_{x}$ form
the Lax-pair, a notion important in the theory of PDE-s)
\begin{align*}
\partial_{t}\psi & =A_{t}\psi\\
\partial_{x}\psi & =A_{x}\psi
\end{align*}
where we treat $\psi$ as unknown, and we try to integrate these equations
for known $A_{\mu}$ functions. The existence of a continuous solution
requires the Schwarz integrability condition to hold, which gives
by differentiating the lines above
\begin{align}
\partial_{t}\partial_{x}\psi & =\partial_{x}\partial_{t}\psi & \Rightarrow\partial_{t}A_{x}\psi+A_{x}\overbrace{\partial_{t}\psi}^{A_{t}\psi} & =\partial_{x}A_{t}\psi+A_{t}\overbrace{\partial_{x}\psi}^{A_{x}\psi}\nonumber \\
 &  & \Rightarrow\;\partial_{t}A_{x}-\partial_{x}A_{t}-[A_{t},A_{x}] & =0.\label{eq:flatness}
\end{align}
This is equivalent to saying that the curvature corresponding to the
connection $A_{\mu}$ vanishes (hence the name ``flatness condition''
for \eqref{eq:flatness}):
\[
[D_{\mu},D_{\nu}]\psi=F_{\mu\nu}\psi=0.
\]
Now the statement is the following: if we manage to find a one parameter
family (labeled by $\lambda$) of functions $A_{t},A_{x}$ such that
\eqref{eq:flatness} is equivalent to the e.o.m. of our model, i.e.
we can ``embed'' the fields and their derivatives of the original
theory (e.g. $\phi(x)$ of \eqref{eq:lagrange}) into these matrices,
and the flatness condition holds for every $\lambda$, then the theory
is integrable. (This ``embedding'' is not unique, it is possible
to achieve it for different $N$ dimensions , and one can also make
a gauge transformation - the curvature remains zero.)

For example, there exists a $\mathfrak{su}(2)$ gauge field for the
sine-Gordon model, written in dimensionless light-cone coordinates
for $\beta=1$ as \cite[p. 200]{APSIII}
\begin{align*}
A_{+} & =\frac{1}{4i\lambda}\begin{pmatrix}\cos\phi & -i\sin\phi\\
i\sin\phi & -\cos\phi
\end{pmatrix},\;A_{-}=i\begin{pmatrix}-\lambda & \frac{1}{2}\partial_{-}\phi\\
\frac{1}{2}\partial_{-}\phi & \lambda
\end{pmatrix}\\
 & \Rightarrow F_{+-}=\frac{i}{2}\begin{pmatrix}0 & \partial_{+}\partial_{-}\phi+\sin\phi\\
\partial_{+}\partial_{-}\phi+\sin\phi & 0
\end{pmatrix}.
\end{align*}
The statement holds because a time-independent functional of the fields
can be constructed, which depends on $\lambda$; and the coefficients
of its Maclaurin series in $\lambda$ gives infinitely many conserved
quantities. 

To see this, we first notice that a matrix solution to \eqref{eq:covariantderivative}
is actually a parallel transport, which is now (due to flatness) does
not depend on the path of transportation, only on its endpoints $(t,x)\leftarrow(t_{0},x_{0})$:
\begin{equation}
\partial_{\mu}U=A_{\mu}(t,x,\lambda)U\;\Rightarrow\;U(t,x;t_{0},x_{0},\lambda)=\mathcal{P}\exp\{\int_{t_{0},x_{0}}^{t,\;x}A_{\mu}(t',x',\lambda)dx'^{\mu}\}.\label{eq:fundamentalmatrix}
\end{equation}
For simplicity we consider the model on compactified space of length
$L$, with PBC. Our time-independent quantity will be the trace of
the so-called monodromy matrix defined as a transport along a closed
space-like loop at time $t$ ($U$ equals  to the identity matrix
only for contractible loops): 
\[
T(t,\lambda)=\mathcal{P}\exp\{\int_{0}^{L}dxA_{x}(t,x,\lambda)\},
\]
which can be shifted to $t'$ by attaching two time-like transports
at its ends like $(t',L)\overset{S}{\leftarrow}(t,L)\overset{T}{\leftarrow}(t,0)\overset{S^{-1}}{\leftarrow}(t',0)$
(see Fig. \eqref{eq:TSTS}). This means 
\begin{equation}
T(t',\lambda)=ST(t,\lambda)S^{-1},\quad S=\mathcal{P}\exp\{\int_{t}^{t'}d\tau A_{t}(\tau,0,\lambda)\}.\label{eq:TSTS}
\end{equation}

\begin{figure}[h]
\begin{centering}
\includegraphics[scale=0.25]{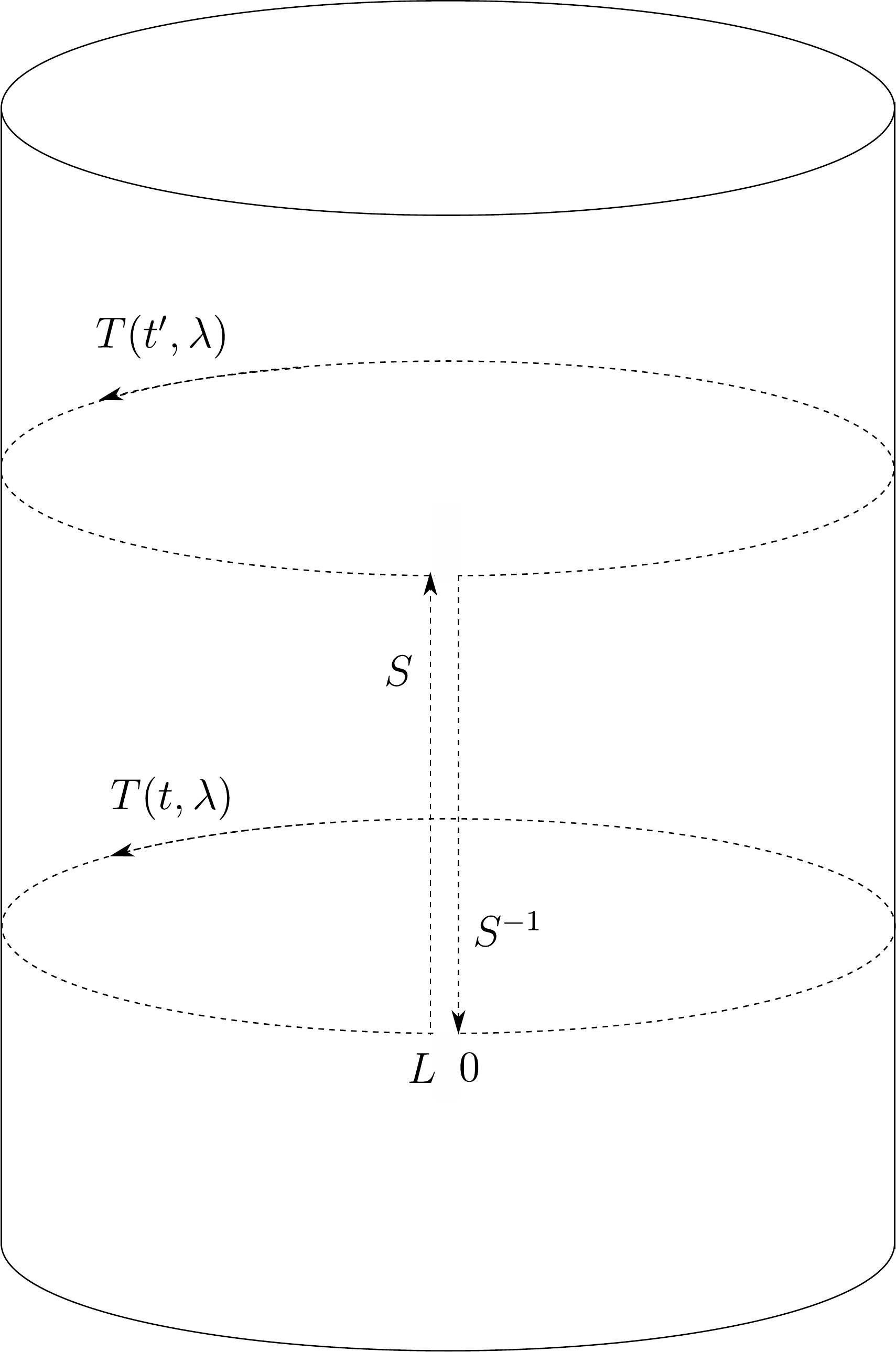}\caption{Visualization of the equality \eqref{eq:TSTS} in terms of paths. }
\par\end{centering}
\label{fig:TSTS}
\end{figure}
We see from \eqref{eq:TSTS} that the trace of $T(t,\lambda)$ is
independent of time by the cyclic property, and so $q(\lambda)\equiv\Tr(T(t,\lambda))$ is
constant. In its expansion  $q(\lambda)=\sum_{n}q_{n}\lambda^{n}$
the $q_{n}[\phi]$-s are conserved charges, where we denoted the fields
of the integrable model collectively as $\phi$. In a concrete model
one could get these charges by solving the PDE \eqref{eq:fundamentalmatrix}
for the matrix $U$. Showing their independence is another - model
sepcific - question  and beyond the scope of this work.

\subsection{Solitons}

\begin{figure}[h]
\begin{centering}
\includegraphics[scale=0.8]{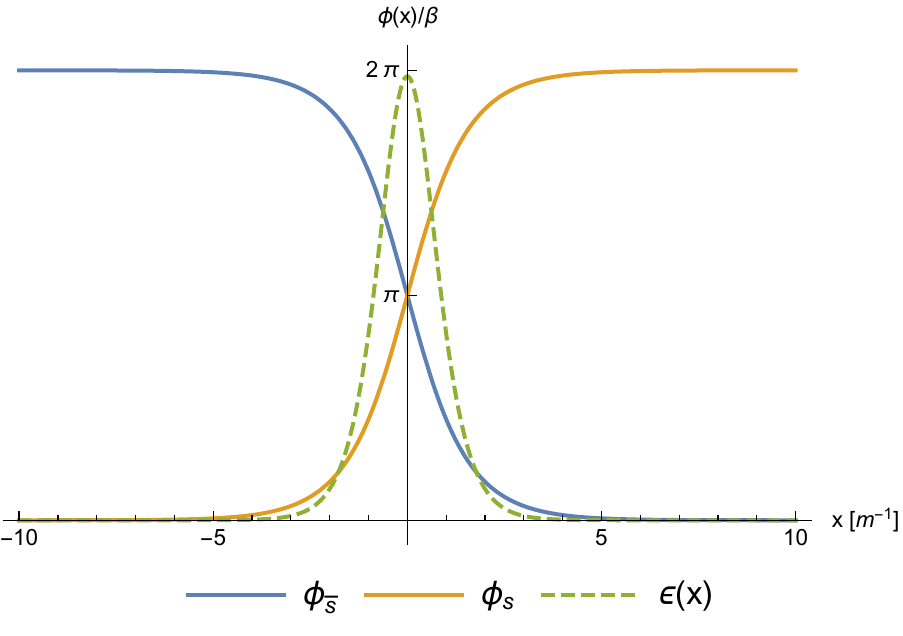}\includegraphics[scale=0.8]{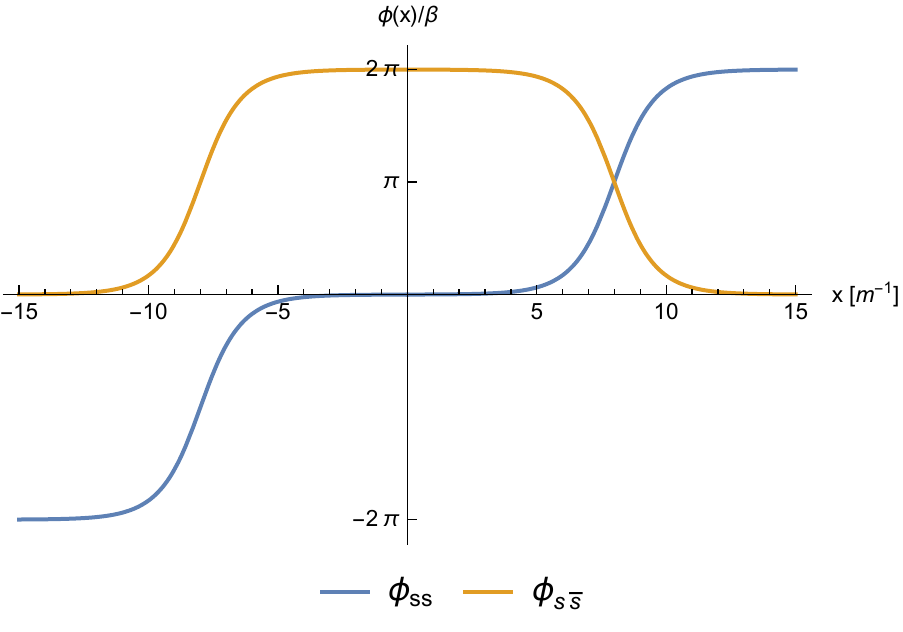}
\par\end{centering}
\caption{Left: Static soliton and anti-soliton. The shape of their energy density
$\epsilon(x)$ is superimposed on the figure in a scale-less way.
Right: Two-soliton and soliton-anti-soliton solutions where the particles
are moving with relative velocity $v=0.1c$ in each; the figure shows
the configurations at $t=100m^{-1}$ after the scattering events.}

\label{fig:sol}
\end{figure}

There are actually interesting solutions to the classical field equations
of the sine-Gordon theory whose behaviour resembles that of relativistic
particles. By making a restriction to time-independent (static) solutions
the finite energy condition \eqref{eq:finiteEcond} allows two types
of field-configurations - solitons ($s$) and anti-solitons ($\bar{s}$),
see \cite[p. 193]{APSIII} and the left of Fig. \ref{fig:sol}   -
interpolating between neighboring minima $2\pi n$ and $2\pi(n+1)$
of the potential, therefore having opposite values of the topological
charge $\tilde{Q}_{0}=\frac{\beta}{2\pi}Q_{0}=\pm1$.

The energy density of these configurations is localized, with a finite
width corresponding to the Compton-length of the mass parameter $m$
in the Lagrangian. However, if we calculate the energy functional
we get a rest mass (since the momentum $p[\phi_{s}]=p[\phi_{\bar{s}}]=0$),
which looks
\begin{equation}
E[\phi_{s}]=E[\phi_{\bar{s}}]=\frac{8m}{\beta^{2}}\equiv M_{s}.\label{eq:solitonmass}
\end{equation}
Since we are in a relativistic theory, by boosting the solutions we
get moving configurations retaining their shape under time evolution;
the name soliton comes from the expression ``solitary wave'' describing
this situation. The width of the ``lump'' in the energy density
suffers Lorentz-contraction, and the 2-momentum vector transforms
accordingly. 

It is possible to obtain other - non-static - solutions of the e.o.m
(with the above mentioned Bäcklund transformation), and the analogy
with particles extends when one analyses them for $t\to\pm\infty$.
There exist a two-soliton (or two-anti-soliton) and a soliton-anti-soliton
type configuration (with $\tilde{Q}_{0}=2$ and $0$ respectively,
see the right hand side of Fig. \ref{fig:sol}), where the two waves
move with a constant (non-zero) relative velocity for asymptotical
times; and after  their (unavoidable) scattering they suffer a time-shift
compared to an undisturbed motion. In the first case this shift is
positive indicating a repulsion, in the second it is negative - an
attraction. The magnitude of the shift depends on the relative velocity;
otherwise the scattering is purely elastic, no momentum transfer happens.

We can get another $\tilde{Q}_{0}=0$ solution by analytically continuing
the $s\bar{s}$-solution making its relative velocity purely imaginary.
The field will become an oscillating function in time, with a localized
envelope. Its interpretation is a new particle in rest (the so-called
``breather'')  with mass
\[
0\leq M_{B}(\omega)\leq2M_{s}
\]
depending on the frequency of oscillation. No such continuation of
the $ss$ or $\bar{s}\bar{s}$ possible. The analogy is clearly a
bound state of a particle and its anti-particle, the case $M_{B}=2M_{s}$
is a weakly bound one, whereas the $M_{B}=0$ is a tight one. 

Any kind of scattering states of the mentioned ``particles'' is
a possible solution of the model with $\tilde{Q}_{0}\in\mathbb{Z}$,
the important properties being that
\begin{itemize}
\item the number of particles will not change,
\item the scatterings happen without exchange of momenta,
\item and that the time-shift a single particle suffers by scattering on
the others is additive, so we can understand the effect of any $n\to n$
process in which our particle participates as if it were a sequence
of $2\to2$ particle scatterings applied on it at distant events.
\end{itemize}

\section{Integrability in quantum field theory}

One can look at the sinh-Gordon model - apart from the infinite number
of $\phi^{2n}$-type interaction terms if we expand the $\cosh$ function
in the potential - as a quantum field theory of a massive scalar field
having a single vacuum at $\langle0\vert\phi\vert0\rangle=0.$ The
usual perturbative approach gives some illustration on how the integrability
property makes its way into the quantum theory.

\begin{figure}[h]
\begin{centering}
\includegraphics[scale=0.3]{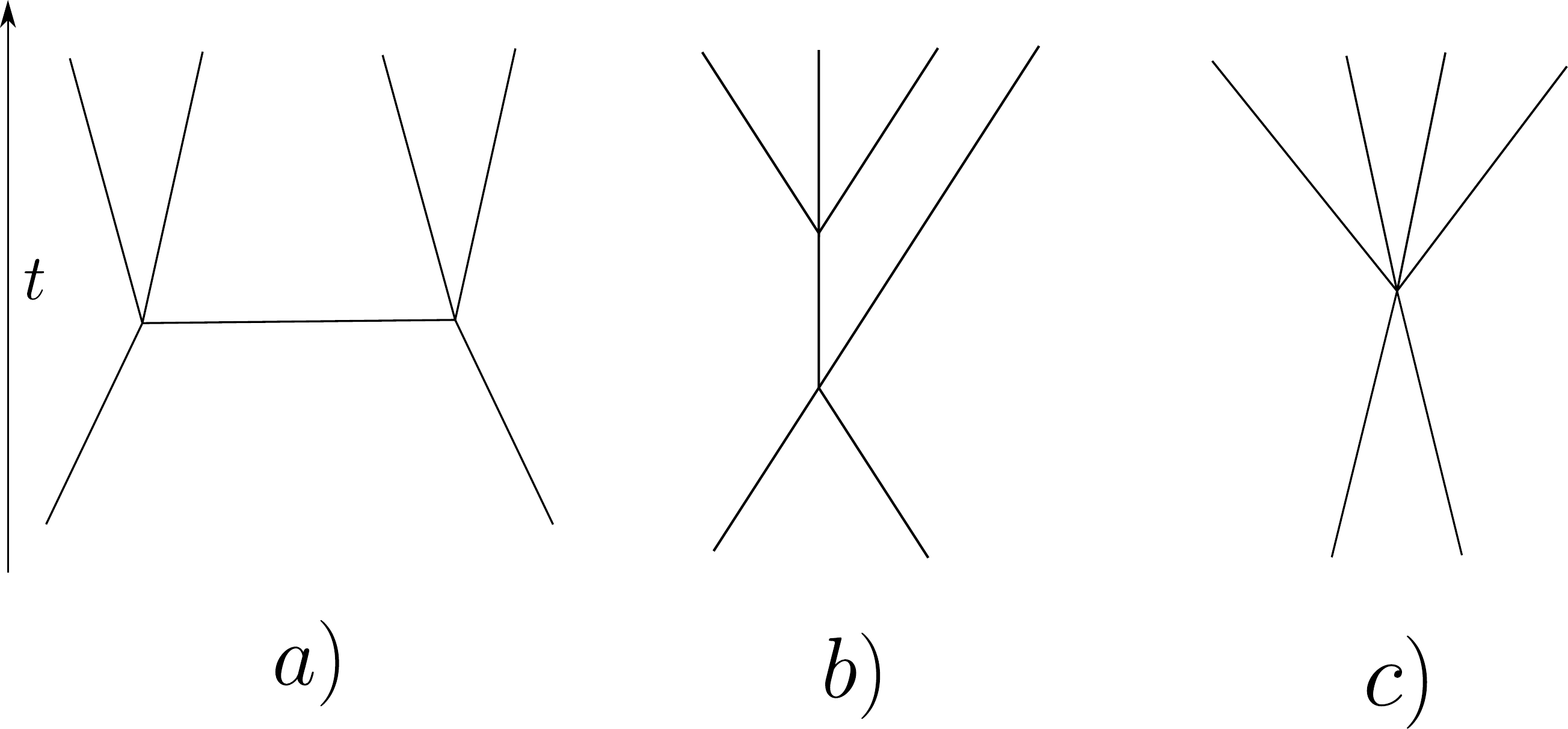}
\par\end{centering}
\caption{Graphs for tree-level contributions. }

\label{fig:graph}
\end{figure}

First, by expanding the Lagrangian in parameter $b$
\[
\mathcal{L}=\frac{1}{2}\partial_{\mu}\phi\partial^{\mu}\phi-\frac{1}{2}m^{2}\phi^{2}-\frac{1}{4!}m^{2}b^{2}\phi^{4}-\frac{1}{6!}m^{2}b^{4}\phi^{6}+\ldots
\]
we consider the scattering amplitude of the $2\to4$ process at tree
level. It can be shown that the contribution coming from the $\phi^{4}$
interaction (see the two type of graphs $(a)$ and $(b)$ in figure
\ref{fig:graph}) is independent of the momenta, and this constant
is exactly canceled by the (obviously) constant contribution of the
$\phi^{6}$ term $(c)$. By looking at $2\to2n-2$ processes, it is
also true, that summing up all the contribution of the interaction
terms up to order $2n$ at tree level, gives vanishing amplitude.

The proper treatment of the $n\to n$ amplitude at tree level however
gives a non-zero amplitude, multiplied by a structure of delta functions
which ensures that the ingoing set of momenta is the same as the outgoing
one. The tree-level calculation of course is a classical result, but
it has been shown, that particle production is also non-existent in
the model for 1-loop, indicating that the quantum theory might exhibit
this property as well. \cite{doreyexact} 

As we see, taking infinitely many interaction terms into account to
analyze sinh-Gordon model seems cumbersome. In case of the sine-Gordon
theory there is even another problem: the soliton mass \eqref{eq:solitonmass}
is non-perturbative $\mathcal{O}(\beta^{-2})$ in the coupling, and
the soliton is not an elementary excitation of the field around a
single minimum of the potential which one could treat via standard
Feynman perturbation theory.\footnote{A semiclassical treatment however is possible for weak couplings;
the spectrum of breathers becomes discrete in the quantum theory.
\cite[ch. 7]{rajaraman}} 

It is however possible to circumvent these problems if we assume that
those infinitely many charges remain conserved quantities at the quantum
level. Their conservation will restrict the initial and final sets
of momenta to be the same, and also require the $n\to n$ particle
processes to factorize into $2\to2$ scatterings; this way the only
S-matrix elements one needs are those for the $2\to2$ events. These
simplifications allow us to determine the S-matrix of the theory not
just perturbatively, but exactly - in an axiomatic framework. The
next subsections will give some motivations how this is built up.

\subsection{Effect of integrability on scattering processes\label{subsec:Effect-of-integrability}}

Let us say we are dealing with theories where only massive particles
exist, and we parametrize the 2-momentum vector with the rapidity
variable $E=m\cosh\theta,\;p=m\sinh\theta$. The ``in'' and ``out''
states, as eigenstates of the exact momentum operator of the interacting
theory can be denoted as $\vert\theta_{1},\ldots,\theta_{N}\rangle_{i_{1}\ldots i_{N},\mathrm{in/out}}$
where $i_{1}\ldots i_{N}$ are labels meaning the different types
of particles in the theory; and the usual notation and normalization
of one-particle states $_{i'}\langle p'\vert p\rangle_{i}=2\pi\;2E\;\delta_{i'i}\delta(p'-p)$
transforms to $_{i'}\langle\theta'\vert\theta\rangle_{i}=2\pi\delta_{i'i}\delta(\theta-\theta')$.

As we saw in subsection \ref{subsec:Higher-spin-charges}, the energy
and momentum are actually equivalent to a pair of spin-1 charges,
and their eigenvalues look like
\[
Q_{\pm1}\vert\theta_{1},\ldots,\theta_{N}\rangle_{i_{1}\ldots i_{N},\mathrm{in/\mathrm{out}}}=\left(m_{i_{1}}e^{\pm\theta_{1}}+\ldots+m_{i_{N}}e^{\pm\theta_{N}}\right)\vert\theta_{1},\ldots,\theta_{N}\rangle_{i_{1}\ldots i_{N},\mathrm{in/\mathrm{out}}}
\]
where $m_{i}$ is the mass of particle type $i$. 

We make an assumption that the higher spin charges commute among themselves\footnote{It is based on the fact that their classical counterparts commute
within Poisson-brackets.}
\[
[Q_{s},Q_{s'}]=0,
\]
therefore with $Q_{\pm1}$ as well; which means they also take their
eigenvalues on the momentum eigenstates. Since they are integrals
of local charge densities - intuitively speaking, they sum up all
the contributions from freely moving particles at asymptotical times
- they are also additive:
\[
Q_{s}\vert\theta_{1},\ldots,\theta_{N}\rangle_{i_{1}\ldots i_{N},\mathrm{in/\mathrm{out}}}=\left(Q_{i_{1}}^{(s)}(\theta_{1})+\ldots+Q_{i_{N}}^{(s)}(\theta_{N})\right)\vert\theta_{1},\ldots,\theta_{N}\rangle_{i_{1}\ldots i_{N},\mathrm{in/\mathrm{out}}},
\]

where $Q_{i_{k}}^{(s)}(\theta_{k})$ is the eigenvalue of $Q_{s}$
on $\vert\theta_{k}\rangle_{i_{k}}$. From the classical transformation
law of charges under Lorentz boosts, we see that in the quantum theory
their eigenvalues should behave as 
\[
Q_{i}^{(s)}(\theta+\Lambda)=Q_{i}^{(s)}(\theta)e^{s\Lambda}
\]
and so we can write them as $Q_{i}^{(s)}(\theta)=q_{i}^{(s)}e^{s\theta}$
where $q_{i}^{(s)}$ is the eigenvalue on a standing particle state
of type ``$i$''.

\begin{figure}[h]
\begin{centering}
\includegraphics[scale=0.7]{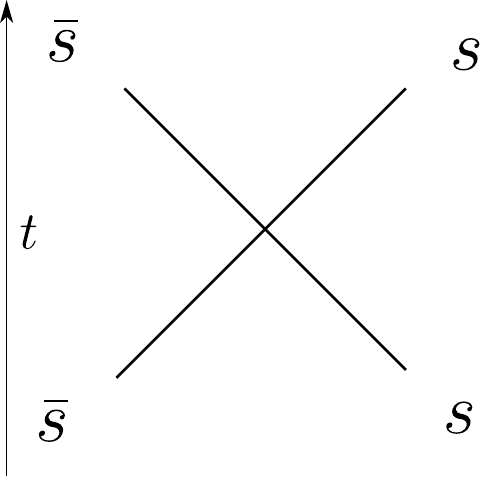}
\par\end{centering}
\caption{Reflection amplitude for the sine-Gordon model where particles are
changing their identities.}

\label{fig:SR}
\end{figure}

Let us take now an $N\to M$ particle process with particle types
and rapidities $\{i\},\{\theta\}$ and $\{i'\},\{\theta'\}$ in the
initial and final states respectively. The requirement that $Q_{s}$
is conserved translates into the constraint that transition can only
take place between those ``in'' and ``out'' states for which the
eigenvalue of $Q_{s}$ is the same. In this case one has to satisfy
the infinite number of equations 
\begin{align}
q_{i_{1}}^{(s)}e^{s\theta_{1}}+\ldots q_{i_{N}}^{(s)}e^{s\theta_{N}} & =q_{i'_{1}}^{(s)}e^{s\theta'_{1}}+\ldots q_{i'_{M}}^{(s)}e^{s\theta'_{M}},\;\forall s\label{eq:infiniteconstraint}
\end{align}
with only a finite number of parameters, which implies 
\begin{align*}
N & =M, & \theta_{k} & =\theta'_{k} & \mathrm{and}\qquad q_{i_{k}}^{(s)} & =q_{i'_{k}}^{(s)} & \forall\left|s\right| & >0, & \forall k
\end{align*}
up to permutations of the particles. This means particle production
is forbidden, the momenta and the masses ($q_{i}^{(\pm1)}=m_{i}$)
of the particles are the same before and after the scattering; and
moreover, the eigenvalues for every higher spin charge coincide. However,
if there are particle types which do not differ in any of these, they
can turn into each other via scattering. For example such (non-diagonal)
scatterings can happen in the sine-Gordon theory, where in the case
of $Q_{0}$ the equation \eqref{eq:infiniteconstraint} only requires
the equality of the total topological charge, and processes like \ref{fig:SR}
are possible.

If we think of particles as wave packets rather than states with definite
momentum, one can associate a position - and so worldlines - to them.
The 2-momentum as a charge generates translations and shifts these
worldlines in the same way for every particle. One can show that the
effect of $\left|s\right|>1$ higher spin charges is similar, but
their influence on the position of a particle depends on its momentum.
Therefore we can make particle trajectories in an $N\to N$ process
- if participants have different momenta - to miss the scattering
event by a symmetry transformation. Of course, in 1+1 dimensions this
means that $\binom{N}{2}$ two-particle collisions will happen instead
arbitrarily far away, and so the scattering amplitude factorizes.
This is the main reason why we chose to work with these theories at
start: in higher dimensions the existence of a generator with higher
rank will render the S-matrix to be trivial, since then we could make
worldlines miss each other completely so no collision happens.

From now on we search for the $2\to2$ amplitudes only.

\subsection{Analytic properties and constraints of two-particle S-matrices\label{subsec:Analytic-properties-and}}

We now collect how the qualitative picture of S-matrix theory modifies
in case of a scattering in 1+1 dimension without momentum transfer.
Instead of separating the no-interaction part and concentrating on
the scattering amplitude as usual
\[
_{\mathrm{out}}\langle f\vert i\rangle_{\mathrm{in}}\equiv S_{fi}=\delta_{fi}+i(2\pi)^{2}\delta^{(2)}(P_{f}-P_{i})T_{fi},
\]
 we now write the S-matrix as ($i,j,k,l$ denoting particle types)\footnote{The amplitude $S_{ij}^{kl}(s)$ becomes zero for those $i,k$ and
$j,l$ which do not satisfy the equality of $q^{(s)}$-s for $\left|s\right|>1$. }
\begin{align}
_{lk,\mathrm{out}}\langle p_{l},p_{k}\vert p_{i},p_{j}\rangle_{ij,\mathrm{in}} & =(2\pi)^{2}(2E_{i})(2E_{j})\delta(p_{k}-p_{i})\delta(p_{l}-p_{j})S_{ij}^{kl}(s),\label{eq:Smxdef}
\end{align}
for $p_{l}\leq p_{k}$ and $p_{i}\geq p_{j}$. This way the usual
momentum-conserving $\delta$-function multiplying the transition
amplitude translates into the $\delta$-structure in \eqref{eq:Smxdef}
because of the coinciding sets of in- and outgoing momenta. The function
$S_{ij}^{kl}$ is parametrized in terms of the only independent relativistic
invariant combined from the on-shell momenta of the participants,
which is now the single Mandelstam-variable $s$, since
\[
s=(p_{i}^{\mu}+p_{j}^{\mu})^{2},\;t=(p_{i}^{\mu}-p_{l}^{\mu})^{2}=2(m_{i}^{2}+m_{j}^{2})-s,\;u=(p_{i}^{\mu}-\overbrace{p_{k}^{\mu}}^{p_{i}^{\mu}})^{2}=0.
\]
The ordering is introduced to exclude the other possible term for
$p_{i}=p_{l},\;p_{j}=p_{k}$, which is equivalent by renaming the
indices.

\begin{figure}[h]
\begin{centering}
\includegraphics[scale=0.5]{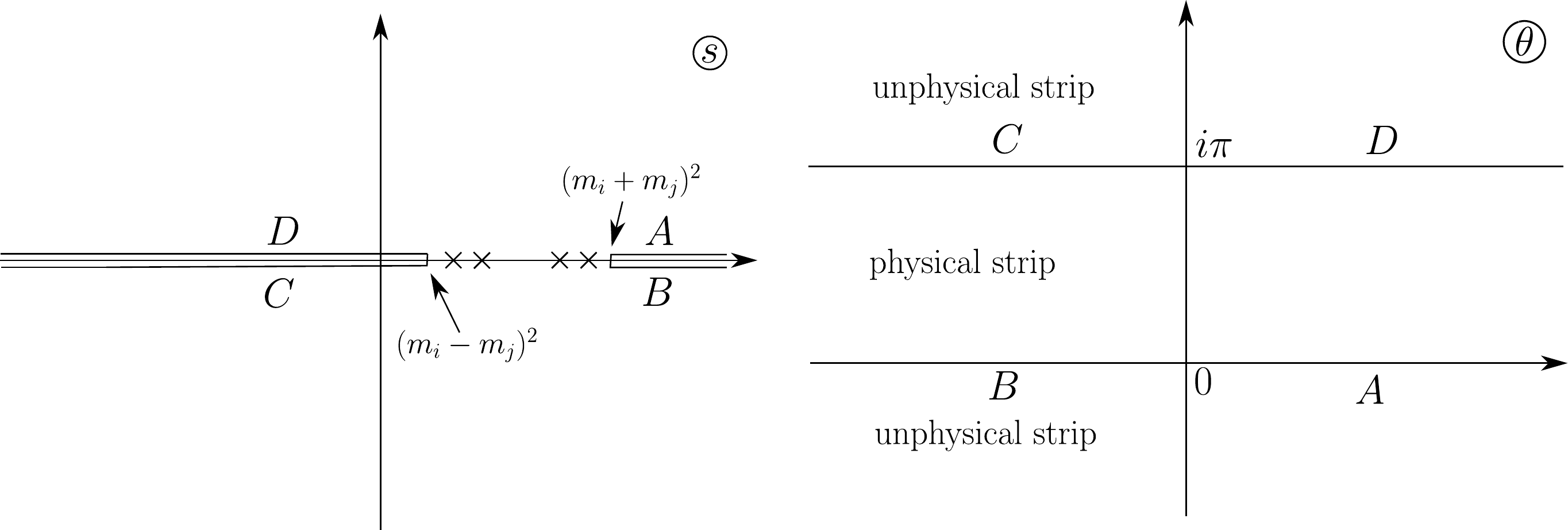}
\par\end{centering}

\caption{Left: The complex plane of the invariant $s$. Physical values come
from approaching $A$, the inverse of the S-matrix is at $B$, and
the physical $t$-channel amplitude reached at $C$. Right: The analytic
structure after the mapping, on the $\theta$ plane.}

\label{fig:mapping}
\end{figure}

On the $s$complex $s$-plane (see figure \ref{fig:mapping}) we have
\begin{itemize}
\item a branch cut on the positive real axis, starting at the threshold
in the center of mass energy $s=(m_{i}+m_{j})^{2}$ above which the
production of the outgoing particles is possible (there are no further
thresholds - as in the case of a general theory - since production
of more than two particles is not possible)
\item another cut on the left for $s\leq(m_{i}-m_{j})^{2}\leftrightarrow t\geq(m_{i}+m_{j})^{2}$
where the $t$-channel process has its 2-particle threshold;
\item poles below the 2-particle threshold on the real axis - usually corresponding
to stable particles (bound states) unreachable by scattering, but
appearing in (either $s$-channel or $t$-channel) virtual processes,
and also as asymptotical states themselves.\footnote{Note that our assumption is - basically for simplicity - that there
are no unstable particles (resonances) for which a pole would appear
(among others) on the first unphysical Riemann sheet connected to
the cut; e.g. by going through the right cut from the upper half plane
we could see a pole close to the real axis with $\re s>(m_{i}+m_{j})^{2}$
and finite (but small) negative imaginary value. This would mean an
increase in the amplitude square for physical $s$-values around the
mass of the resonance due to the presence of the pole. The imaginary
part of the pole position is related to the finite life-time, which
disappears in case of stable bound states between the cuts.}
\end{itemize}

We reach the physical values of the amplitude - as usual -  by approaching
the real axis from the upper half plane above the 2-particle threshold:
\begin{align*}
S_{ij}^{kl}(s)\vert_{\mathrm{phys}} & =S_{ij}^{kl}(s+i\epsilon) & \mathrm{for}\quad & s\geq(m_{i}+m_{j})^{2},
\end{align*}
where $\epsilon>0$ is infinitesimal. This is consistent with the
Feynman prescription of the propagator, which modifies every mass
as $m^{2}\to m^{2}-i\epsilon$, and therefore moves the cut below
the real axis a little - allowing the evaluation of the amplitude
for real $s$. The $S_{ij}^{mn}$ -s are real analytic functions by
assumption (consult with \cite[p. 17]{eden} in which case it is true)
\[
\left(S_{ij}^{mn}(s)\right)^{*}=S_{ij}^{mn}(s^{*})
\]
and so the unitarity of the S-matrix - which condition is valid only
for physical values - (summation over same upper and lower indices
implied)
\begin{align*}
S_{mn}^{kl}(s+i\epsilon)\left(S_{ij}^{mn}(s+i\epsilon)\right)^{*} & =\delta_{i}^{k}\delta_{j}^{l} & \mathrm{if}\quad & s\geq(m_{i}+m_{j})^{2},
\end{align*}
gives that one can reach the physical value for the inverse process
from below:
\begin{align*}
\left(S^{-1}\right)_{ij}^{kl}(s)\vert_{\mathrm{phys}} & =S_{ij}^{kl}(s-i\epsilon) & \mathrm{for}\quad & s\geq(m_{i}+m_{j})^{2}.
\end{align*}
The crossing symmetry states that the physical values of the crossed
amplitude can be expressed as the original amplitude at unphysical
domains. For a type of crossing which is allowed by kinematics (the
bar in $\bar{j},\bar{l}$ means the antiparticle):
\begin{align*}
S_{i\bar{l}}^{k\bar{j}}(s+i\epsilon) & =S_{ij}^{kl}(\overbrace{2(m_{i}^{2}+m_{j}^{2})-(s+i\epsilon)}^{t-i\epsilon}) & \mathrm{if}\quad & s\geq(m_{i}+m_{j})^{2}\leftrightarrow t\leq(m_{i}-m_{j})^{2},
\end{align*}
which shows we get the first by approaching the left ($t$-channel)
branch cut of the latter from below. These properties are summarized
on the upper half of figure \ref{fig:mapping}.

As before, the rapidity variable will simplify matters here as well.
The relations between the Mandelstams and the rapidities of particles
$i,j$ are\footnote{This choice of $t(\theta)$ is consistent with the $s\to\theta$ mapping.}
\begin{align}
s(\theta) & =m_{i}^{2}+m_{j}^{2}+2m_{i}m_{j}\cosh\theta, & \theta & =\theta_{i}-\theta_{j}\nonumber \\
t(\theta) & =s(i\pi-\theta)\nonumber \\
\theta & =\ln\left(1+\Delta_{+}+i\sqrt{\Delta_{+}\Delta_{-}}\right), & \Delta_{\pm} & =\pm\frac{s-(m_{i}\pm m_{j})^{2}}{2m_{i}m_{j}}\label{eq:mapping}
\end{align}
where the last line (which comes by inverting the first) with the
usual choice of branch cuts of the logarithm and the square root maps
the physical Riemann sheet of the variable $s$ onto the strip $0\leq\im\theta\leq\pi$
of the $\theta$ complex plane in the way presented by the lower half
of figure \ref{fig:mapping}. This map opens up the cuts and so the
unphysical sheets are mapped onto the strips $n\pi\leq\im\theta\leq(n+1)\pi,\;n\in\mathbb{Z}/\{0\}$.
Therefore the S-matrix is continuous at the edges of the strips, and
so a meromorphic function in $\theta$, having poles only on the imaginary
axis.

The real analyticity, unitarity and crossing properties now read
\begin{align*}
\left(S_{ij}^{kl}(\theta)\right)^{*} & =S_{ij}^{kl}(-\theta^{*})\\
S_{mn}^{kl}(\theta)S_{ij}^{mn}(-\theta) & =\delta_{i}^{k}\delta_{j}^{l}\\
S_{i\bar{l}}^{k\bar{j}}(\theta) & =S_{ij}^{kl}(i\pi-\theta)
\end{align*}
where the first implies that the matrix is real on the imaginary axis;
and the second and third - so far a condition only at physical values,
i.e. for $\theta>0$ - now can be continued to the whole complex plane.
Note that by the unitarity, where there are poles in the physical
strip we see zeros on the unphysical strip $-\pi\leq\im\theta\leq0$,
and vice versa (there can be zeros in the physical strip as well).

Another relations can be deduced for discrete symmetries, which we
just show (the ones for $P$ and $T$ are more obvious thinking in
the center of mass frame of the $ij$ system):
\begin{align*}
S_{ij}^{kl}(\theta)=\begin{cases}
S_{\bar{i}\bar{j}}^{\bar{k}\bar{l}}(\theta) & C\\
S_{ji}^{lk}(\theta) & P\\
S_{lk}^{ji}(\theta) & T
\end{cases}.
\end{align*}
There is still one condition which S-matrices should satisfy, coming
from the equivalence of two different factorizations of $3\to3$ processes
(following the discussion at the end of subsection \ref{subsec:Effect-of-integrability}
about shifting particle trajectories): 
\[
S_{cb}^{lm}(\theta_{12})S_{ia}^{cn}(\theta_{13})S_{jk}^{ba}(\theta_{23})=S_{ba}^{ml}(\theta_{23})S_{ck}^{na}(\theta_{13})S_{ij}^{cb}(\theta_{12}),
\]
which is the so-called Yang-Baxter equation.

In the next subsection a method is presented with which one can find
solutions to all these functional relations. Before we proceed, we
make a big simplification and restrict ourselves from now on to theories
where the mass spectrum is non-degenerate ($m_{i}\neq m_{j}$ for
$i\neq j$), and so
\begin{itemize}
\item particles are necessarily the same as their antiparticles ($m_{i}=m_{\bar{i}}\;\Rightarrow\;i=\bar{i}$)
\item the S-matrix is diagonal $S_{ij}^{kl}(\theta)=S_{ij}^{kl}(\theta)\delta_{i}^{k}\delta_{j}^{l}$
(no summation) since already the spin-one charge $m_{i}=q_{i}^{(1)}$
is different for each particle: they cannot turn into other particles
with same $\left|s\right|>0$ charges.
\end{itemize}
The Yang-Baxter equation in this case becomes an identity, and all
other functional relations can be summarized as (we omit upper indices
since outgoing particles are now the same as ingoing ones)
\begin{align}
S_{ji}(\theta) & =S_{ij}(\theta) & P-\mathrm{symmetry}\nonumber \\
S_{ij}(\theta)^{*} & =S_{ij}(-\theta^{*}) & \mathrm{real\;analiticity}\nonumber \\
S_{ij}(\theta)^{-1} & =S_{ij}(-\theta) & \mathrm{unitarity}\nonumber \\
S_{i\bar{j}}(\theta)\equiv S_{ij}(\theta) & =S_{ij}(i\pi-\theta) & \mathrm{crossing}\label{eq:analyticprops}
\end{align}
By applying unitarity and crossing successively two times one sees
that the S-matrix is periodic in the imaginary direction $S_{ij}(\theta+2\pi i)=S_{ij}(\theta)$,
so there is only a single unphysical strip (we choose $-\pi\leq\im\theta<0$).On
the $s$-plane this means that it does not matter around which (the
left or right) branch point we continue analytically, we arrive at
the same Riemann sheet.

From the unitarity and real analyticity together, one can easily see
that for real $\theta$-s $S_{ij}$ is just a phase shift
\begin{align*}
S_{ij}(\theta) & =S_{ij}(0)e^{i\delta_{ij}(\theta)} & \mathrm{if}\;\theta & \in\mathbb{R},
\end{align*}
with $S_{ij}(0)=\pm1$ and $\delta_{ij}(-\theta)=-\delta_{ij}(\theta)$.

For indistinguishable particles, the two cases for $S_{ii}(0)$ and
the statistics of the particles combine in the following way \cite{zamolodchikovTBA}:
\begin{itemize}
\item $S_{ii}(0)=-1$: Exchanging particles with the same rapidity gives
a minus, which is only consistent with an anti-symmetric wavefunction;
and so bosons - with symmetric wavefunction - surely cannot occupy
the same state with the same rapidity, their momenta must differ.
They behave like fermions, and therefore the situation is called ``fermionic''.
Then, by following this logic, multiple fermions can have the same
momentum, they are ``bosonic''.
\item $S_{ii}(0)=1:$ The situation is reversed, bosons and fermions behave
like it is ``expected''; however the only known diagonally scattering
model for this case is that of the free boson with $S(\theta)\equiv1$.
\cite[p. 665]{mussardo}
\end{itemize}

\section{Bootstrap principle\label{sec:Bootstrap-principle}}

Since the analytic properties of the S-matrices of diagonally scattering
theories - i.e. the functions $S_{ij}(\theta)$ - are quite simple
(there are only poles and zeros on the imaginary axis), one can try
to search for functions satisfying \eqref{eq:analyticprops}. By writing
the conditions in terms of the variable $x=e^{\theta}$, and looking
for a solution as a ratio of polynomials in $x$ one finds by increasing
their order that the simplest S-matrix is (apart from the trivial
ones $S(\theta)=\pm1$)
\begin{align}
\frac{x^{2}-2i\alpha x-1}{x^{2}+2i\alpha x-1} & \leftrightarrow S(\theta)=\frac{\sinh\theta-i\alpha}{\sinh\theta+i\alpha}, & \alpha & \in\mathbb{R}.\label{eq:shGSmatrix}
\end{align}
From the perturbative analysis of the sinh-Gordon model one can guess
that this S-matrix describes the theory if $\alpha=\sin\left(\frac{\pi b^{2}}{8\pi+b^{2}}\right)$
- at least for small values of $b$. Here $0<a<1$, and therefore
the denominator has no zeros for $0\leq\im\theta\leq\pi$ in contrast
to the numerator; the roles are switched for $-\pi\leq\im\theta\leq0$.
As mentioned, by the unitarity $S(-\theta)=S(\theta)^{-1}$, a pole
in one strip means a zero in the other strip at the conjugate position.

The absence of poles in the physical strip signals that there are
no bound states in the theory, only a single ``elementary'' particle.
For negative values of $\alpha$ (i.e. for $ib\to\beta$), there are
two poles in $0\leq\im\theta\leq\pi$.\footnote{Then for $\alpha<0$ this S-matrix describes the scattering of the
lightest breather in the $\tilde{Q}_{0}=0$ topological sector of
the sine-Gordon theory.}

The function \eqref{eq:shGSmatrix} is a building block for S-matrices
like \eqref{eq:analyticprops}; a product of these
\begin{equation}
S_{ij}(\theta)=\prod_{p\in P_{ij}}\frac{\sinh\theta-i\sin(\pi p)}{\sinh\theta+i\sin(\pi p)},\label{eq:Smxproduct}
\end{equation}
- where the $p$ real parameters are coming from a discrete set $P_{ij}$
- can describe every S-matrix in a model with diagonal scattering
of self-conjugate particles \cite{mitra1977}. Still, one should fix
these sets somehow. 

The main assumption (``maximal analyticity'' of the S-matrix) is
that there are no poles in the physical strip which cannot be explained
by the mass spectrum of the theory. One treats all types of particles
in the model on an equal footing; since bound states are assumed to
be stable and to appear as asymptotical states. If a self-consistent
system is found, where one can associate a particle to every physical
pole in every S-matrix between the particles, their quantum field
theory is considered to be solved exactly.\footnote{In reality there is another source for these poles - the so-called
anomalous thresholds - which complicates this picture. An example
for the treatment of these in the sine-Gordon model is being presented
in \cite{coleman1978}.}

The simplest case is a first-order pole due to a bound-state. If the
pole in $S_{ij}(\theta)$ on the imaginary axis corresponds to the
bound state ``$k$'', and its position is at $\theta=iu_{ij}^{k}$,
we can get the mass of particle $k$ from the expression on $s(\theta)$:
\begin{equation}
m_{k}^{2}=s(iu_{ij}^{k})=m_{i}^{2}+m_{j}^{2}+2m_{i}m_{j}\cos u_{ij}^{k},\label{eq:masstriangle}
\end{equation}
where $u_{ij}^{k}$ is the so-called fusion angle of the (unphysical)
$ij\to k$ process.

\begin{figure}[h]
\begin{centering}
\includegraphics[scale=0.8]{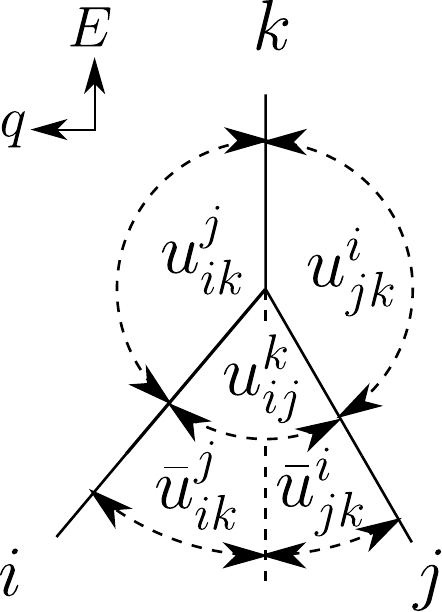}\caption{The so-called Coleman-Thun diagram on the Euclidean $(E,q)$ plane.
The complex rapidities of the particles are $\theta_{i}=i\bar{u}_{ik}^{j},\;\theta_{j}=-i\bar{u}_{jk}^{i},\;\theta_{k}=0$.
}
\par\end{centering}
\label{fig:coleman}
\end{figure}

The meaning of this pole and its residue is yet unclear. Those singularities
of $n$-point functions which happen at unphysical values in their
external momenta, can be summarized by diagrams (see \cite{coleman1978})
where all the lines represent on-shell particles, with imaginary momenta
$p\to iq,\;p^{\mu}=(E,iq)\;\Rightarrow\;E^{2}+q^{2}=m^{2}$. These
particles interact at vertices (where the complex 2-momentum vectors
have to satisfy conservation) with on-shell vertex functions as couplings.
So in the $E+iq$ complex plane one can draw lines of length $m$,
with such angles that the sum of the vectors starting from a vertex
is zero. The conservation equation for an $ij\to k$ vertex in terms
of the complex variable $E+p=E+iq$ (see figure \ref{fig:coleman}):
\begin{equation}
m_{i}e^{i\bar{u}_{ik}^{j}}+m_{j}e^{-i\bar{u}_{jk}^{i}}=m_{k},\label{eq:massbootstrap}
\end{equation}
is just (formally) the conservation of $Q_{+1}$ between states ``$\vert\theta+i\bar{u}_{ik}^{j},\theta-i\bar{u}_{jk}^{i}\rangle_{ij,\mathrm{in}}$''
and $\vert\theta\rangle_{k,\mathrm{out}}$ ; where $\bar{u}_{ab}^{c}=\pi-u_{ab}^{c}$,
and for the meaning of the angles $u_{ik}^{j},u_{jk}^{i}$ see figure
\ref{fig:coleman}.

\begin{figure}[h]

\begin{centering}
\includegraphics[scale=0.6]{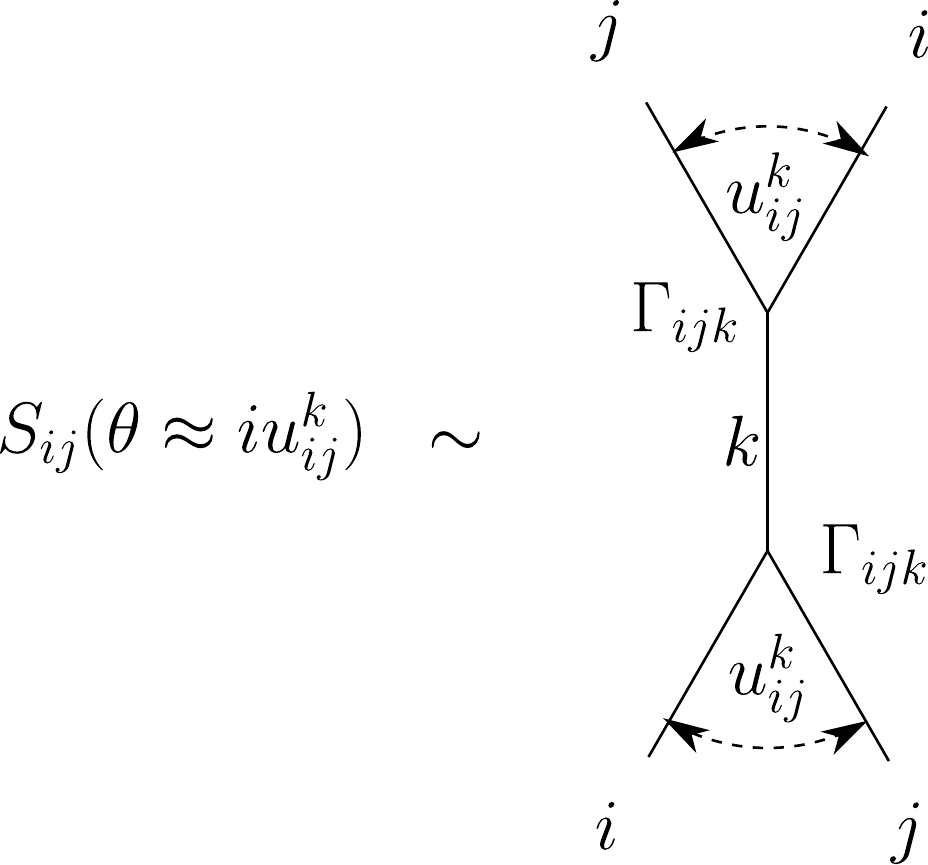}\caption{The simple pole corresponding to an on-shell bound state propagator
in the $s$-channel. }
\par\end{centering}
\label{fig:schannel}
\end{figure}

\begin{figure}[h]
\begin{centering}
\includegraphics[scale=0.6]{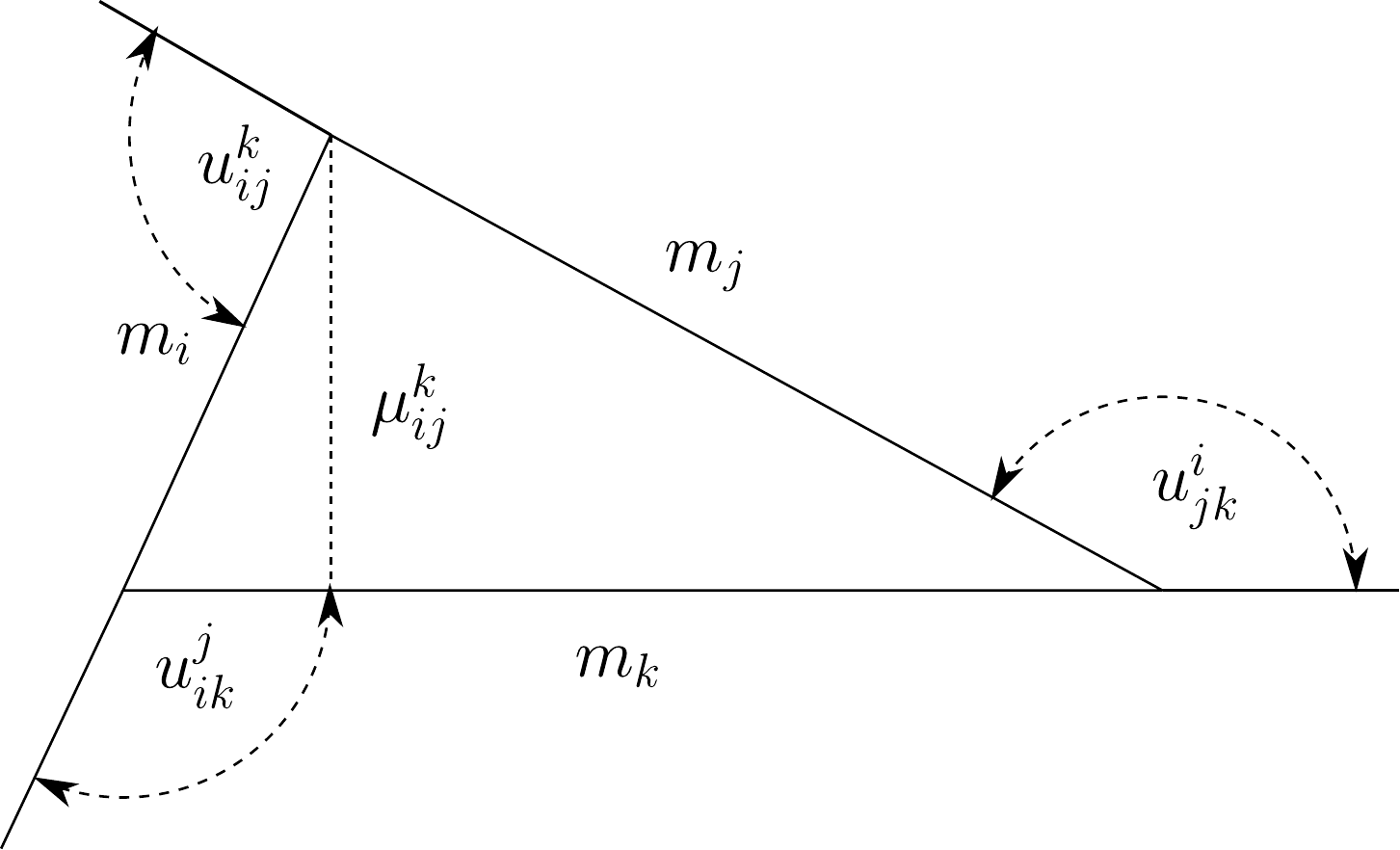}
\par\end{centering}
\caption{The mass triangle. The relevance of the $\mu_{ij}^{k}$ height will
be explained in subsection \ref{subsec:TBA-for-models}. }

\label{fig:masstriangle}
\end{figure}

In the S-matrix - by looking at an amputated version of figure \ref{fig:schannel},
the singularity comes from the internal propagator which becomes on-shell;
and the residue is the square of the $\Gamma_{ijk}$ on-shell vertex
function, which is a constant in $1+1$ dimension for the number of
independent invariants being zero. There is also a $t$-channel process
to this bound state so the vicinity of the corresponding pair of poles
looks like (the sign comes via $S_{ij}(i\pi-\theta)=S_{ij}(\theta)$):
\[
S_{ij}(\theta)\simeq\begin{cases}
\frac{i\Gamma_{ijk}^{2}}{\theta-iu_{ij}^{k}}+\ldots & s\text{-channel}\\
\frac{-i\Gamma_{ijk}^{2}}{\theta-i\bar{u}_{ij}^{k}}+\ldots & t\text{-channel}
\end{cases},
\]
and there are such pairs also in $S_{jk}$ and $S_{ik}$ at $\theta=iu_{jk}^{i},i\bar{u}_{jk}^{i}\;\leftrightarrow\;s,t=m_{i}^{2}$
and $\theta=iu_{ik}^{j},i\bar{u}_{ik}^{j}\;\leftrightarrow\;s,t=m_{j}^{2}$
respectively due to bound states $i$ and $j$, if $\Gamma_{ijk}\neq0$.
The relation \eqref{eq:masstriangle} resembling the law of cosines
between the fusion angles and the masses has a geometrical interpretation:
they are the outer angles and sides of the so-called mass triangle
shown in figure \ref{fig:masstriangle}. 

\begin{figure}[h]

\begin{centering}
\includegraphics[scale=0.5]{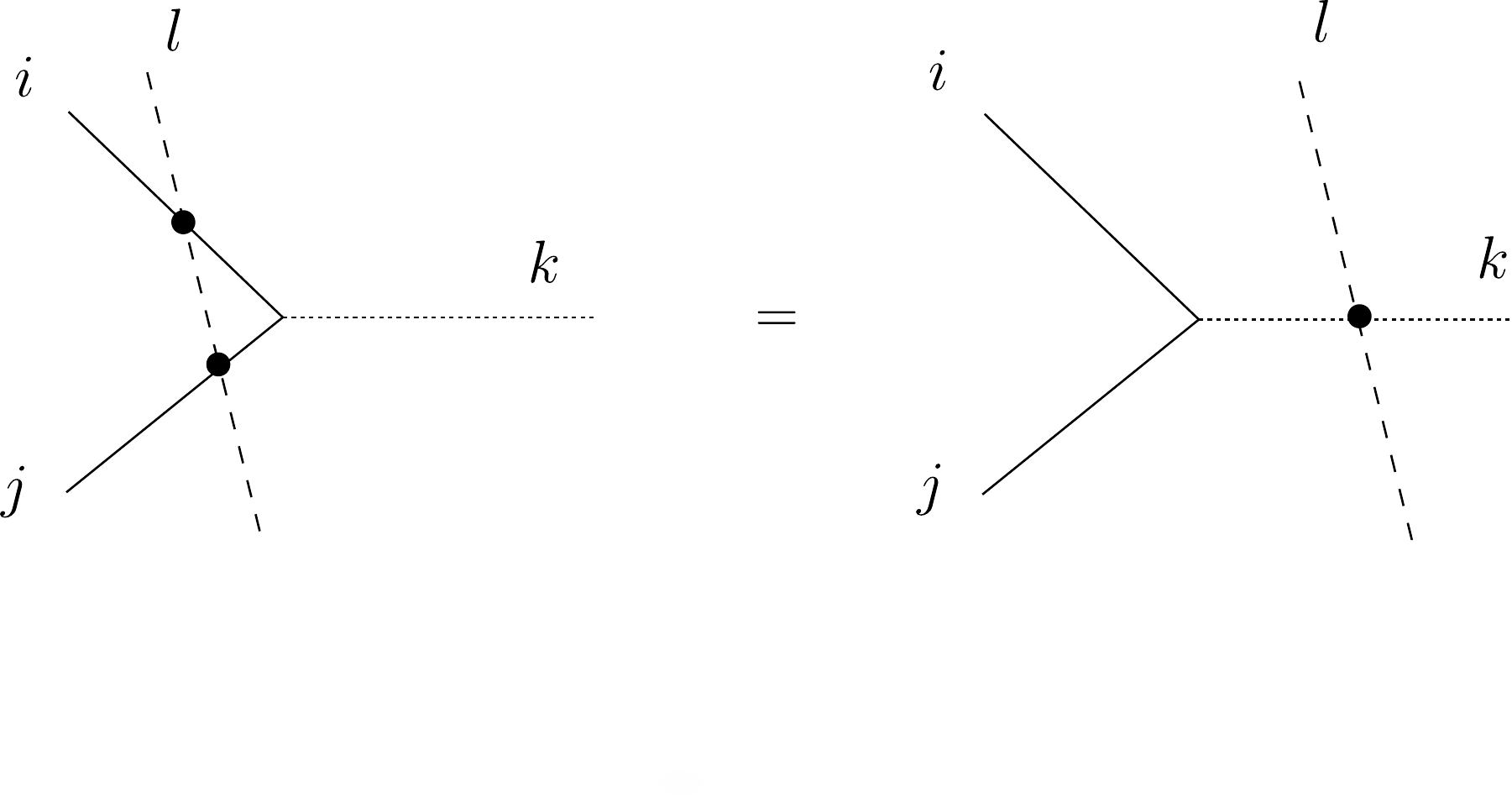}\caption{Illustration to the bootstrap equation. }
\par\end{centering}
\label{fig:bootstrap}
\end{figure}

We can now state the bootstrap principle: scattering on a bound state
is equivalent to scattering on its constituents. One can read off
an equation for the S-matrices from figure \ref{fig:bootstrap}, showing
the equivalence of scatterings before and after the fusion due to
shifting the trajectory of particle $l$ by the symmetry of higher
spin charges:
\begin{equation}
S_{lk}(\theta)=S_{li}(\theta+i\bar{u}_{ik}^{j})S_{lj}(\theta-i\bar{u}_{jk}^{i}),\label{eq:bootstrap}
\end{equation}
which argument is now rather formal - one should treat this as an
axiom. With the help of this equation one can determine the sets of
parameters in \eqref{eq:Smxproduct}.

We can know sketch the steps for solving a model by bootstrap. With
an initial guess on the S-matrix $S_{aa}$ of the particle with the
lowest mass $m_{a}$ in the theory one applies equation \eqref{eq:bootstrap}
\begin{align*}
S_{ab}(\theta) & =S_{aa}(\theta+i\bar{u}_{ab}^{a})S_{aa}(\theta-i\bar{u}_{ab}^{a})\\
S_{bb}(\theta) & =S_{ab}(\theta+i\bar{u}_{ab}^{a})S_{ab}(\theta-i\bar{u}_{ab}^{a}),
\end{align*}
where we assumed that a pole of yet unknown origin in the physical
strip of $S_{aa}$ is at a fusion angle $u_{aa}^{b}$ for a new particle
$b$. The physical poles in the new $S_{ab},S_{bb}$ functions demand
an explanation in terms of new bound states $c,d,\ldots$.\footnote{This may become complicated; higher order poles can appear and some
of the singularities can explained by anomalous thresholds only. } By repeating the steps for these, poles in their scattering matrices
may correspond to already known particles, or new ones. If at the
end the procedure closes in a consistent way, our initial guess was
appropriate.

For example, it may happen, that already $b=a$, and then two $a$-type
particle fuses into one. The fusion angle is necessarily $u_{aa}^{a}=\frac{2\pi}{3}$.
The simplest example for this, if we take \eqref{eq:shGSmatrix} at
the value $\alpha=-\sin(\frac{2\pi}{3})$. This is the S-matrix of
the single particle in the so-called Lee-Yang model \cite{cardymussardo}, which we will
use later in this work for calculations.

\section{Form factors}

A quantum field theory is considered to be solved if we have not only
the S-matrix, but also the correlation functions of local operators
at hand. These can be expressed in the spectral representation as
an infinite sum. The 2-point function of a local (hermitian) operator
$\mathcal{O}$ looks as (for a model with only a single particle type
for simplicity):
\begin{align}
 & \langle\mathcal{O}(t,x)\mathcal{O}(0,0)\rangle=\label{eq:correlationfunction}\\
= & \sum_{N=0}^{\infty}\int\frac{d\theta_{1}\ldots d\theta_{N}}{N!(2\pi)^{N}}\langle0\vert\mathcal{O}(0,0)\vert\theta_{1},\ldots,\theta_{N}\rangle_{\mathrm{in}}{}_{\mathrm{in}}\langle\theta_{N},\ldots,\theta_{1}\vert\mathcal{O}(0,0)\vert0\rangle e^{-iE_{N}t+ip_{N}x}.\nonumber 
\end{align}

where
\[
\begin{pmatrix}E_{N}\\
p_{N}
\end{pmatrix}=\begin{pmatrix}\sum_{i=1}^{N}m\cosh\theta_{i}\\
\sum_{i=1}^{N}m\sinh\theta_{i}
\end{pmatrix}
\]
is the 2-momentum of the $N$-particle ``in''-state. For each term
in this sum an exact expression may be available due to the factorization
of S-matrices, and it has good convergence properties - understood
rather as a series in Euclidean field theory. Since the Euclidean
version is rotationally invariant, the imaginary time direction is
equivalent to any other, and so 
\[
\langle\mathcal{O}(r,0)\mathcal{O}(0,0)\rangle_{E}=\langle\mathcal{O}(x_{1},x_{2})\mathcal{O}(0,0)\rangle_{E}
\]
 if $r=\sqrt{x_{1}^{2}+x_{2}^{2}}$. The exponent in \eqref{eq:correlationfunction}
then gets replaced by $e^{-mr\sum_{i=1}^{N}\cosh\theta_{i}}$, where
$e^{-mr}$ is clearly a small expansion parameter for the IR ($r\to\infty$)
limit. The understanding of the behaviour of the sum in the UV ($r\to0$)
limit is more involved (see \cite[p. 700]{mussardo}).

Introducing now a formal notation will be beneficial for practical
purposes. Note that an ordering on the ``in'' and ``out'' energy-momentum
eigenstates is natural,
\begin{align*}
\vert\theta_{1},\ldots,\theta_{N}\rangle_{i_{1}\ldots i_{N}}=\begin{cases}
\vert\theta_{1},\ldots,\theta_{N}\rangle_{i_{1}\ldots i_{N},\mathrm{in}} & \mathrm{if}\;\theta_{1}>\ldots>\theta_{N}\\
\vert\theta_{1},\ldots,\theta_{N}\rangle_{i_{1}\ldots i_{N},\mathrm{out}} & \mathrm{if}\;\theta_{1}<\ldots<\theta_{N}
\end{cases},
\end{align*}
since in this way the rapidities in the kets appear in the same order
as the wave packets of particles will take their positions on the
$x$-axis for asymptotical times.\footnote{In this case we should integrate only over domains where rapidities
are ordered in \eqref{eq:correlationfunction}, and drop ``$N!$''
.}

To define (formally) the state $\vert\theta_{1},\ldots,\theta_{N}\rangle_{i_{1}\ldots i_{N}}$
 for a general order of the arguments we need the following consideration:
since we are in a theory of massive particles, interactions are only
short-ranged; and so - apart from times at which scatterings happen
- the particles are moving freely. These intermediate states are then
similar to the asymptotical ones, differing from them by the 2-particle
scattering phases for those collisions which already happened (or
about to happen). Their definition is then completed by requiring
\begin{equation}
\vert\ldots,\theta_{k},\theta_{k+1},\ldots\rangle_{\ldots i_{k}i_{k+1}\ldots}=S_{i_{k}i_{k+1}}(\theta_{k}-\theta_{k+1})\vert\ldots,\theta_{k+1},\theta_{k},\ldots\rangle_{\ldots i_{k+1}i_{k}\ldots},\label{eq:perm_on_state}
\end{equation}
for switching some neighboring arguments - where only the states with
fully ordered rapidities are asymptotical.\footnote{It is possible to promote this formal notation and construct an algebra
of creation and annihilation operator-like objects called the Zamolodchikov-Faddeev
algebra. In addition to their canonical commutation relation, by changing
the order of the creators or annihilators among themselves we pick
up a scattering phase ($\theta_{12}=\theta_{1}-\theta_{2}$)
\begin{align*}
A_{i}(\theta_{1})A_{j}(\theta_{2}) & =S_{ij}(\theta_{12})A_{j}(\theta_{2})A_{i}(\theta_{1})\\
A_{i}(\theta_{1})A_{j}^{\dagger}(\theta_{2}) & =S_{ij}(\theta_{21})A_{j}^{\dagger}(\theta_{2})A_{i}(\theta_{1})+2\pi\delta_{ij}\delta(\theta_{1}-\theta_{2}).
\end{align*}
One gets the different states by acting on the vacuum with creation
operators $\vert\theta_{1},\ldots,\theta_{N}\rangle_{i_{1}\ldots i_{N}}\equiv A_{i_{1}}^{\dagger}(\theta_{1})\ldots A_{i_{N}}^{\dagger}(\theta_{N})\vert0\rangle$
\cite[p. 690]{mussardo}.}

In $n$-point functions of operators (after multiple insertions) quantities
like
\begin{equation}
F_{j_{1}\ldots j_{M},i_{1}\ldots i_{N}}^{\mathcal{O}}(\theta_{1}^{'},\ldots,\theta_{M}^{'}\vert\theta_{1},\ldots,\theta_{N})={}_{\mathrm{out},j_{1}\ldots j_{M}}\langle\theta_{1}^{'},\ldots,\theta_{M}^{'}\vert\mathcal{O}(0,0)\vert\theta_{1},\ldots,\theta_{N}\rangle_{i_{1}\ldots i_{N},\mathrm{in}}\label{eq:FFdef}
\end{equation}
appear\footnote{Here the rapidities in the ``bra'' appear in reversed order as it
is the adjoint of the ``ket'' $\vert\theta'_{M},\ldots,\theta'_{1}\rangle_{j_{M}\ldots j_{1},\mathrm{out}}$.
In terms of the annihilation operators, $_{j_{1}\ldots j_{M}}\langle\theta_{1}^{'},\ldots,\theta_{M}^{'}\vert\equiv\langle0\vert A_{j_{1}}(\theta_{1}^{'})\ldots A_{j_{M}}(\theta_{M}^{'})$.} which one calls form factors (FF-s).\footnote{After their resemblance to momentum-space matrix elements in the theory
of strong interactions at low energies containing the information
about the (less tangible) microscopical physics.}

These are functions of the invariants (generalized Mandelstam variables)
one can combine from in- and outgoing momenta. Just like the physical
S-matrix is defined as a boundary value in the complex $s$ plane
along a branch cut; the physical value of these matrix elements is
defined similarly via Feynman prescription. \cite[sec. 3]{karowskiexact}

Then, one can make a transition from the invariants to the original
rapidity variables with appropriate complex mappings similar to \eqref{eq:mapping}.
This way - although in the definition \eqref{eq:FFdef} the arguments
are ordered reals - one achieves the analytical continuation of the
FF for complex values of the arguments - and does so for such real
values,  which are not ordered anymore as well \cite[sec. 3]{karowskiexact}.
One can even show that permuting the arguments gives exactly those
scattering phases as in \eqref{eq:perm_on_state} and then \eqref{eq:FFdef}
can be extended for matrix elements where the operator is sandwiched
between the above mentioned intermediate states, not just asymptotical
ones.

Form factors like \eqref{eq:FFdef} can be expressed in terms of the
elementary ones defined as (now already for general ordering of the
real rapidities)
\[
F_{i_{1}\ldots i_{N}}^{\mathcal{O}}(\theta_{1},\ldots,\theta_{N})=\langle0\vert\mathcal{O}(0,0)\vert\theta_{1},\ldots,\theta_{N}\rangle_{i_{1}\ldots i_{N}},
\]
via crossing relations. Within the LSZ reduction formalism one can
derive these relations (starting from \eqref{eq:FFdef} and applying
the reduction process on creation and annihilation operators of ``in''
and ``out'' asymptotical fields), where generally disconnected terms
appear for cases when some of the outgoing momenta coincides with
an ingoing one.

For the simplest case of one outgoing particle with rapidity $\theta'$,
which is assumed to be far from any ingoing rapidity except $\theta$,
the crossing looks like as \cite[app. A]{karowskiexact}:
\begin{align}
F_{j,ii_{1}\ldots i_{k}\ldots i_{N}}^{\mathcal{O}}(\theta'\vert\theta,\theta_{1},\ldots,\theta_{N}) & =2\pi\delta_{ij}\delta(\theta'-\theta)F_{i_{1}\ldots i_{N}}^{\mathcal{O}}(\theta_{1},\ldots,\theta_{N})+\label{eq:crossing}\\
 & \quad+F_{jii_{1}\ldots i_{N}}^{\mathcal{O}}(\theta'+i\pi-i\epsilon,\theta,\theta_{1},\ldots,\theta_{N})\nonumber \\
 & =2\pi\delta_{ij}\delta(\theta'-\theta)F_{i_{1}\ldots i_{N}}^{\mathcal{O}}(\theta_{1},\ldots,\theta_{N})\prod_{k=1}^{N}S_{ii_{k}}(\theta-\theta_{k})+\nonumber \\
 & \quad+F_{ii_{1}\ldots i_{N}j}^{\mathcal{O}}(\theta,\theta_{1},\ldots,\theta_{N},\theta'-i\pi+i\epsilon),\nonumber 
\end{align}
i.e. one can get two equivalent reductions. Here the $\epsilon$ prescriptions
in the connected part (the second term in each line) are consistent
with the above mentioned definition of the physical values.

\begin{figure}[h]

\centering{}\includegraphics[scale=0.5]{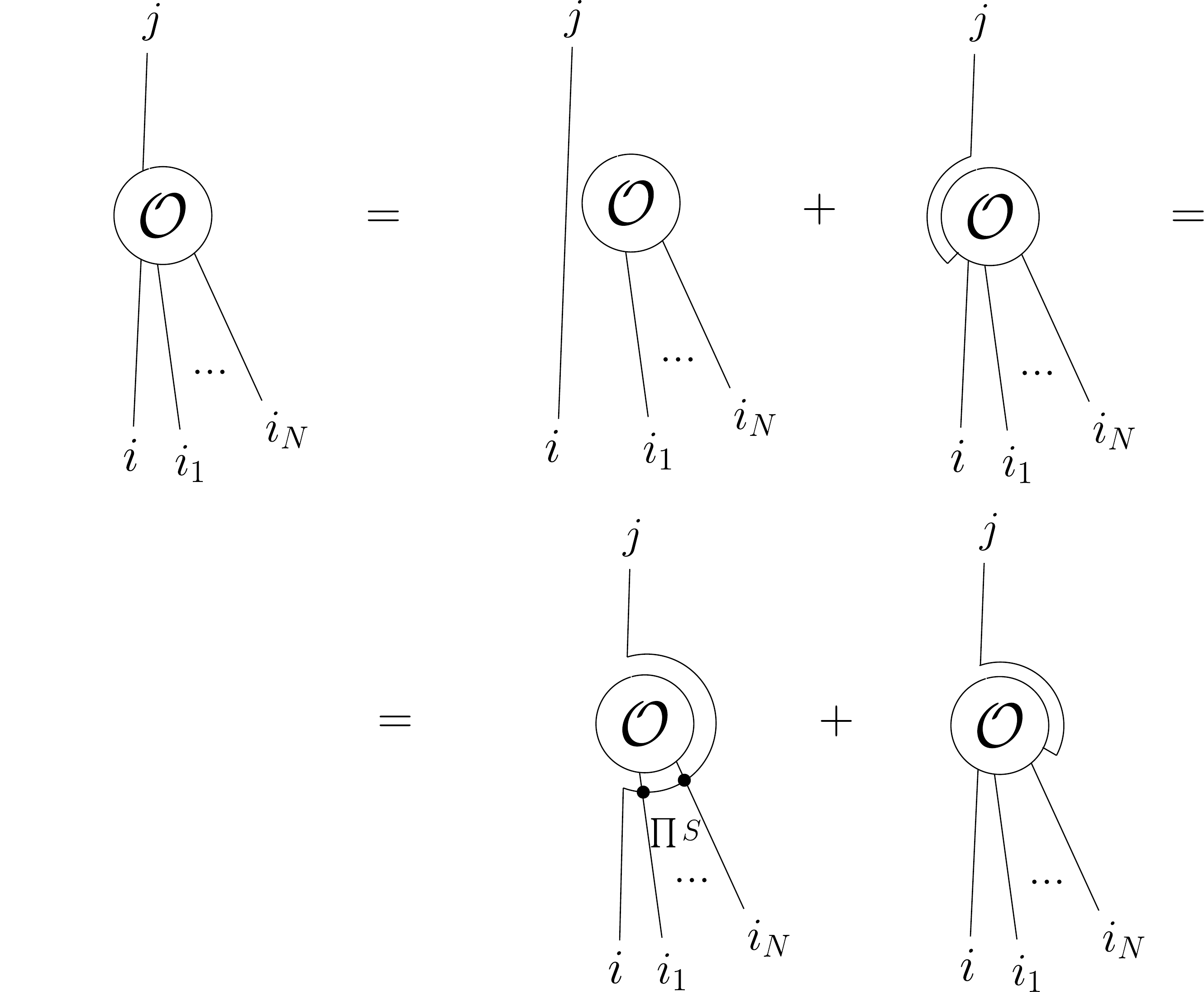}\caption{Intuitive pictorial representation of the crossing formula \eqref{eq:crossing}.
The $i\pi$ shift is indicated by turning the direction of the particle
line.}
\end{figure}

Note that since the disconnected parts differ, the equality of the
two lines above (in the distributional sense) can be satisfied only
if there are (pole) singularities in the connected parts. That is,
writing these poles as
\[
\frac{1}{\theta'-\theta\pm i\epsilon}=\mathcal{P}\frac{1}{\theta'-\theta}\mp i\pi\delta(\theta'-\theta),
\]
one can derive their residue from the equality \eqref{eq:crossing}
by collecting terms proportional to $\delta(\theta'-\theta)$. These
are the so-called kinematical poles, they appear in elementary FF-s
whenever the difference of two arguments is $i\pi$.

By requiring maximal analyticity again it is assumed that - apart
from the kinematical ones - every singularity in FF-s comes from a
pole of the S-matrix, thus all of them has physical origin.

The analytical properties of the elementary FF-s are summarized by
relations called the form factor axioms \cite[ch. 9]{mussardo1992}:
\begin{enumerate}
\item Permutation axiom\footnote{Comes directly from \eqref{eq:perm_on_state}}
\begin{equation}
F_{i_{1}\ldots i_{k}i_{k+1}\ldots i_{N}}^{\mathcal{O}}(\theta_{1},\ldots,\theta_{k},\theta_{k+1},\ldots,\theta_{N})=S(\theta_{k}-\theta_{k+1})F_{i_{1}\ldots i_{k+1}i_{k}\ldots i_{N}}^{\mathcal{O}}(\theta_{1},\ldots,\theta_{k+1},\theta_{k},\ldots,\theta_{N})\label{eq:PERM}
\end{equation}

\begin{center}
\includegraphics[scale=0.5]{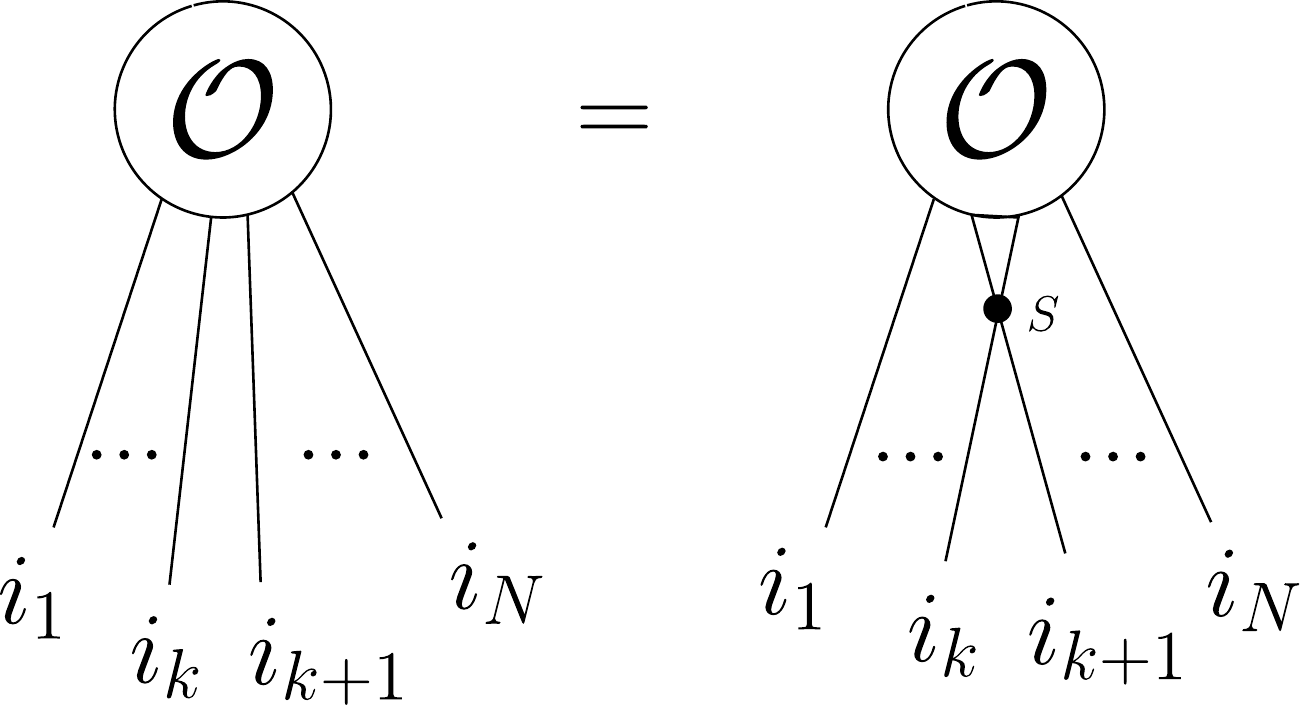}
\par\end{center}
\item Periodicity axiom\footnote{By comparing the connected parts in \eqref{eq:crossing} far from
the $\theta'-\theta$ pole one arrives at $F_{j\ldots}^{\mathcal{O}}(\theta'+i\pi,\ldots)=F_{\ldots j}^{\mathcal{O}}(\ldots,\theta'-i\pi)$.
By $\theta'\to\theta'+i\pi$ we find the form of the relation presented
in the axiom. }:
\begin{equation}
F_{i_{1}i_{2}\ldots i_{N}}^{\mathcal{O}}(\theta_{1}+2i\pi,\theta_{2},\ldots,\theta_{N})=F_{i_{2}\ldots i_{N}i_{1}}^{\mathcal{O}}(\theta_{2},\ldots,\theta_{N},\theta_{1})\label{eq:PER}
\end{equation}

\begin{center}
\includegraphics[scale=0.5]{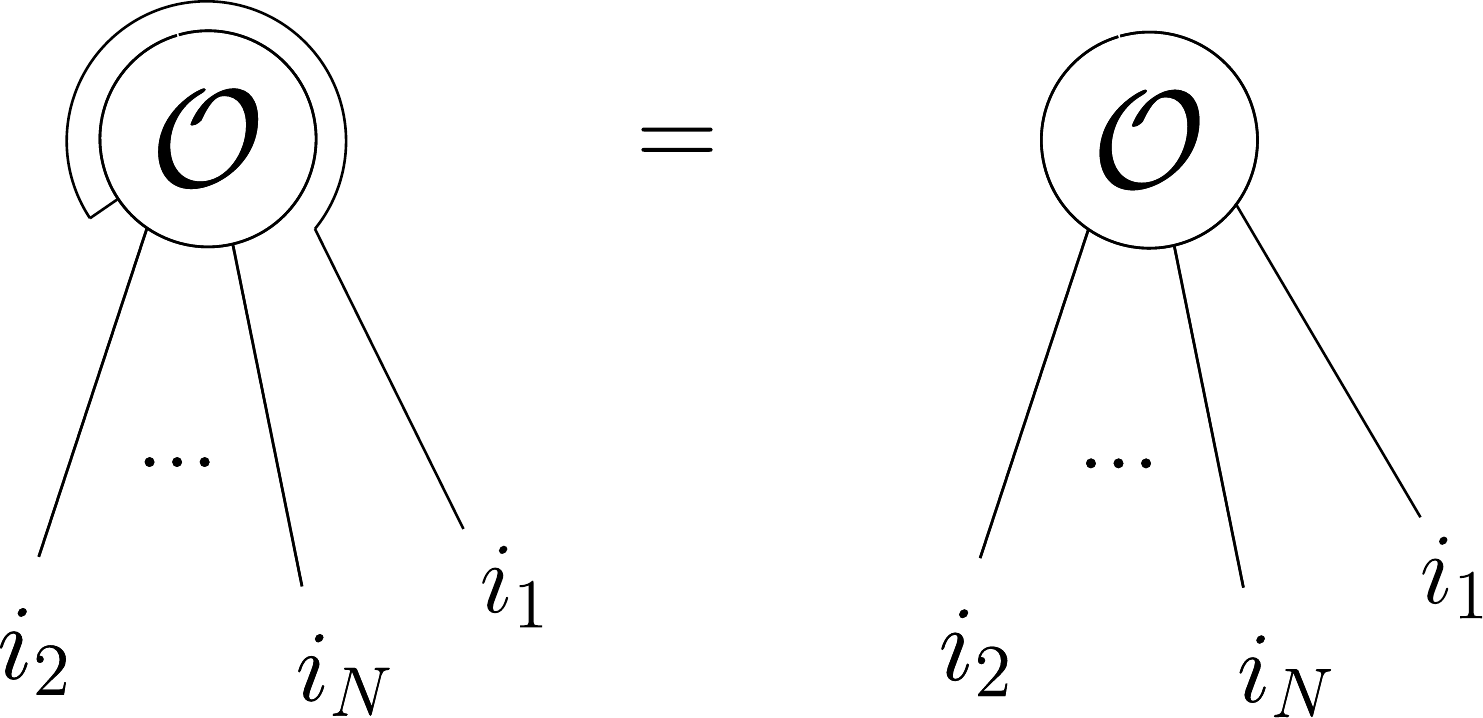}
\par\end{center}
\item Kinematical axiom\footnote{The residue comes from the equality of the two crossing relations
in the distributional sense as mentioned above.}:
\begin{equation}
\Res_{\theta'=\theta}F_{jii_{1}\ldots i_{N}}^{\mathcal{O}}(\theta'+i\pi,\theta,\theta_{1},\ldots,\theta_{N})=i\delta_{ij}\left(1-\prod_{k=1}^{N}S(\theta-\theta_{k})\right)F_{i_{1}\ldots i_{N}}^{\mathcal{O}}(\theta_{1},\ldots,\theta_{N})\label{eq:KIN}
\end{equation}

\begin{center}
\includegraphics[scale=0.5]{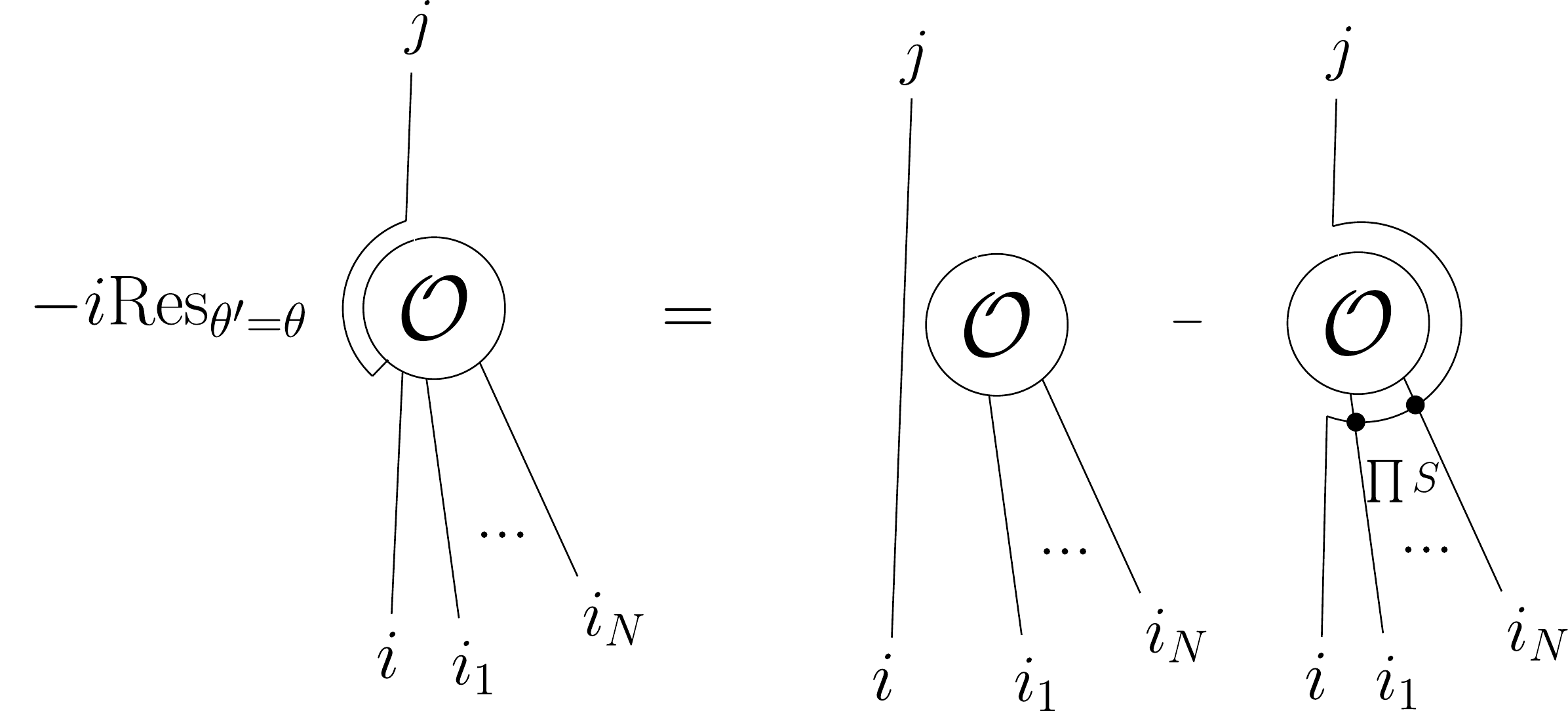}
\par\end{center}
\item Dynamical axiom\footnote{As in the case of the S-matrix, this pole appears for non-physical
imaginary momenta, where the $s$ Mandelstam variable for the particles
``$i$'' and ``$j$'' becomes $m_{k}^{2}$. The singularity comes
from the on-shell propagator, whereas the residue is the on-shell
vertex function (appearing only once for the ``fusion'') times the
form factor containing the particle ``$k$''. Note that this is
the simplest case: for higher order poles of the S-matrix the corresponding
singularity in the form-factor is also higher order.}:

If the S-matrix has a simple bound state pole with residue 
\[
\Res_{\theta=iu_{ij}^{k}}S(\theta)=i\Gamma_{ijk}^{2},
\]
the form factor inherits it with residue $i\Gamma_{ijk}$, since only
the fusion part of the process takes place: 
\begin{equation}
\Res_{\theta'=\theta}F_{iji_{1}\ldots i_{N}}^{\mathcal{O}}(\theta'+i\bar{u}_{ik}^{j},\theta-i\bar{u}_{jk}^{i},\theta_{1},\ldots,\theta_{N})=i\Gamma_{ijk}F_{ki_{1}\ldots i_{N}}^{\mathcal{O}}(\theta,\theta_{1},\ldots,\theta_{N})\label{eq:DYN}
\end{equation}

\begin{center}
\includegraphics[scale=0.5]{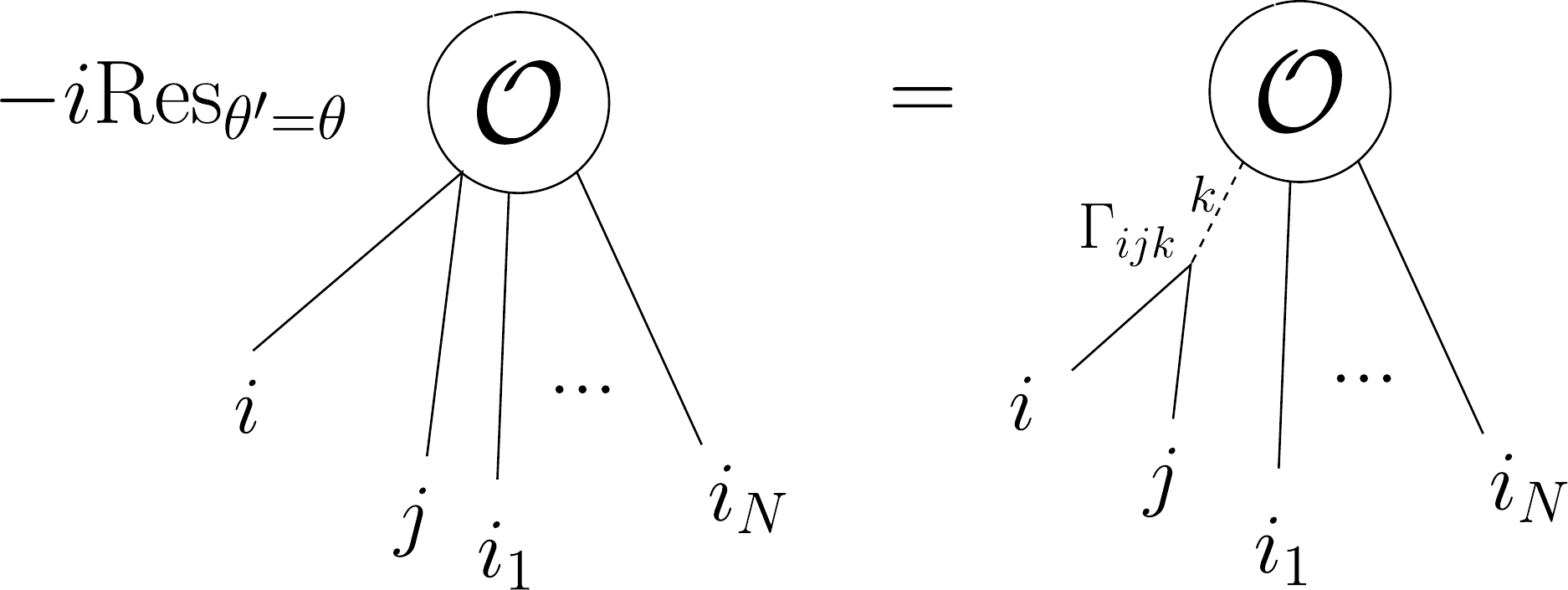}
\par\end{center}
\item Lorentz transformation:
\begin{equation}
F_{i_{1}i_{2}\ldots i_{N}}^{\mathcal{O}}(\theta_{1}+\Lambda,\theta_{2}+\Lambda,\ldots,\theta_{N}+\Lambda)=e^{s\Lambda}F_{i_{1}i_{2}\ldots i_{N}}^{\mathcal{O}}(\theta_{1},\theta_{2},\ldots,\theta_{N}),\label{eq:LOR}
\end{equation}
where the operator $\mathcal{O}$ transforms with spin $s$.
\end{enumerate}
Note that the last property means that for scalar ($s=0$) operators
the form factor is a function of the differences of its arguments
$\theta_{kl}=\theta_{k}-\theta_{l}$, since we can take $\Lambda=-\theta_{j}$
for any $j$. From now on, we make a restriction for scalar operators
in this work. 

One can try to find exact solutions satisfying these axioms by using
an ansatz provided by the first two axioms. For a model with a single
particle type this looks as\footnote{The index $N$ of the form factor $F_{N}^{\mathcal{O}}$ now refers
to the number of particles rather than the types.}  \cite{karowskiweisz1978}
\[
F_{N}^{\mathcal{O}}(\theta_{1},\ldots,\theta_{N})=\prod_{i<j}F_{\mathrm{min}}(\theta_{ij})K_{N}^{\mathcal{O}}(\theta_{1},\ldots,\theta_{N}),
\]
where the $F_{\mathrm{min}}(\theta)$ minimal form factor satisfies
the permutation and periodicity axioms for a two-particle form factor
$F_{2}^{\mathcal{O}}(\theta_{1},\theta_{2})$, with $\theta=\theta_{1}-\theta_{2}$:
\begin{align*}
F_{\mathrm{min}}(\theta) & =S(\theta)F_{\mathrm{min}}(-\theta)\\
F_{\mathrm{min}}(i\pi+\theta) & =F_{\mathrm{min}}(i\pi-\theta),
\end{align*}
and has no poles and zeros in the physical strip $0\leq\im\theta\leq\pi$. 

Substituting the ansatz back to these axioms shows that the $K_{N}^{\mathcal{O}}$
functions satisfy permutation and periodicity with $S=1$, and so
they are symmetric and $2i\pi$-periodic in the $\theta_{1},\ldots,\theta_{N}$
variables; therefore being symmetric functions of the variables $x_{k}=e^{\theta_{k}}$.
They also contain all the poles, and since the positions of the poles
do not depend on the operator we can write
\[
K_{N}^{\mathcal{O}}=\frac{Q_{N}^{\mathcal{O}}}{D_{N}},
\]
where the denominator is only a model dependent polynomial responsible
for the poles, and $Q_{N}^{\mathcal{O}}$ is another one containing
the information on the operator.

The kinematical axiom relates $N$-particle form factors with $(N+2$)-particle
ones, and so does the dynamical axiom for $(N+1)$-particle FF-s.
These translate into recursive equations for the $Q_{N}^{\mathcal{O}}$
functions, starting from $N=1$ \cite{mussardo}. The starting point
$Q_{1}^{\mathcal{O}}$ for these recursions depends on the operator,
and can be deduced from physical arguments.

\nomenclature{e.o.m}{equation of motion}

\nomenclature{d.o.f.}{degrees of freedom}

\nomenclature{FF}{form factor}

\nomenclature{BY}{Bethe-Yang}

\nomenclature{TBA}{Thermodynamical Bethe Ansatz}

\nomenclature{LO}{leading order}

\nomenclature{LY}{Lee-Yang}

\nomenclature{SLYM}{Scaling Lee-Yang Model}

\lhead[\chaptername~\thechapter]{\rightmark}

\rhead[\leftmark]{}

\lfoot[\thepage]{}

\cfoot{}

\rfoot[]{\thepage}

\chapter{Finite volume effects\label{chap:Finite-volume-effects}}

In the previous chapter we sketched the whole program for solving
an IQFT in an exact manner for the case where the one-dimensional
world this theory lived in was infinite. For reasons enumerated in
the introduction we would also like to know the solution to our theory
in a finite world. The simplest case is when the space is made into
a circle of circumference $L$.

We take the quantum field itself with periodic boundary condition,
which leads to the fact that the total momentum of a (multiparticle)
momentum-eigenstate $\vert P\rangle$ gets quantized:
\[
\phi(t,x+L)=\phi(t,x)\;\Rightarrow\;\langle0\vert e^{-i\hat{P}L}\phi(t,x)e^{i\hat{P}L}\vert P\rangle\overset{!}{=}\langle0\vert\phi(t,x)\vert P\rangle\;\Rightarrow\;e^{iPL}=1
\]
and so $PL=2\pi n$, where $n$ is integer. For one-particle states
with mass $m$ this simply means, that their momentum is.
\begin{equation}
pL=2\pi n.\label{eq:freeqc}
\end{equation}
For multiparticle systems, the quantization of the total momentum
remains the same, however the quantization conditions of individual
particles get modified slightly compared to a 1-particle state due
to the scatterings on other particles and virtual processes (see later
in section \ref{sec:Asymptotical-Bethe-Ansatz}, and \ref{sec:Thermodynamical-Bethe-Ansatz}).
These corrections disappear when $L\to\infty$, and the energies will
look like that of a free particle:
\[
p_{i}L=2\pi n_{i},\quad E_{n_{1},\ldots,n_{N}}=\sum_{i=1}^{N}\sqrt{m_{i}^{2}+\left(\frac{2\pi n_{i}}{L}\right)^{2}}.
\]
The discrete spectrum will also go to a continuous one in this limit. 

\begin{figure}[h]
\begin{centering}
\includegraphics[scale=0.5]{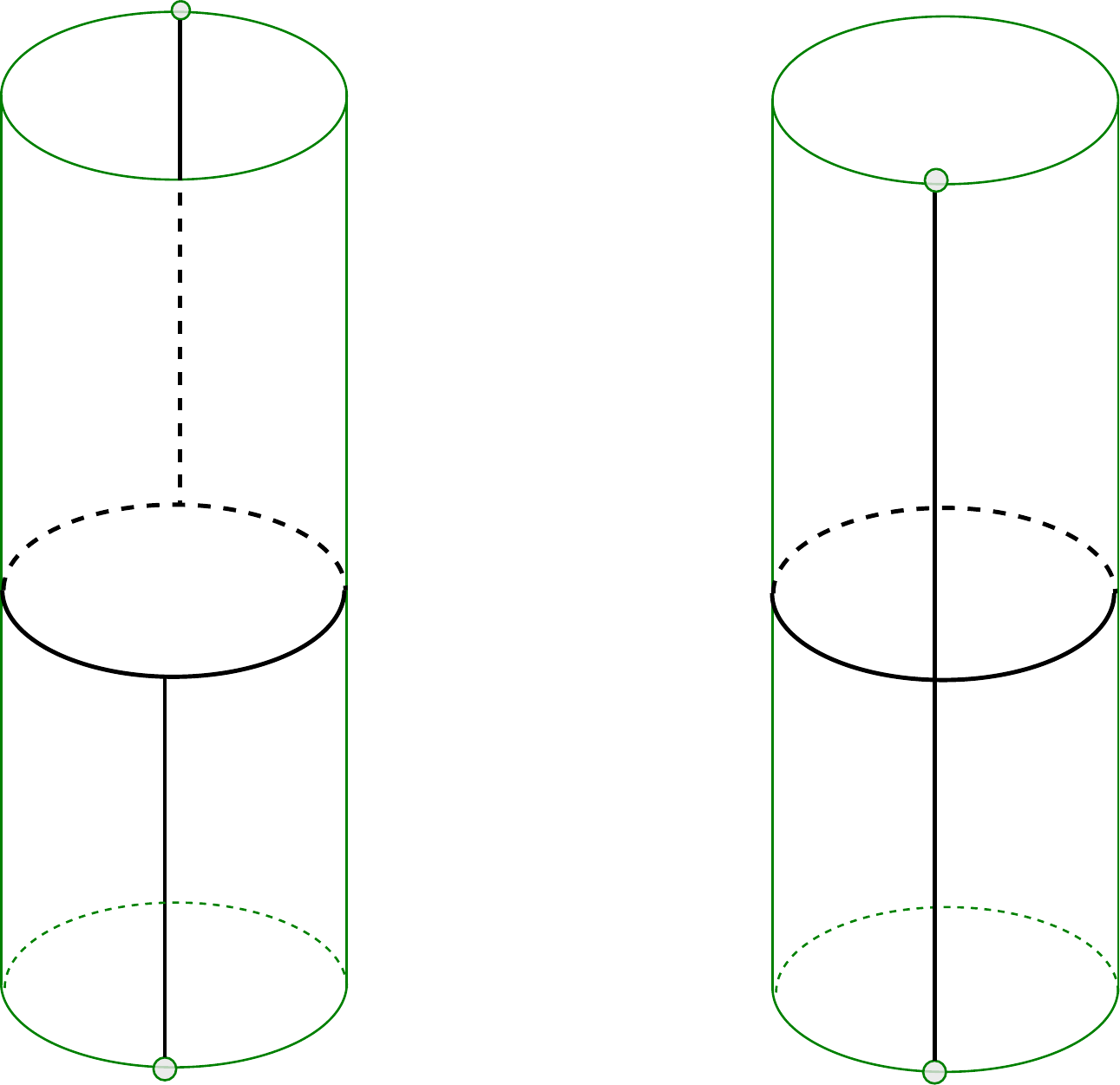}
\par\end{centering}
\caption{Left: Lüscher's $\mu$-term correction to the mass of a standing particle
comes from a virtual process where the particle disintegrates into
its virtual (mirror-)particle constituents which fuse into another
particle after wrapping the space-time cylinder. Right: The $F$-term
correction comes from another virtual process where a virtual particle-anti-particle
pair annihilates only after going around the space, whilst they scatter
on the physical particle. }

\label{fig:muF}
\end{figure}

For finite, but large $L$, the mentioned corrections become polynomial
in the variable $L^{-1}$ for the energy compared to the free case.
This is the regime of the so-called asymptotical Bethe ansatz, which
takes the effect of the particles' scattering on each other into account.
By the exact S-matrix of the theory, this already knows about those
virtual processes which present in the infinite volume theory. However,
on compactified space also topologically non-trivial virtual processes
can happen. These will be responsible for $\mathcal{O}(e^{-\mathrm{const.}\!L})$
exponential corrections\footnote{Since $e^{-\frac{1}{x}}$ is not analytic in $x$, the $L^{-1}=x$
being the small parameter in the expansion these corrections cannot
be expressed with polynomials. }, which become dominant for smaller volumes. This is the so-called
Lüscher domain. In figure \ref{fig:muF} the two type of processes
are shown, corresponding to the $\mu$- and $F$-terms introduced
in paper \cite{luscherI} for the lowest exponential mass-corrections
of the standing particle \cite[eq. 2.1, 2.2]{pozsgaymu}.

The thermodynamical Bethe ansatz presented in section \eqref{sec:Thermodynamical-Bethe-Ansatz}
sums up all these corrections coming from virtual processes in an
exact way, and so the spectrum is available for any $L$.

The volume dependence is not as well understood in case of the the
form factors yet. There are such exact results based on the TBA for
diagonal form factors (see section \ref{sec:Form-factors-in} for
their definition), however for non-diagonal ones only the polynomial,
and the $\mu$-type corrections are known. In the present work we
deal with the $F$-term correction (and its relation to the $\mu$-term)
for non-diagonal form factors which was proposed in the preprint \cite{preprint}.

The $L\to0$ limit explores the short range behaviour of the QFT,
and so corresponds to a UV fixpoint. There the theory is scale invariant
and equivalent to a conformal field theory, for which the spectrum
is well-known \cite[p. 657]{mussardo}. The original theory for finite
$L$ lies away from the fixpoint in a relevant direction. By adding
a term to the CFT action which perturbs the theory in this relevant
direction in operator space - a so-called (integrable) deformation
\cite[p. 529]{mussardo}, the corrections to the energy levels and
matrix elements of local operators (i.e. form factors) can be approximated
for any (not too large) $L$ value numerically with the so-called
truncated conformal space approach (TCSA). This gives an opportunity
to compare the above mentioned analytic results with data from the
TCSA method and  check their validity.

\section{Asymptotical Bethe Ansatz\label{sec:Asymptotical-Bethe-Ansatz}}

Let us take a diagonal scattering theory with particle types $a,b,c,\ldots$
and scattering phases\footnote{Note that $S_{ab}(0)=\pm1$ is now implied in the phase $\delta_{ab}$.}
$S_{ab}(\theta)=e^{i\delta_{ab}(\theta)}$ on a space-time cylinder
with circumference $L$.

If we have a state with $N$ particles, and this finite size $L\gg N\xi$
is large enough to allow the particles to be far from each other compared
to the interaction length $\xi\sim m_{a}^{-1}$ - characterized by
the lowest mass $m_{a}$ of the theory; then there are - as for infinite
volume - regions in the configuration space of the particles where
they are all well separated and propagating freely. Also in a diagonal
scattering the identities of the particles do not change, and they
keep their momenta.

This way one can follow what happens with a single particle along
its trajectory. Intuitively speaking, we can think of it as a quantum
mechanical particle on a circle, in the potential of the other particles
(for the proper treatment with multiparticle wavefunctions see \cite{zamolodchikovTBA}).
The periodicity condition on the wavefunction of this particle then
looks:
\begin{equation}
e^{ip_{j}L}\prod_{k\neq j}e^{i\delta_{i_{j}i_{k}}(\theta_{j}-\theta_{k})}=1,\quad i_{j},i_{k}\in\{a,b,c,\ldots\}\label{eq:expBY}
\end{equation}
where the first factor accumulates the translation in the domains
where the particle is far from the others - basically the whole circle;
and it also picks up the product of scattering phases by going through
the localized interaction regions of all the other particles. These
are the so-called Bethe-Yang equations. In their logarithmic form:
\begin{equation}
m_{j}L\sinh\theta_{j}+\sum_{k\neq j}\delta_{i_{j}i_{k}}(\theta_{j}-\theta_{k})=2\pi n_{j}\;\forall j,\quad n_{j}\in\mathbb{Z},\label{eq:logBY}
\end{equation}
the ``fermionic'' case mentioned in \ref{subsec:Analytic-properties-and}
requires $n_{j}\neq n_{k}$; the ``bosonic'' allows coinciding quantum
numbers and rapidities.

By equating the r.h.s. with the (free) one-particle momentum, the
correction is of order $p_{j}-p_{j}^{(\mathrm{free})}=\mathcal{O}(L^{-1})$,
since the scattering phases do not depend on the volume. This means
$\mathcal{O}(L^{-2})$ in the energy. Note that by summing up all
the BY equations, all the phases drop out due to unitarity $\delta_{ab}(\theta)+\delta_{ab}(-\theta)=0$,
and the total momentum quantization is restored. The energy of the
whole system is just the sum of the individual energies, expressed
with the rapidities which satisfy the BY equations
\[
E_{n_{1},\ldots,n_{N}}(L)=\sum_{j=1}^{N}m_{j}\cosh\theta_{j}(n_{1},\ldots,n_{N};L).
\]

\begin{figure}[h]
\begin{centering}
\includegraphics{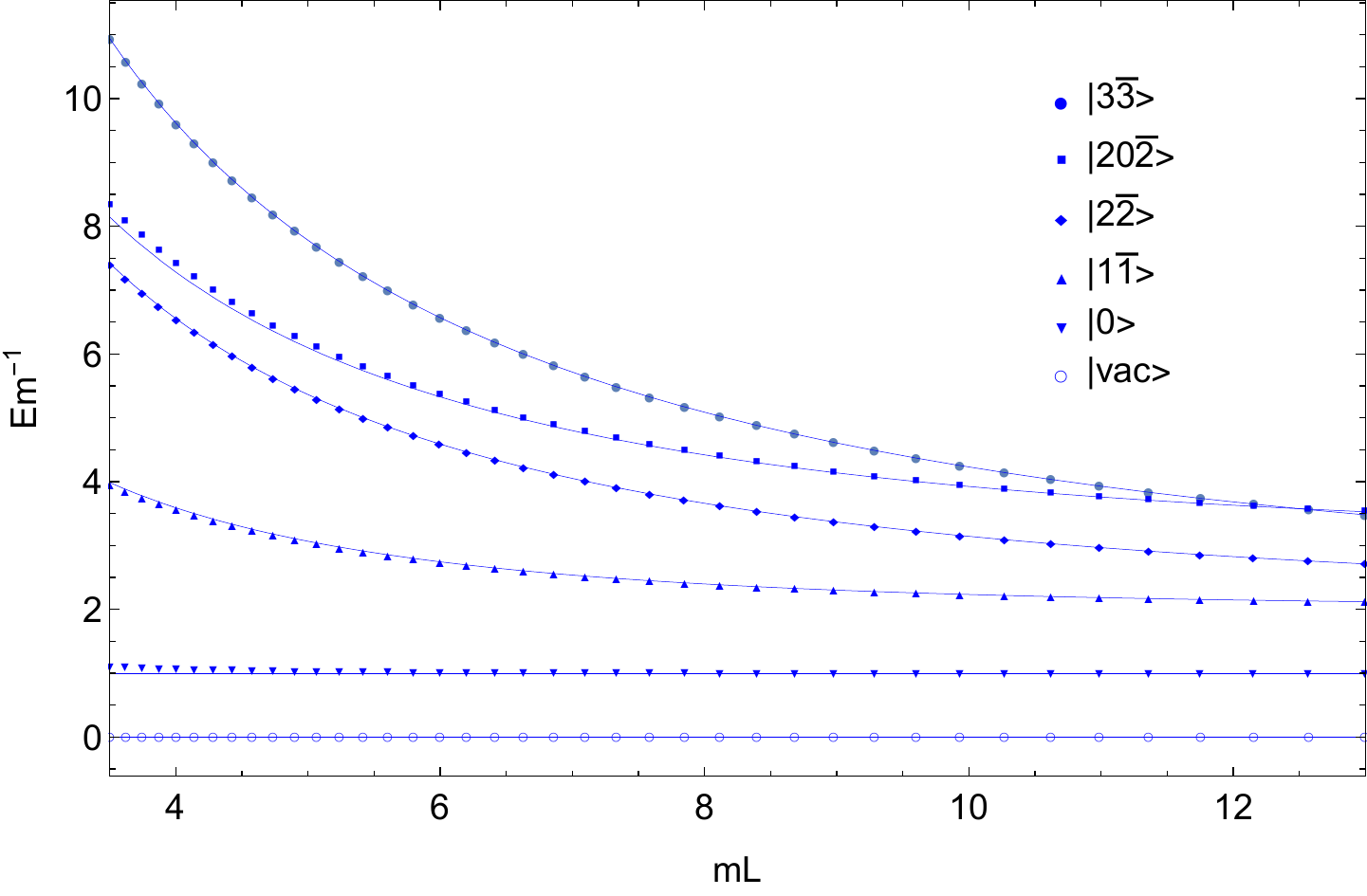}\caption{The energy spectrum of the scaling Lee-Yang model. The discrete points
represent the result of the TCSA method mentioned in the introduction
of this chapter (i.e. the exact energy values, since the numerical
uncertainty is small \cite{preprint}), while the solid lines are
the corresponding energies obtained from the solution to the BY equations
at every $L$ for one, two- and three-particle states with quantum
numbers $n_{1}=0\leftrightarrow\vert0\rangle,\quad n_{1},n_{2}=1,-1\leftrightarrow\vert1\bar{1}\rangle,\quad\mathrm{etc.}$,
and total momentum zero. The difference of the two is $\protect\Ordo(e^{-\mathrm{const.}L})$
and in this range it is small, therefore we subtract this polynomial
BY approximation of the spectrum in the following, to visualize and
compare the several exponential corrections which we are curious about.}
\par\end{centering}
\label{LYspectrum}
\end{figure}

\section{Thermodynamical Bethe Ansatz\label{sec:Thermodynamical-Bethe-Ansatz}}

In this section we first calculate the exact ground state energy of
our diagonally scattering model with the help of the asymptotical
Bethe ansatz, and then show some exact solutions for the excited states
with particle content for specific models. For simplicity, we consider
only a single particle type with mass $m$ and S-matrix $S(\theta)$.

\subsection{Mirror model and the ground state energy\label{subsec:Mirror-model-and}}

\begin{figure}[h]

\begin{centering}
\includegraphics[scale=0.33]{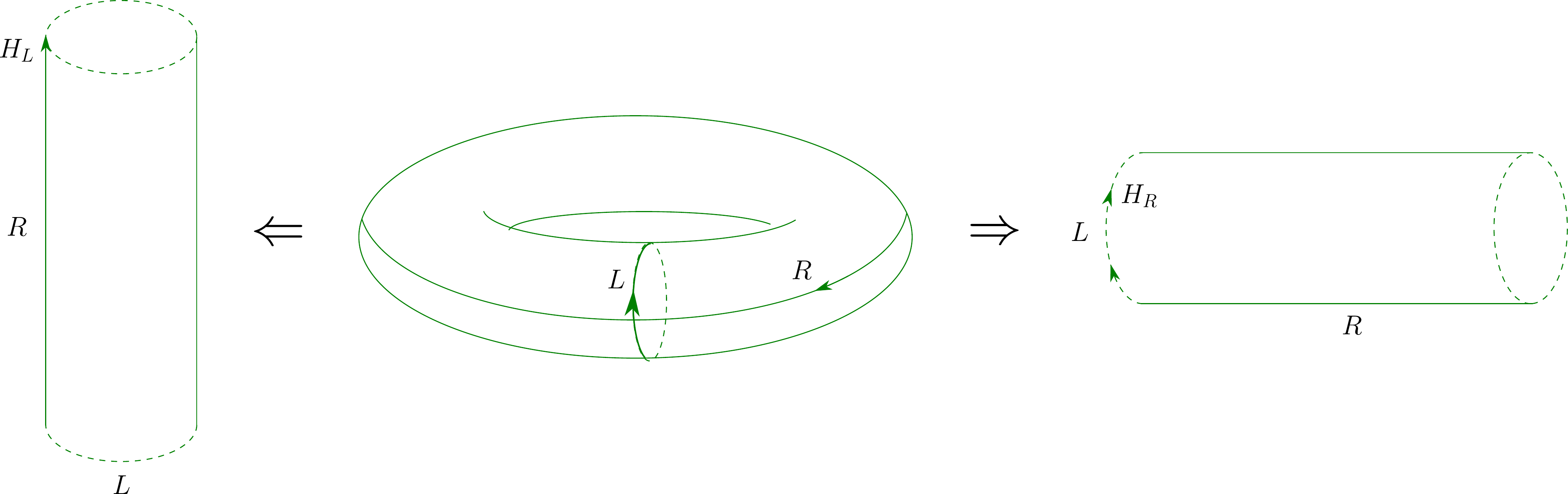}\caption{The original and the mirror model from the finite volume finite temperature
Euclidean theory. The arrows show the direction in which the Hamiltonians
$H_{L},H_{R}$ generate the time translation.}
\par\end{centering}
\label{fig:torus}
\end{figure}

In the Euclidean field theory which is equivalent to our 1+1 dimensional
QFT the time and space direction is equivalent due to rotational symmetry.
If we take the theory on a torus with circumferences $R\gg L$ (see
Fig. \ref{fig:torus}), we can quantize along either of the two circles,
and get the same partition function:\footnote{According to which circle we choose to be the Euclidean time direction,
it is called $R$- and $L$-channel quantization. }
\begin{equation}
Z(R,L)=\Tr(e^{-RH_{L}})=\Tr(e^{-LH_{R}}),\label{eq:partitionfunc}
\end{equation}

where the first form corresponds to a system of size $L$ with Hamiltonian
$H_{L}$ at inverse temperature $R$, and the second to another system
with the roles of $L$ and $R$ exchanged. The $R\to\infty$ limit
then describes two different Minkowskian theories after Wick-rotating
back with respect to the two different choices of the imaginary time
axis.\footnote{For the correct form of these Wick-rotations see \cite{tongeren}}

The one corresponding to the first trace of \eqref{eq:partitionfunc}
is a zero-temperature QFT in a finite volume of size $L$, whereas
the other is a thermal QFT with inverse temperature $\beta\equiv T^{-1}\equiv L$
in infinite volume - the so-called mirror model. The partition function
in the limit gives
\begin{equation}
Z(R,L)\overset{R\to\infty}{\longrightarrow}e^{-RE_{0}(L)}\label{eq:groundstateZ}
\end{equation}
for the first one and so the ground state energy we need will become
available to us if we are able to calculate $Z(R,L)$ from the second
trace. This can be achieved by the BY equations\footnote{Now we use them for the mirror model (!)},
since in the thermodynamical limit both the spatial extent $R$ of
the mirror model and the number of particles $N$ goes to infinity
whilst their ratio is constant; and if this ratio $R/N\gg\xi=m^{-1}$,
then the arguments of the previous section hold. 

As one can see from the free quantization condition easily, the smallest
rapidity difference (for changing the quantum number by one) decreases
as $\propto(mR)^{-1}$ and so in the limit one can practically define
a continuous density $\rho(\theta)$ as the number of particles in
some interval $[\theta,\theta+\Delta\theta]$.

The scattering phase sum in \eqref{eq:logBY} becomes a convolution:
\begin{align}
2\pi n_{j} & =mR\sinh\theta_{j}+\int_{-\infty}^{\infty}d\theta'\delta(\theta_{j}-\theta')\rho(\theta')\label{eq:convBY}
\end{align}
where $\delta(\theta)=-i\ln S(\theta)$.

Of course not every level - which is a solution to the BY equations
- is occupied by a particle, therefore one can introduce the density
of ``holes'' (unoccupied levels) $\bar{\rho}(\theta)$. The number
of particles and holes together is the number of levels in the above
mentioned interval, and so an infinitesimal version of \eqref{eq:convBY}
can be derived:
\begin{equation}
2\pi(\rho(\theta)+\bar{\rho}(\theta))=mR\cosh\theta+\int_{-\infty}^{\infty}d\theta'\varphi(\theta-\theta')\rho(\theta'),\;\varphi(\theta)\equiv\partial_{\theta}\delta(\theta),\label{eq:TBAconstraint}
\end{equation}
which gives a constraint among the densities.

Many microscopic realizations of the densities $\rho,\bar{\rho}$
are possible in the interval of length $\Delta\theta$, namely we
can choose $\rho\Delta\theta$ levels from $(\rho+\bar{\rho})\Delta\theta$
to be occupied in $\Omega$ number of ways if we stick to the ``fermionic''
case where levels can be occupied only once: 
\[
\Omega=\frac{[(\rho+\bar{\rho})\Delta\theta]!}{[\rho\Delta\theta]![\bar{\rho}\Delta\theta]!}\;\Rightarrow\;S[\rho,\bar{\rho}]=\int_{-\infty}^{\infty}d\theta\{(\rho+\bar{\rho})\ln(\rho+\bar{\rho})-\rho\ln\rho-\bar{\rho}\ln\bar{\rho}\}.
\]
The $S[\rho,\bar{\rho}]$ entropy functional sums up the contributions
from all intervals. From statistical physics we know that the most
probable configuration of the densities minimizes the free energy
function
\[
F[\rho,\bar{\rho}]=E[\rho]-\overbrace{T}^{1/L}S[\rho,\bar{\rho}];\quad E[\rho]=m\int_{-\infty}^{\infty}d\theta\cosh\theta\rho(\theta)
\]
of the system, and due to the sharpness of the probability distribution
around this minimizing configuration in the partition function only
$F[\rho_{\min},\bar{\rho}_{\min}]$ gives the dominant contribution
\begin{equation}
Z(R,L)\overset{R\to\infty}{\longrightarrow}Z(\beta)=e^{-\overbrace{\beta}^{L}\overbrace{F[\rho_{\min},\bar{\rho}_{\min}]}^{Rf(L)}},\label{eq:mirrorrmodelZ}
\end{equation}
where $f(L)$ is the free energy density per unit length of the mirror
model at temperature $L^{-1}$.

By matching \eqref{eq:groundstateZ} and \eqref{eq:mirrorrmodelZ}
we get
\[
E_{0}(L)=Lf(L).
\]
To calculate $\rho_{\min}$ and $\bar{\rho}_{\min}$ one should vary
$F[\rho,\bar{\rho}]$ while $\rho$ and $\bar{\rho}$ satisfies the
constraint \eqref{eq:TBAconstraint}. The solution can be expressed
in a simple way if one introduces a new function $\epsilon(\theta)$,
the so-called pseudoenergy as $\rho(\theta)/\bar{\rho}(\theta)=e^{-\epsilon(\theta)}$
\begin{align}
\epsilon(\theta) & =mL\cosh\theta-\int_{-\infty}^{\infty}\frac{d\theta'}{2\pi}\varphi(\theta-\theta')\ln(1+e^{-\epsilon(\theta')})\label{eq:GSpseudoenergy}\\
E_{0}(L) & =-m\int_{-\infty}^{\infty}\frac{d\theta}{2\pi}\cosh\theta\ln(1+e^{-\epsilon(\theta)}),\label{eq:GSEnergy}
\end{align}
where the first line is the minimum condition reformulated - the (ground
state) TBA equation. This is a non-linear integral equation for $\epsilon(\theta)$,
and its solution - substituted back to the minimal free energy - gives
$E_{0}(L)$ as an integral shown above. 

In the large volume expansion of \eqref{eq:GSpseudoenergy}, the first
term is $\propto L$, and so $e^{-\epsilon(\theta)}$ is a small parameter.
By using $\ln(1+x)\approx x$, the leading order to \eqref{eq:GSpseudoenergy}
comes by dropping the integral term completely: $\epsilon^{(0)}(\theta)=mL\cosh\theta$;
and then the ground state energy is:
\begin{equation}
E_{0}(L)\approx-\frac{1}{2\pi}\int_{-\infty}^{\infty}d\left(m\sinh\theta\right)e^{-mL\cosh\theta}=-\frac{1}{2\pi}\int_{-\infty}^{\infty}dpe^{-LE(p)}.\label{eq:OrdomL}
\end{equation}
This can be calculated directly from the trace of the mirror model
also, where from the sum over the complete system of states we keep
only the vacuum and the one-particle ones (due to large $L$):
\[
Z(R,L)=\langle0|e^{-LH_{R}}\vert0\rangle+\sum_{n}\langle p_{n}\vert e^{-LH_{R}}\vert p_{n}\rangle+\ldots=1+\sum_{n}e^{-LE_{n}(R)}+\ldots
\]
The sum becomes an integral $\sum_{n}\rightarrow R\int\frac{dp}{2\pi}$
in the $R\to\infty$ limit, and by
\[
E_{0}(L)=-\lim_{R\to\infty}\frac{\ln Z(R,L)}{R}
\]
we get the same expression as before. The integral in \eqref{eq:OrdomL}
behaves as $\propto e^{-mL}\sqrt{m/L}$ for large volumes. The interpretation
of this leading exponential $\mathcal{O}(e^{-mL})$ correction is
that a pair of virtual (mirror-)particles appear by vacuum polarization,
and one of them goes around the $L$-sized circle before getting annihilated
by the other \cite{bajnokthesis}.

\subsection{Excited states\label{subsec:Excited-states}}

The mirror model trick in the previous section worked only for the
ground state energy. But we know that the BY equations - for the original
model - give a good description of excited states with real particle
content for asymptotical $L$ volumes. To proceed, we need a similar
set of equations like \eqref{eq:GSpseudoenergy} and \eqref{eq:GSEnergy}
which sum up the effects of virtual processes, and in the BY regime
match the expected polynomial corrections.

The excited state TBA equations can be obtained by an analytic continuation
in the $L$ parameter. The reason is that on the $L$ complex plane
there are branch points in $E_{0}(L)$, and by going around them to
different sheets and then back to real $L$ values one can reach the
higher energy levels.\footnote{A simplified but detailed explanation on this can be found in \cite{tongeren}.}
During the continuation, a singularity of $\ln(1+e^{-\epsilon(\theta)})$
in the $\theta$ plane crosses the real axis - and so the integration
contour of \eqref{eq:GSEnergy}. Continuity requires the contour to
be deformed around the crossing singularity; nevertheless we get the
same result by separating the residue - with the correct sign - and
taking the integral over the real axis again. The singularities which
are important to us happen where the logarithm has a zero. The residue
for this type of singularity is understood after a partial integration
(the boundary term is assumed to be zero):
\[
g(z_{0})=0\;\Rightarrow\;\int_{-\infty}^{\infty}dzf'(z)\ln g(z)=-\int_{-\infty}^{\infty}f(z)\frac{g'(z)}{g(z)}\;\Rightarrow\;\Res_{z=z_{0}}f(z)\frac{g'(z)}{g(z)}=f(z_{0}),
\]
where the $f,g$ are complex functions. One concludes that the value
of the residue does not depend on $g$ since its derivative drops
out. The role of $g(z)$ is played by $1+e^{-\epsilon(\theta)}$,
and that of $f(z)$ by $\sinh\theta$ in \eqref{eq:GSEnergy}, and
by $\delta(\theta)=-i\ln S(\theta)$ in \eqref{eq:GSpseudoenergy}.
Let us assume that the position of the singularities after the continuation
in the sinh-Gordon theory are at $\bar{\theta}_{j}+\frac{i\pi}{2},\;\bar{\theta}_{j}\in\mathbb{R}$,
where $\{\bar{\theta}_{j}\}_{j=1}^{N}$ are the rapidities\footnote{Not to be confused with notation $\bar{u}_{ij}^{k}=\pi-u_{ij}^{k}$
for fusion angles. The bar means here instead, that $\bar{\theta}$-s
are the solutions to the quantization conditions. } of the real particles in the $N$-particle excited state \cite{teschner}.
After separating the residues the continued TBA equation will look
like:
\begin{align}
\epsilon(\theta\vert\{\theta\}) & =mL\cosh\theta+\sum_{j}\ln S(\theta-\theta_{j}-\frac{i\pi}{2})-\label{eq:shGpseudoenergy}\\
 & \quad-\int_{-\infty}^{\infty}\frac{d\theta'}{2\pi}\varphi(\theta-\theta')\ln(1+e^{-\epsilon(\theta'|\{\theta\})}),\nonumber 
\end{align}
where the $\{\theta\}\equiv\{\theta_{j}\}_{j=1}^{N}$ are ``dummy''
variables whose introduction as temporary substitutes for $\bar{\theta}_{j}$
will clarify the notation later in this work.

The positions of the $\bar{\theta}_{j}$ rapidities are determined
in a self-consistent way: the argument of the logarithm is zero where\footnote{The $\frac{i\pi}{2}$ shift can be intuitively understood as the relation
between mirror and real particles \cite[p. 27]{tongeren}. }
\[
e^{\epsilon(\bar{\theta}_{j}+\frac{i\pi}{2}\vert\{\bar{\theta}_{j}\})}=-1,
\]
which resembles the exponential form \eqref{eq:expBY} of the BY equations,
and in the large $L$ limit indeed reduces to them. By taking the
logarithm:
\begin{align*}
\epsilon(\bar{\theta}_{k}+\frac{i\pi}{2}\vert\{\bar{\theta}\}) & =imL\sinh\bar{\theta}_{k}+\ln S(0)+\sum_{j\neq k}\ln S(\bar{\theta}_{kj})-\\
 & \quad-\int_{-\infty}^{\infty}\frac{d\theta'}{2\pi}\varphi(\bar{\theta}_{k}+\frac{i\pi}{2}-\theta')\ln(1+e^{-\epsilon(\theta'|\{\bar{\theta}\})})=i\pi(2n_{k}+1).
\end{align*}
Since $S(0)=-1$, the phase-shift $\delta$ has a discontinuity at
zero due to the branch-cut of the logarithm, and its value is $\pm\pi$
depending on from which direction we approach zero. This does not
matter here, because we could might as well define the r.h.s. with
$2n_{k}-1$. We assume the $\ln S(0)$ in the above equation to be
$i\pi$, and therefore we define a function
\begin{align}
Q_{k}(\{\theta\})\equiv-i\epsilon(\theta_{k},\{\theta\})-\pi & =\overbrace{mL\sinh\theta_{k}-i\sum_{j\neq k}\ln S(\theta_{kj})}^{Q_{k}^{(0)}(\{\theta\})}+\label{eq:Qk}\\
 & \quad+i\int_{-\infty}^{\infty}\frac{d\theta'}{2\pi}\varphi(\theta_{k}+\frac{i\pi}{2}-\theta')\ln(1+e^{-\epsilon(\theta'|\{\theta\})})\nonumber 
\end{align}
with which the quantization conditions $Q_{k}(\{\bar{\theta}\})=2\pi n_{k}$
differ from the BY equations by the integral term only.\footnote{One may wonder why the integral term in \eqref{eq:Qk} is non-vanishing
even for one-particle states, for which we know from \eqref{eq:freeqc}
that the free particle quantization condition $p_{n}L=2\pi n$ is
exact. This means that $p_{n}\neq m\sinh\bar{\theta}$, which is indeed
the case: just as the energy, the total momentum also contains an
integral term.} The energy of the excited state with $\{n\}\equiv\{n_{k}\}_{k=1}^{N}$
is 
\[
E_{\{n\}}(L)=\sum_{j}m\cosh\bar{\theta}_{j}-m\int_{-\infty}^{\infty}\frac{d\theta}{2\pi}\cosh\theta\ln(1+e^{-\epsilon(\theta\vert\{\bar{\theta}_{j}\})}).
\]
Dropping the integral term both in $Q_{k}$ and $E_{\{n\}}$ gives
back the asymptotical Bethe ansatz 
\begin{align*}
Q_{k}^{(0)}(\{\bar{\theta}^{(0)}\}) & =2\pi n_{k}, & E_{\{n\}}^{(0)}(L) & =\sum_{j}m\cosh\bar{\theta}_{j}^{(0)}
\end{align*}
where by $(0)$ we mean that this is a leading order approximation.

\subsection{Lüscher's F-term corrections\label{subsec:L=0000FCscher's-F-term-corrections}}

In this subsection we try to interpret the lowest exponential corrections
to the excited states in terms of virtual processes. To keep every
$\mathcal{O}(e^{-mL})$ correction in the energy we need to iterate
\eqref{eq:shGpseudoenergy} once further than we did for the ground
state
\[
\epsilon^{(1)}(\theta\vert\{\theta\})=\overbrace{mL\cosh\theta+\sum_{j}\ln S(\theta-\theta_{j}-\frac{i\pi}{2})}^{\epsilon^{(0)}(\theta|\{\theta\})}-\int_{-\infty}^{\infty}\frac{d\theta'}{2\pi}\varphi(\theta-\theta')e^{-\epsilon^{(0)}(\theta'|\{\theta\})},
\]
and use the corresponding quantization condition:
\begin{align*}
Q_{k}^{(1)}(\{\bar{\theta}^{(1)}\}) & =2\pi n_{k}, & Q_{k}^{(1)}(\{\theta\}) & =Q_{k}^{(0)}(\{\theta\})+\partial_{k}\delta\Phi(\{\theta\}),
\end{align*}
where we substituted $\epsilon^{(0)}(\theta|\{\theta\})$ into the
integral term of \eqref{eq:Qk} and the derivative of 
\begin{equation}
\delta\Phi(\{\theta\})=\int_{-\infty}^{\infty}\frac{d\theta'}{2\pi}e^{-\epsilon^{(0)}(\theta'\vert\{\theta\})}=\int_{-\infty}^{\infty}\frac{d\theta'}{2\pi}\prod_{j}S(\theta_{j}+\frac{i\pi}{2}-\theta')e^{-mL\cosh\theta'}\label{eq:deltaPhi}
\end{equation}
appeared. Comparing the two quantization conditions
\[
0=Q_{k}^{(1)}(\{\bar{\theta}^{(1)}\})-Q_{k}^{(0)}(\{\bar{\theta}^{(0)}\})=\delta\bar{\theta}_{l}\partial_{l}Q_{k}^{(0)}(\{\bar{\theta}^{(0)}\})+\partial_{k}\delta\Phi(\{\bar{\theta}^{(0)}\})+\Ordo(e^{-2mL})
\]
one gets that the corrections $\delta\bar{\theta}_{l}=\bar{\theta}_{l}^{(1)}-\bar{\theta}_{l}^{(0)}$
to the rapidities are of order $\Ordo(e^{-mL})$, and so the energy
\begin{equation}
E_{\{n\}}^{(1)}(L)=m\sum_{j}\cosh\bar{\theta}_{j}^{(1)}-m\int_{-\infty}^{\infty}\frac{d\theta}{2\pi}\,\cosh\theta\,\prod_{j}S(\bar{\theta}_{j}^{(0)}+\frac{i\pi}{2}-\theta)e^{-mL\cosh\theta}+\Ordo(e^{-2mL})\label{eq:EnergyFterm}
\end{equation}
contains corrections to $E_{\{n\}}^{(0)}(L)$ from both the sum and
the integral term. Note that in the latter it is enough to use the
$\bar{\theta}^{(0)}$-s instead of the $\bar{\theta}^{(1)}$-s since
the difference is $\Ordo(e^{-2mL})$. 

\begin{figure}[h]
\begin{centering}
\includegraphics[width=5cm]{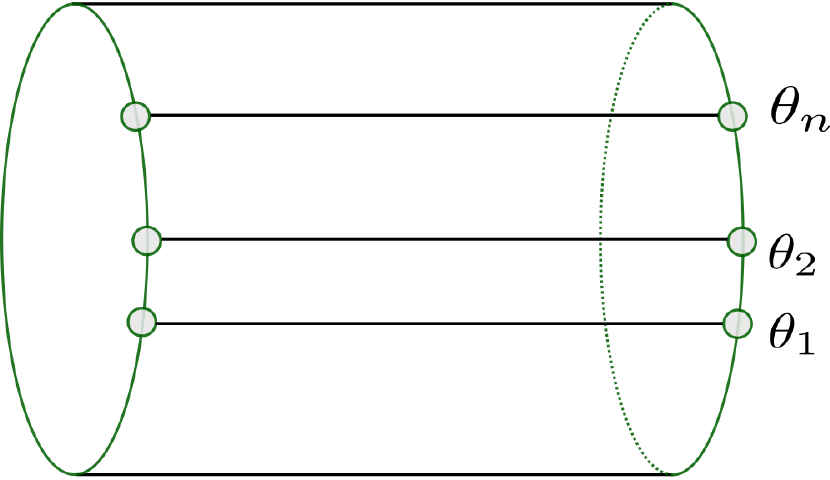}
\par\end{centering}
\caption{A particle with rapidity $\theta_{j}$ in the $R$-channel behaves
as a defect on which the mirror particles scatter in the $L$-channel.
Source: \cite{preprint}}

\label{fig:expart}
\end{figure}

As for the ground state we may again support these results by interpreting
the direct expansion of the trace over mirror model states. Now the
real ($R$-channel) particles with worldlines illustrated in Fig.
\ref{fig:expart} become non-local objects in the $L$-channel and
act as impurities on which the mirror particles scatter. In the partition
function their effect is taken into account formally with the so-called
defect operator $D$ (see \cite{defect}) which acts on one-particle
states $\vert p_{n}\rangle,\;p_{n}L\equiv mL\sinh\theta=2\pi n$ by
picking up a so-called transmission factor $T(\theta)$:
\[
\langle p_{n}\vert D(\theta_{j})\vert p_{n}\rangle=T(\theta)=S(\theta_{j}+\frac{i\pi}{2}-\theta),
\]
which in this case is the scattering phase on a - originally - $R$-channel
particle with rapidity $\theta_{j}$. The partition function for an
$N$-particle excited state in the $L$-channel is then (up to one-particle
states)
\[
Z_{N}(R,L)=\Tr(\prod_{j}D(\bar{\theta}_{j}^{(0)})e^{-LH_{R}})=1+\sum_{n}\prod_{j}S(\bar{\theta}_{j}^{(0)}+\frac{i\pi}{2}-\theta(n))e^{-LE_{n}(R)}+\ldots,
\]
and by extracting the lowest energy from $Z_{N}(R,L)$ with the $R\to\infty$
limit as we did for the real ground state we get the same integral
as the second term in \eqref{eq:EnergyFterm}.

Intuitively the corrections to the first term in the energy \eqref{eq:EnergyFterm}
comes from the rapidity changes of the real particles due to the single
mirror particle scattering on them; whilst the integral term is the
contribution of the same virtual particle (see the right side of figure
\ref{fig:muF}). These corrections are called the $F$-terms.

\subsection{TBA for models with fusion and the $\mu$-term\label{subsec:TBA-for-models}}

In models with bound states the type of singularities of the integrands
mentioned in \ref{subsec:Excited-states} come in groups for each
real particle \cite{pozsgayexactII}. For example the Lee-Yang TBA
equation \cite{doreyI} 
\begin{align*}
\epsilon(\theta\vert\{\theta_{\pm}\}) & =mL\cosh\theta+\sum_{j}\left(\ln S(\theta-\theta_{j+}-\frac{i\pi}{2})+\ln S(\theta-\theta_{j-}-\frac{i\pi}{2})\right)-\\
 & \quad-\int_{-\infty}^{\infty}\frac{d\theta'}{2\pi}\varphi(\theta-\theta')\ln(1+e^{-\epsilon(\theta'\vert\{\theta_{\pm}\})})
\end{align*}
($\{\theta_{\pm}\}\equiv\{\theta_{j\pm}\}_{j=1}^{N}$, and they are
generic variables as $\theta_{j}$ were - although they are complex)
contains a pair of them at $\bar{\theta}_{j\pm}+\frac{i\pi}{2}$,
where $\bar{\theta}_{j\pm}=\bar{\theta}_{j}\pm i\bar{u}_{j}\;\bar{\theta}_{j},\bar{u}_{j}\in\mathbb{R}$
for particle $j$ in the excited state with $\bar{\theta}_{j}$. This
means that a single particle shows up as its constituents similarly
as in the bootstrap equation
\begin{align}
S(\theta) & =S(\theta+iu)S(\theta-iu); & u & \equiv\frac{\pi}{3}\label{eq:LYbootstrap}
\end{align}
where the S-matrix has poles like \footnote{The consistency with notations in section \ref{sec:Bootstrap-principle}
is $u_{aa}^{a}\equiv2u;\;\bar{u}_{aa}^{a}\equiv u;\;\Gamma_{aaa}\equiv\Gamma$}
\begin{equation}
S(\theta)\simeq\begin{cases}
\frac{i\Gamma^{2}}{\theta-2iu}+\ldots & s\text{-channel}\\
\frac{-i\Gamma^{2}}{\theta-iu}+\ldots & t\text{-channel}
\end{cases}.\label{eq:LYSmxpoles}
\end{equation}
However as the finite volume rapidities $\bar{\theta}_{j}$, the $\bar{u}_{j}$-s
will also depend on $L$, and $\bar{u}_{j}\to u$ only in the limit
$L\to\infty$. The quantization conditions involve these constituents,
and so there are two times as many of them:
\begin{align}
Q_{k\pm}(\{\bar{\theta}_{\pm}\}) & =2\pi n_{k\pm}, & Q_{k\pm}(\{\theta_{\pm}\}) & =-i\epsilon(\theta_{k\pm}+\frac{i\pi}{2}\vert\{\theta_{\pm}\})-\pi.\label{eq:qcexact}
\end{align}
For later use we define
\begin{align}
Q_{k}(\{\theta_{\pm}\}) & =Q_{k+}(\{\theta_{\pm}\})+Q_{k-}(\{\theta_{\pm}\})\quad\mathrm{and}\label{eq:Q}\\
\tilde{Q}_{k}(\{\theta_{\pm}\}) & =Q_{k+}(\{\theta_{\pm}\})-Q_{k-}(\{\theta_{\pm}\})\label{eq:antiQ}
\end{align}
where $Q_{k}$-s  will approach those BY equations for asymptotical
volumes, in which every particle is represented only by itself, not
as a bound state of two. The energy also contains the two sets of
residues while remaining real:
\begin{equation}
E_{\{n_{\pm}\}}(L)=m\sum_{j}(\cosh\bar{\theta}_{j+}+\cosh\bar{\theta}_{j-})-m\int_{-\infty}^{\infty}\frac{d\theta'}{2\pi}\,\cosh\theta'\,\ln(1+e^{-\epsilon(\theta'\vert\{\bar{\theta}_{\pm}\})}).\label{eq:Energyexact}
\end{equation}
The possibility of fusion allows other type of virtual processes where
mirror particles propagate macroscopically around the finite one-dimensional
world. As shown on the left side of figure \ref{fig:muF}, a standing
real particle dissociates into two mirror particles who fuse into
another real one after ``wrapping'' the cylinder. The correction
in the energy due to this process is stronger than the $F$-type one.
To see this we drop the integral terms again but now this approximation
will still contain exponential corrections, not only polynomial ones
as in the case of the shG \nomenclature{shG}{sinh-Gordon} model,
where fusion was missing.

The quantizations without the integral terms (the $(\mu)$ in superscript
refers to the correction $\Ordo(e^{-\mu L})$, where $\mu$ is a constant
of mass dimension introduced later on):
\begin{align}
Q_{k\pm}^{(0)}(\{\bar{\theta}_{\pm}^{(\mu)}\}) & =2\pi n_{k\pm},\label{eq:mutermqc}\\
Q_{k\pm}^{(0)}(\{\theta_{\pm}\})\; & =mL\sinh\theta_{k\pm}-i\ln S(\theta_{k\pm,k\mp})-i\sum_{j,\sigma:j\neq k,\sigma=\pm}\ln S(\theta_{k\pm,j\sigma})\nonumber 
\end{align}
where $\theta_{k\rho,j\sigma}=\theta_{k\rho}-\theta_{j\sigma},\;\rho,\sigma\in\{+,-\}$.
We decorated the functions $Q_{k\pm}^{(0)}$ with superscript $(0)$
since they look like analytically continued BY equations for $2N$
particles with complex rapidities $\theta_{j\pm}$, however the real
parts $\bar{\theta}_{j}^{(\mu)}$ of their solutions $\bar{\theta}_{j\pm}^{(\mu)}=\bar{\theta}_{j}^{(\mu)}\pm i\bar{u}_{j}^{(\mu)}$
contain $\Ordo(e^{-\mu L})$ terms compared to $\bar{\theta}_{j}^{(0)}$,
where the latter appear in
\begin{align*}
Q_{k}^{(0)}(\{\bar{\theta}^{(0)}\}) & =2\pi n_{k}, & n_{k} & \equiv n_{k+}+n_{k-}, & n_{k},n_{k\pm} & \in\mathbb{Z}
\end{align*}
$Q_{k}^{(0)}$-s now being the limits\footnote{By using (the logarithm of) the bootstrap equation \eqref{eq:LYbootstrap}. }
of the $Q_{k}$-s defined above in the BY regime where all $\Ordo(e^{-\mathrm{const.}\!L})$
terms are suppressed:
\begin{align*}
Q_{k}^{(0)}(\{\theta\}) & =\lim_{\epsilon\to0}\{Q_{k+}^{(0)}(\{\theta_{j}\pm i(u+\epsilon)\}_{j=1}^{N})+\\
 & +Q_{k-}^{(0)}(\{\theta_{j}\pm i(u+\epsilon)\}_{j=1}^{N})\}=\\
 & =mL\sinh\theta_{k}-\sum_{j:j\neq k}\ln S(\theta_{k,j}), & \theta_{k,j} & =\theta_{k}-\theta_{j}.
\end{align*}
 Substituting $\bar{u}_{j}^{(\mu)}=u+\delta\bar{u}_{j}$ into \eqref{eq:mutermqc}
one sees that the imaginary part of
\begin{align*}
Q_{k\pm}^{(0)}(\{\bar{\theta}_{\pm}^{(\mu)}\}) & =mL\cos u\sinh\bar{\theta}_{k}^{(0)}\pm imL\sin u\cosh\bar{\theta}_{k}^{(0)}-\\
 & -i\ln S(\pm2i(u+\delta\bar{u}_{k}))-\ldots+\Ordo(\delta\bar{u})=\\
 & =2\pi n_{k\pm}
\end{align*}
disappears only if the pole (or zero) of the S-matrix at $+2iu$ ($-2iu$)
compensates all the other terms of the imaginary part:
\[
mL\sin u\cosh\bar{\theta}_{k}^{(0)}-\ln\frac{\Gamma^{2}}{2\delta\bar{u}_{k}}-\ldots=0.
\]
Therefore the LO correction to the fusion angle is\footnote{Note that we can substitute $\bar{\theta}_{j}^{(\mu)}\to\bar{\theta}_{j}^{(0)}$
everywhere since the difference is of the same magnitude as $\delta\bar{u}_{k}$,
and so negligible in the expression of $\delta\bar{u}_{k}$ itself.} $\left|\delta\bar{u}_{k}\right|=\left|\frac{\Gamma^{2}}{2}(\ldots)\right|e^{-(m\sin u)L\cosh\bar{\theta}_{k}^{(0)}}$,
which - since $\cosh x>1\;\forall x\in\mathbb{R}$ - is a $\Ordo(e^{-\mu L})$
quantity with
\[
\mu=m\sin u<m.
\]
The meaning of $\mu\equiv\mu_{aa}^{a}$ is shown on figure \ref{fig:masstriangle},
it is the height of the mass triangle, and as we see this so-called
$\mu$-term is stronger than the $F$-term, which will be $\Ordo(e^{-mL})$
for the LY case as well. To derive the full expression for the LO
of $\delta\bar{u}_{k}$ (see \cite{pozsgaymu})
\[
\delta\bar{u}_{k}=(-1)^{n_{k}}\frac{\Gamma^{2}}{2}e^{-mL\sin u\cosh\bar{\theta}_{k}^{(0)}}\prod_{j:j\neq k}\sqrt{\frac{S(\bar{\theta}_{k,j}^{(0)}+iu)}{S(\bar{\theta}_{k,j}^{(0)}-iu)}},
\]
one needs both quantization equations for $Q_{k\pm}^{(0)}$ to hold
simultaneously. It is useful however to introduce quantities $\delta u_{k\pm}$
which are determined by the exponentialized versions of these, individually:
\begin{equation}
e^{iQ_{k\pm}^{(0)}(\{\theta_{j}\pm i(u+\delta u_{j\pm}\}_{j=1}^{N})}=1\;\Rightarrow\;\delta u_{k\pm}(\{\theta\})=\frac{\Gamma^{2}}{2}e^{\pm imL\sinh(\theta_{k}\pm iu)}\prod_{j:j\neq k}S(\theta_{k,j}\pm iu)^{\pm1}.\label{eq:dupm}
\end{equation}
At the solutions of the asymptotical Bethe ansatz $\{\bar{\theta}^{(0)}\}$,
they are the same, since the $\delta\bar{u}_{k}$ has to solve both
conditions simultaneously:
\[
\delta\bar{u}_{k}=\delta u_{k+}(\{\bar{\theta}^{(0)}\})=\delta u_{k-}(\{\bar{\theta}^{(0)}\}),
\]
otherwise $\left[\delta u_{k+}(\{\theta\})\right]^{*}=\delta u_{k-}(\{\theta\})$,
as we see from \eqref{eq:dupm}, and the real analyticity and unitarity
of $S$. Taking \eqref{eq:Q}, dropping the integral terms, and expanding
it in the correction of the fusion angle up to $\mathcal{O}(e^{-\mu L})$
one gets
\begin{equation}
Q_{k}^{(\mu)}(\{\bar{\theta}^{(\mu)}\})=2\pi n_{k},\quad Q_{k}^{(\mu)}(\{\theta\})=Q_{k}^{(0)}(\{\theta\})+\partial_{k}\sum_{j}(\delta u_{j+}(\{\theta\})+\delta u_{j-}(\{\theta\})).\label{eq:rewrittenqc}
\end{equation}
This formula comes actually by rewriting the mentioned expansion in
terms of the functions $\delta u_{k\pm}$. The quantization conditions
\eqref{eq:mutermqc} and \eqref{eq:rewrittenqc} are equivalent. The
energy of the particles at this order is clearly
\begin{equation}
E_{\{n\}}^{(\mu)}(L)=m\sum_{j,\sigma}\cosh\bar{\theta}_{j\sigma}^{(\mu)}=m\sum_{j}\cosh\bar{\theta}_{j}^{(\mu)}-2\overbrace{m\sin u}^{\mu}\sum_{j}\cosh\bar{\theta}_{j}^{(0)}\delta\bar{u}_{j},\label{eq:Energymuterm}
\end{equation}
where one should remember that $\bar{\theta}_{j}^{(\mu)}$-s contain
$\Ordo(e^{-\mu L})$ corrections too. The proofs to the claims in
this section can be found in appendix \ref{chap:Derivation-of-the}.
Some of the equalities understood as the two sides are equal only
up to $\mathcal{O}(e^{-2\mu L})$ differences. Indicating them is
not necessary, since $2\mu=\sqrt{3}m>m$ in this case, and so $F$-terms
are larger.

\subsection{Relation between $F$- and $\mu$-terms\label{subsec:Relation-between--}}

\begin{figure}[h]
\begin{centering}
\includegraphics[width=10cm]{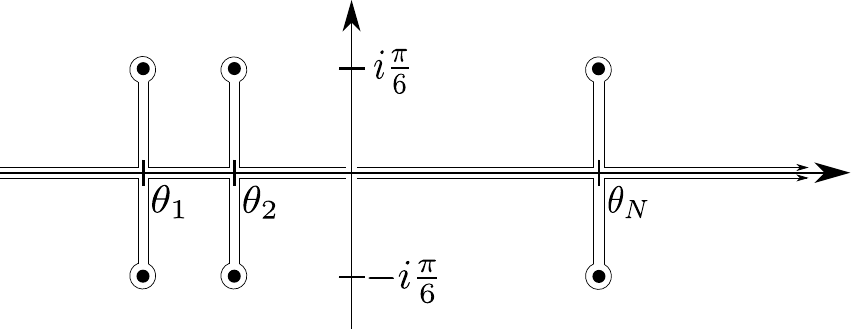}
\par\end{centering}
\caption{The two integration contours whose contributions averaged gives both the $\mu$- and the $F$-term. Source: \cite{preprint} }

\label{fig:contour}
\end{figure}

The first iteration to the Lee-Yang TBA looks as
\begin{align}
\epsilon^{(1)}(\theta\vert\{\theta_{\pm}\}) & =\overbrace{mL\cosh\theta+\sum_{j,\sigma}\ln S(\theta-\theta_{j\sigma}-\frac{i\pi}{2})}^{\epsilon^{(0)}(\theta\vert\{\theta_{\pm}\})}-\label{eq:BSTBA}\\
 & \quad-\int_{-\infty}^{\infty}\frac{d\theta'}{2\pi}\varphi(\theta-\theta')\overbrace{\prod_{j,\sigma}S(\theta_{j\sigma}+i\frac{\pi}{2}-\theta')e^{-mL\cosh(\theta')}}^{e^{-\epsilon^{(0)}(\theta'\vert\{\theta_{\pm}\})}}.\nonumber 
\end{align}
This contains $\Ordo(e^{-(\mu+m)L})$ terms which we drop, since what
interests us is the $\Ordo(e^{-mL})$ only. Therefore in the integral
term we take $\theta_{j\pm}=\theta_{j}\pm iu$, i.e. we neglect the
correction to the fusion angle and use \eqref{eq:LYbootstrap}. Similarly
to \eqref{eq:rewrittenqc} one can substitute $\theta_{j\pm}=\theta_{j}\pm i(u+\delta u_{j\pm})$
into $\epsilon^{(0)}(\theta\vert\{\theta_{\pm}\})$ and expand it
in the $\delta u_{j\pm}$ - since in the quantization condition both
of them becomes $\delta\bar{u}_{j}$. What we get is
\begin{align*}
\epsilon^{(1)}(\theta\vert\{\theta\}) & =\overbrace{mL\cosh\theta+\sum_{j}\ln S(\theta-\theta_{j}-\frac{i\pi}{2})}^{\epsilon^{(0)}(\theta\vert\{\theta\})}+\\
 & \quad+\sum_{j}\varphi(\theta-iu-\theta_{j}-\frac{i\pi}{2})\delta u_{j-}(\{\theta\})-\\
 & \quad-\sum_{j}\varphi(\theta+iu-\theta_{j}-\frac{i\pi}{2})\delta u_{j+}(\{\theta\})-\\
 & \quad-\int_{-\infty}^{\infty}\frac{d\theta'}{2\pi}\varphi(\theta-\theta')\prod_{j}S(\theta_{j}+i\frac{\pi}{2}-\theta')e^{-mL\cosh(\theta')}.
\end{align*}
This gives immediately the following quantization conditions\footnote{One could equivalently use $\epsilon^{(1)}(\theta\vert\{\theta_{\pm}\})$
to write a pair of quantization conditions for each particle.} 
\begin{align}
Q_{k}^{(1)}(\{\bar{\theta}^{(1)}\}) & =2\pi n_{k},\label{eq:qcLYOmL}\\
Q_{k}^{(1)}(\{\theta\}) & =Q_{k}^{(0)}(\{\theta\})+\partial_{k}\sum_{j}(\delta u_{j+}(\{\theta\})+\delta u_{j-}(\{\theta\}))+\partial_{k}\delta\Phi(\{\theta\}),\nonumber 
\end{align}
where $\delta\Phi$ is defined just like in \eqref{eq:deltaPhi} for
the shG model. The energy is then
\begin{align}
E_{\{n\}}^{(1)}(L) & =m\sum_{j}\cosh\bar{\theta}_{j}^{(1)}-2\mu\sum_{j}\cosh\bar{\theta}_{j}^{(0)}\delta\bar{u}_{j}-\nonumber \\
 & \quad-m\int_{-\infty}^{\infty}\frac{d\theta}{2\pi}\,\cosh\theta\,\prod_{j}S(\bar{\theta}_{j}^{(0)}+i\frac{\pi}{2}-\theta)e^{-mL\cosh\theta}.\label{eq:EnergymuFterm}
\end{align}
What we promised in the title of this subsection is a correspondence
between the integral (the $F$-term) and the terms proportional to
the correction of the fusion angle (the $\mu$-terms). We can think
of their relation purely from mathematical point of view. The poles
in
\[
e^{-\epsilon^{(0)}(\theta\vert\{\theta\})}=\prod_{j}S(\theta_{j}+i\frac{\pi}{2}-\theta)e^{-mL\cosh\theta}
\]
(which factor always appears in the integrands) come from the two
S-matrix poles at $2iu=2i\pi/3$ and $iu=i\pi/3$ with opposite residues
(see \eqref{eq:LYSmxpoles}), which means the integrand has poles
at $\theta=\theta_{j}\pm i\frac{\pi}{6}$ with
\begin{equation}
\text{Res}_{\theta=\theta_{j}\pm i\frac{\pi}{6}}e^{-\epsilon^{(0)}(\theta\vert\{\theta\})}=\pm2i\delta u_{j\mp}(\{\theta\}).\label{eq:Res}
\end{equation}
As one can easily deduce from a comparison of \eqref{eq:qcLYOmL},
\eqref{eq:deltaPhi}, and \eqref{eq:Res}, if we average the contribution
of the two contours shown in figure \ref{fig:contour}, i.e. we apply
the residue theorem as 
\begin{equation}
\frac{1}{2}\times2\pi i\times\sum_{j}\left\{ \text{Res}_{\theta=\theta_{j}-i\frac{\pi}{6}}\frac{f(\theta)}{2\pi}e^{-\epsilon^{(0)}(\theta\vert\{\theta\})}-\text{Res}_{\theta=\theta_{j}+i\frac{\pi}{6}}\frac{f(\theta)}{2\pi}e^{-\epsilon^{(0)}(\theta\vert\{\theta\})}\right\} \label{eq:residuerule}
\end{equation}
where $f(\theta)$ is some regular function at the poles in the integrand;
we get the correct $\mu$-terms, and this will work for the energy
as well with $f(\theta)=\cosh\theta$.

\begin{figure}[h]
\begin{centering}
\includegraphics{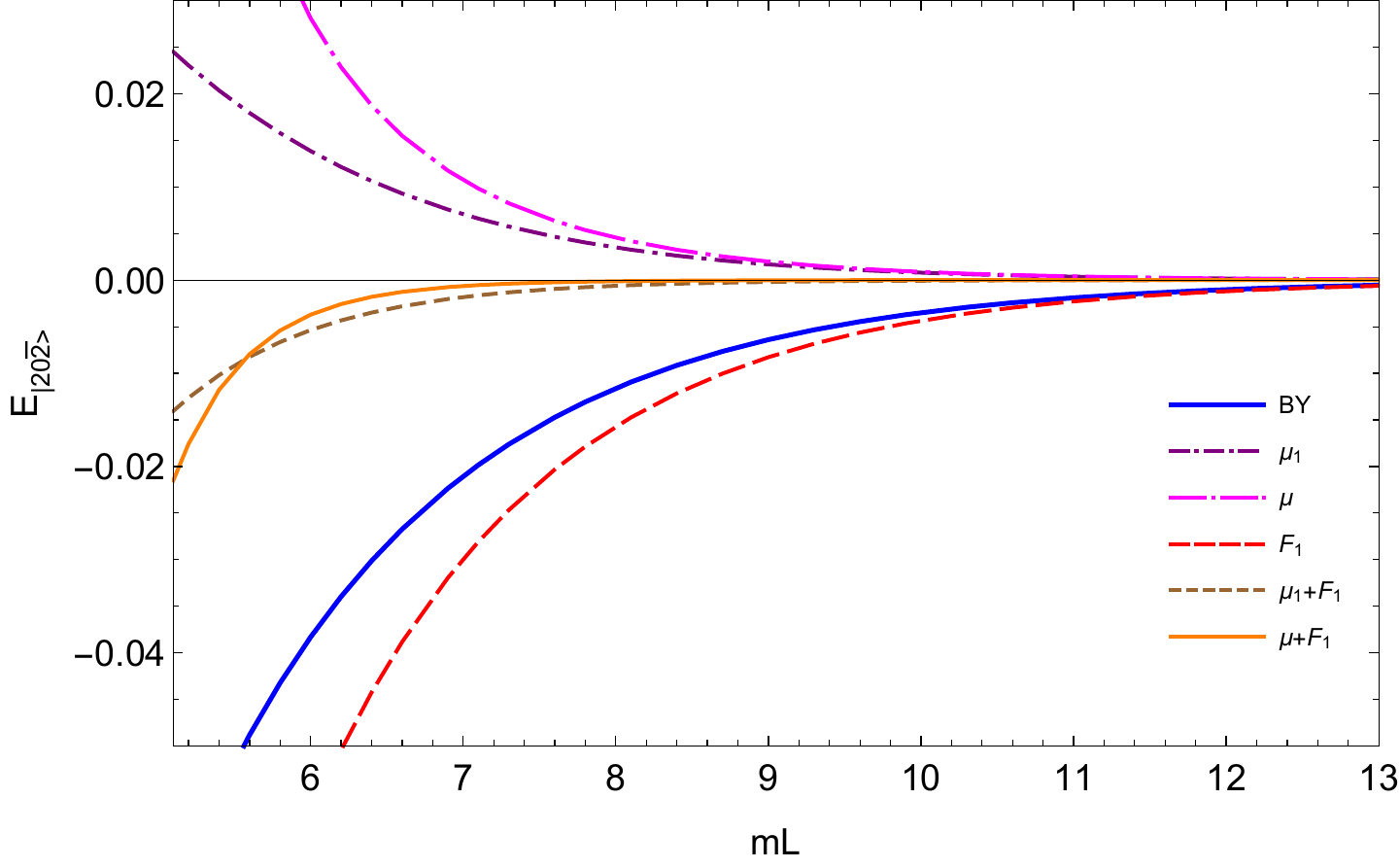}
\par\end{centering}
\caption{Lüscher-corrections to the three-particle state with quantum numbers
$\{2,0,-2\}$ in SLYM. The exact (TCSA) values are subtracted, and
so the better an approximation is, the faster it converges to zero.
From \eqref{eq:Energymuterm} we get ``$\mu_{1}$'', the leading
order. ``$F_{1}$'' is the F-term correction only which we get by
using \eqref{eq:EnergyFterm} as if we were in a fusionless theory,
and one sees that even the BY correction is better than this. However formula \eqref{eq:EnergymuFterm} denoted as ``$\mu_{1}+F_{1}$'' gives a more precise
result. The ``$\mu$'' and ``$\mu+F_{1}$'' curves display what
we would get by using \eqref{eq:qcexact} and \eqref{eq:Energyexact}
without the integral terms (``$\mu$'') and with the first iteration
of the integral terms (``$\mu+F_{1}$''), and \textit{not} expanding
the energy formula in the leading fusion angle corrections $\delta\bar{u}_{k}$.
This method sums up the higher order terms which grow with some power of $e^{-\mu L}$, and gives a slightly better result. Source: \cite{preprint}}

\label{fig:LYenexp}
\end{figure}

On figure \eqref{fig:LYenexp} we compare the different energy corrections
dealt with in this section.

\section{Form factors in finite volume\label{sec:Form-factors-in}}

In the previous section we calculated the energy of the finite volume
states with the help of infinite volume data, namely the S-matrix.
The same idea should apply to finite volume correlation functions,
and so we try to express them via infinite volume (elementary) form
factors.\footnote{Partly for simplicity, partly since the generalizations of some results
are not available we keep thinking either in terms of the shG or the
LY model. In derivations we still use the abstract level of the axiomatic
framework, not the explicit formulas of the S-matrices and the form
factors. }

The introduction of finite volume $N$-particle states $\vert n_{1},\ldots n_{N}\rangle_{L}=|\{n\}\rangle_{L}\equiv\vert\{\bar{\theta}\}\rangle_{L}=\vert\bar{\theta}_{1},\ldots,\bar{\theta}_{N}\rangle_{L}$
- where $n$-s are the quantum numbers indexing the state and $\bar{\theta}$-s
are the corresponding solutions to the exact quantization conditions
- allows one to make a similar insertion as in \eqref{eq:correlationfunction},
and define the finite volume form factors. For the Euclidean correlator:
\begin{equation}
\langle\mathcal{O}(r,0)\mathcal{O}(0,0)\rangle_{L}=\sum_{N=0}^{\infty}\sum_{\{n\}}\left|\langle0\vert\mathcal{O}(0,0)\vert\{n\}\rangle_{L}\right|^{2}e^{-rE_{\{n\}}(L)}.\label{eq:FVcorr}
\end{equation}
As was shown in \cite{pozsgaytakacsI},  the finite volume form factor
\[
\langle\bar{\vartheta}_{1},\dots,\bar{\vartheta}_{M}\vert\mathcal{O}(0,0)\vert\bar{\theta}_{1},\dots,\bar{\theta}_{N}\rangle_{L}\equiv\langle\{m\}\vert\mathcal{O}\vert\{n\}\rangle_{L}
\]
 i.e. the operator between states with quantum numbers $\langle\{m\}_{l=1}^{M}\vert$
and $\vert\{n\}_{k=1}^{N}\rangle$ in the BY regime is
\begin{equation}
\langle\{m\}\vert\mathcal{O}\vert\{n\}\rangle_{L}=\frac{F_{N+M}(\{\bar{\vartheta}^{(0)}+i\pi\},\{\bar{\theta}^{(0)}\})}{\sqrt{\prod_{i<j}S(\bar{\vartheta}_{j,i}^{(0)})\rho_{M}^{(0)}(\{\bar{\vartheta}^{(0)}\})\prod_{i<j}S(\bar{\theta}_{i,j}^{(0)})\rho_{N}^{(0)}(\{\bar{\theta}^{(0)}\})}}+\mathcal{O}(e^{-\mu L}),\label{eq:FVFFBY}
\end{equation}
where
\begin{itemize}
\item $F_{N+M}(\{\vartheta+i\pi\},\{\theta\})$\footnote{Note that here - and everywhere else, where the order of argument
counts - we mean the notation $\{\theta\}\equiv\theta_{1},\ldots,\theta_{N}$
as a sequence of arguments rather than a set (!) } is the (connected part) of the crossed infinite volume form factor
$F_{N,M}(\vartheta_{1},\ldots,\vartheta_{M}\vert\theta_{1},\ldots,\theta_{N})=\langle\vartheta_{1},\ldots,\vartheta_{M}\vert\mathcal{O}(0,0)\vert\theta_{1},\ldots,\theta_{N}\rangle$,
\item $\rho_{N}^{(0)}(\{\theta\})=\det\left[\partial Q^{(0)}\right](\{\theta\}),\quad\left[\partial Q^{(0)}\right]_{ij}\equiv\partial_{i}Q_{j}^{(0)}$
is the Jacobian determinant of the functions $Q_{j}^{(0)}(\{\theta\})$,
which is basically the density of states given in the rapidity variables.
This comes from a change of variables when the integral in \eqref{eq:correlationfunction}
is expressed as a sum $\sum_{\{n\}}$ like in \eqref{eq:FVcorr} to
relate the two formula. The square root is due to the fact that this
way one relates the absolute squares of the form factors, and so 
a $\sqrt{\rho_{N}^{(0)}}$ factor gets associated to each of them
   (see \cite{pozsgaytakacsI});
\item the product of S-matrices for every pair of particles make the r.h.s
orderless in the rapidities which one can easily see by using the
permutation axiom \eqref{eq:PERM} on $F_{N+M}$ - this is because
exchanging two quantum numbers should be meaningless on the l.h.s.
The finite volume states are then related (at least up to exponential
corrections) to the infinite volume ones as
\[
\vert\{n\}\rangle_{L}=\frac{1}{\sqrt{\prod_{i<j}S(\bar{\theta}_{i,j}^{(0)})\rho_{N}^{(0)}(\{\bar{\theta}^{(0)}\})}}\vert\bar{\theta}_{1}^{(0)},\dots,\bar{\theta}_{N}^{(0)}\rangle;
\]
\item and everything depends on the quantum numbers through the solutions
of the quantization conditions $Q_{k}^{(0)}(\{\bar{\theta}^{(0)}\})=2\pi n_{k}$
and $Q_{k}^{(0)}(\{\bar{\vartheta}^{(0)}\})=2\pi m_{k}$ .
\end{itemize}
The limitations of the above formula is that if we pick a rapidity
from each side, and their numerical values coincide, i.e. $\bar{\vartheta}_{i}^{(0)}=\bar{\theta}_{j}^{(0)}$
for some $i,j$, then the crossed form factor in the numerator becomes
infinite due to the kinematical pole \eqref{eq:KIN}. This can happen
for special cases only, since $m_{i}=n_{j}$ for a single pair of
quantum numbers does not imply the equality of the rapidities for
every $L$, because the presence of other particles modify them through
the BY equations on the two sides differently. If however the two
states are equal ($M=N,\;\{m\}=\{n\}$ and so the form factor is a
diagonal matrix element) the calculation of $\langle\{n\}\vert\mathcal{O}\vert\{n\}\rangle_{L}$
requires special treatment. 

For these diagonal finite volume form factors the formulas analogous
to \eqref{eq:FVFFBY} (and so correct for the BY regime only) usually
contain infinite volume elementary form factors, from which the kinematical
singularities are subtracted in some way - so-called connected form
factors \cite{pozsgaytakacsII}. It was shown in \cite{bajnokchao}
that these formulas can be also achieved as taking an appropriate
limit of \eqref{eq:FVFFBY}. 

Based on the ground state TBA an exact result exists on the volume
dependence of the vacuum expectation value $\langle0\vert\mathcal{O}\vert0\rangle_{L}$
- known as the Leclair-Mussardo formula - which actually involves
the above mentioned connected form factors (see \cite{leclairmussardo}).
With the help of these and the TBA for excited states an exact result
also exist for general diagonal form factors in finite volume \cite{pozsgayexactI,pozsgayexactII}.

For non-diagonal form factors (where the two sets of quantum numbers
are not the same $\{n\}\neq\{m\}$) similar exact formulas are not
available yet. The leading exponential corrections, the $\mu$-terms
first appeared in \cite{pozsgaymu}, and came from the idea that in
models with fusion particles are represented by a pair of constituents
with complex rapidities as was presented for the LY model in the previous
section. The $F$-term was investigated with field-theoretical methods
in the mirror model for one-particle form factors $\langle0\vert\mathcal{O}\vert\bar{\theta}^{(1)}\rangle_{L}$
in \cite{fieldtheoretical}, where $\bar{\theta}^{(1)}$ is the correction
explained in subsec. \ref{subsec:L=0000FCscher's-F-term-corrections}. In
the paper \cite{preprint} this result is extended to the general
multiparticle case $\langle\{\bar{\vartheta}^{(1)}\}\vert\mathcal{O}\vert\{\bar{\theta}^{(1)}\}\rangle_{L}$
formally. The connection between $F$- and $\mu$-terms presented
in subsec. \ref{subsec:Relation-between--} can be shown also for the corresponding
formulas of the form factors, and this correspondence may be - apart
from being a good check on the validity of the $F$-term result -
a good starting point for analyzing higher order corrections in the
future. In the remaining part of my thesis I intend to give a detailed
derivation of this relation in the LY model.

\lhead[\chaptername~\thechapter]{\rightmark}

\rhead[\leftmark]{}

\lfoot[\thepage]{}

\cfoot{}

\rfoot[]{\thepage}

\chapter{Lüscher-corrections of non-diagonal form factors\label{chap:L=0000FCscher-corrections-of-non-diago}}

In this chapter we motivate the derivation of  the $\mu$- and the
$F$-term corrections in case of a general non-diagonal form factor,
like $\langle\{m\}\vert\mathcal{O}\vert\{n\}\rangle_{L},\;\{n\}\neq\{m\}$.
The model in which we derive the $\mu$-terms is the LY, and it is
the shG in case of the $F$-terms - since for the latter they are
the only exponential corrections. But then the same $\mathcal{O}(e^{-mL})$
terms are thought to be valid in the LY model - at least if we drop
$\mathcal{O}(e^{-(\mu+m)L})$ contributions like we did in \ref{subsec:Relation-between--};
and with the same choice of the contour what we used there we should
get them both.

The steps for showing the equivalence of the residues with the $\mu$-terms
originating from the direct approach presented in \ref{sec:-term-correction}
is pointed out along the discussions, although it is only done explicitly
for elementary form factors $\langle0\vert\mathcal{O}\vert\{n\}\rangle_{L}$.\footnote{The $\mu$-corrections coming from the in- and outgoing sides will
not mix, and the same is true for the residues of the $F$-term integrand:
the poles corresponding to the $N$- or $M$-particle set does not
contribute to the $\mu$-terms of the other set; so the generality
of the calculation is not spoiled.} Technicalities are left for the appendices again.

\section{$\mu$-term correction\label{sec:-term-correction}}

For the energy, to derive the $\Ordo(e^{-\mu L})$ correction we only
needed a pair of analytically continued BY equations for each particle
(see \eqref{eq:mutermqc}), in which they were represented with a
pair of rapidities complex conjugate to each other, and with imaginary
parts close to the situation where particles with these complex rapidities
fuse into an on-shell bound state particle.\footnote{This is called bound state quantization in the literature.}
The idea elaborated in \cite{pozsgaymu} is then to use only the formula
\eqref{eq:FVFFBY} valid in the region of polynomial corrections for
the form factors too, within which the particles present themselves
with two such constituents just discussed.

For only a single particle this means that the formula
\[
\langle0\vert\mathcal{O}\vert\theta_{+},\theta_{-}\rangle_{L}^{(0)}=\frac{F_{2}(\theta_{+},\theta_{-})}{\sqrt{S(\theta_{+}-\theta_{-})\rho_{2}^{(0)}(\theta_{+},\theta_{-})}}
\]
gives the finite volume form factor up to the $\mu$-term\footnote{In \cite{pozsgaymu} it is claimed that actually much more is true:
this method sums up all higher order terms which are powers of $e^{-\mu L}$. }
\[
\langle0\vert\mathcal{O}\vert\bar{\theta}^{(\mu)}\rangle_{L}=\langle0\vert\mathcal{O}\vert\bar{\theta}_{+}^{(\mu)},\bar{\theta}_{-}^{(\mu)}\rangle_{L}^{(0)}+\Ordo(e^{-mL})
\]
where $Q_{\pm}^{(0)}(\bar{\theta}_{+}^{(\mu)},\bar{\theta}_{-}^{(\mu)})=2\pi n_{\pm}$,
and $\bar{\theta}_{\pm}^{(\mu)}=\bar{\theta}^{(\mu)}\pm i(u+\delta\bar{u})$.
To get the $\Ordo(e^{-\mu L})$ terms explicitly, we expand the above
formula in $\delta\bar{u}$.\footnote{It should be mentioned that $\bar{\theta}^{(\mu)}$ also contains
$\delta\bar{u}$ corrections compared to $\bar{\theta}^{(0)}$, however
we apply our expansion formally only for the explicitly appearing
$\delta\bar{u}$-s, since this way we can keep our calculations simple} The two-particle form factor is close to its dynamical pole \eqref{eq:DYN}
\[
F_{2}(\theta'+iu,\theta-iu)\simeq\frac{i\Gamma}{\theta'-\theta}F_{1}(\theta)\;\Rightarrow F_{2}(\bar{\theta}_{+}^{(\mu)},\bar{\theta}_{-}^{(\mu)})=\frac{\Gamma}{2\delta\bar{u}}F_{1}(\bar{\theta}^{(\mu)})+F_{1}^{b}(\bar{\theta}^{(\mu)})+\Ordo(\delta\bar{u}),
\]
where $F_{1}^{b}$ is the finite part of the expansion\footnote{It is symmetrically evaluated, $F_{1}^{b}(\theta)\equiv F_{2}^{\left[0\right]}(\overset{\bullet}{\theta}+iu,\overset{\bullet}{\theta}-iu)$
see App. \ref{chap:Finite-parts-of} for the explanation and the notation.} . The S-matrix in the denominator is also on its bound state pole
\[
S(2i(u+\delta\bar{u}))=\frac{\Gamma^{2}}{2\delta\bar{u}}+S_{0}+\Ordo(\delta u),
\]
and one can show for the number density $\rho_{2}^{(0)}(\bar{\theta}_{+}^{(\mu)},\bar{\theta}_{-}^{(\mu)})$
that it also starts with a $\delta\bar{u}^{-1}$ pole. We see that
from the whole expression these poles altogether drop out and the
$\Ordo(1)$ parts in the $\delta\bar{u}$ expansions get promoted
to be the $\mu$-terms after multiplying both the numerator and denominator
with $\delta\bar{u}$. To present the explicit result we now turn
to the multiparticle case derived in App. \ref{chap:Derivation-of-the}. 

We start again with \eqref{eq:FVFFBY} where by a regrouping of the
S-matrices 
\begin{equation}
\langle0\vert\mathcal{O}\vert\{\theta_{\pm}\}\rangle_{L}^{(0)}=\frac{F_{2N}(\theta_{1+},\theta_{1-},\ldots,\theta_{N+},\theta_{N-})}{\sqrt{\prod\limits _{k}S(\theta_{k+,k-})\rho_{2N}^{(0)}(\{\theta_{\pm}\})\prod\limits _{i<j,\rho,\sigma}S(\theta_{i\rho,j\sigma})}},\label{eq:mutermtemplate}
\end{equation}
and this formula should be taken at $\{\bar{\theta}_{\pm}^{(\mu)}\}$
where $Q_{k\pm}^{(0)}(\{\bar{\theta}_{\pm}^{(\mu)}\})=2\pi n_{k\pm}$
to give $\langle0\vert\mathcal{O}\vert\{\bar{\theta}^{(\mu)}\}\rangle_{L}^{(\mu)}$,
i.e. the finite volume form factor up to $\Ordo(e^{-\mu L})$. 

To compensate the $N$-fold dynamical singularity in $F_{2N}$ (which
drops out as indicated above with the similar factor $\prod_{k=1}^{N}\delta\bar{u}_{k}^{-1}$of
the denominator) we expand the fraction with $\prod_{k}2\Gamma^{-1}\delta\bar{u}_{k}$:
\begin{equation}
\prod_{k}\left(\frac{2\delta\bar{u}_{k}}{\Gamma}\right)F_{2N}(\{\bar{\theta}_{\pm}^{(\mu)}\})=F_{N}(\{\bar{\theta}^{(\mu)}\})+\sum_{k}\left(\frac{2\delta\bar{u}_{k}}{\Gamma}\right)F_{N,k}^{b}(\{\bar{\theta}^{(0)}\})+O(\delta\bar{u}^{2}),\label{eq:F2N}
\end{equation}
where the $F_{N,k}^{b}$ are similar to $F_{1}^{b}$; it is a form
factor in which only the $k^{\mathrm{th}}$ particle is represented
with its constituents - the others present themselves with a single
real rapidity - and this dynamical pole for $k$ gets subtracted:
$F_{N,k}^{b}(\{\theta\})=F_{N+1}^{\left[0\right]}(\theta_{1},\ldots,\overset{\bullet}{\theta}_{k}+iu,\overset{\bullet}{\theta}_{k}-iu,\ldots,\theta_{N})$.\footnote{See App. \ref{chap:Finite-parts-of} for this notation.}

As stated before, the first factor in the denominator contains also
the same singularity (squared) as the numerator, and it is shown in
App. \ref{chap:Derivation-of-the} that its expansion is 
\begin{equation}
\prod_{k}\left(\frac{2\delta\bar{u}_{k}}{\Gamma}\right)^{2}S(2i(u+\delta\bar{u}_{k}))\rho_{2N}^{(0)}(\{\bar{\theta}_{\pm}^{(\mu)}\})=\rho_{N}^{\left(\mu\right)}(\{\bar{\theta}^{(\mu)}\})\left(1+\sum_{k}\partial_{k}Q_{k}^{(0)}(\{\bar{\theta}^{(0)}\})\delta\bar{u}_{k}\right),\label{eq:densities}
\end{equation}
where the density of states $\rho_{N}^{\left(\mu\right)}$ corresponds
to the quantizations \eqref{eq:rewrittenqc}:
\begin{align*}
\rho_{N}^{(\mu)}(\{\theta\}) & =\text{det}\left[\partial Q^{(\mu)}\right](\{\theta\}),\\
\partial_{i}Q_{j}^{(\mu)}(\{\theta\}) & =\partial_{i}Q_{j}^{(0)}(\{\theta\})+\partial_{i}\partial_{j}\sum_{k}(\delta u_{k+}(\{\theta\})+\delta u_{k-}(\{\theta\})).
\end{align*}
Since we have seen before that in \eqref{eq:qcLYOmL} the $\mu$-terms
come as the residues of $\delta\Phi$, the correspondence of $\rho_{N}^{(\mu)}$
and $\rho_{N}^{(1)}(\{\theta\})=\text{det}\left[\partial Q^{(1)}\right](\{\theta\})$
is already clear, the derivatives in the Jacobian, and taking the
determinant will not influence the $F$- and $\mu$-term relation.

The remaining factor contains a purely imaginary correction:
\[
\prod_{i<j}S(\bar{\theta}_{i+,j+}^{(\mu)})S(\bar{\theta}_{i+,j-}^{(\mu)})S(\bar{\theta}_{i-,j+}^{(\mu)})S(\bar{\theta}_{i-,j-}^{(\mu)})=\prod_{i<j}S(\bar{\theta}_{i,j}^{(\mu)})\left(1-i\Delta\varphi(\bar{\theta}_{i,j}^{(0)})\left(\delta\bar{u}_{i}+\delta\bar{u}_{j}\right)\right),
\]
where $i\Delta\varphi(\theta)=\varphi(\theta+2iu)-\varphi(\theta-2iu)$.

With the help of these expansions the full expression can be recollected
as
\begin{equation}
\langle0\vert\mathcal{O}\vert\{n\}\rangle_{L}=\frac{F_{N}(\{\bar{\theta}^{(\mu)}\})+\delta^{(\mu)}F_{N}(\{\bar{\theta}^{(0)}\})}{\sqrt{\prod\limits _{k<j}S(\bar{\theta}_{k,j}^{(\mu)})\rho_{N}^{(\mu)}(\{\bar{\theta}^{(\mu)}\})}}+\Ordo(e^{-mL}),\label{eq:mutermrecollection}
\end{equation}
where $\delta^{(\mu)}F_{N}(\{\bar{\theta}^{(0)}\})$ contains all
those corrections, which do not belong to the norm corresponding to
the quantization condition truncated at the appropriate order $(\mu)$,
but rather to the infinite volume form factor:
\begin{align}
\delta^{(\mu)}F_{N}(\{\bar{\theta}^{(0)}\}) & =\sum_{k}\left\{ \frac{2}{\Gamma}F_{N,k}^{b}(\{\bar{\theta}^{(0)}\})-\frac{1}{2}\partial_{k}Q_{k}^{(0)}(\{\bar{\theta}^{(0)}\})F_{N}(\{\bar{\theta}^{(0)}\})\right\} \delta\bar{u}_{k}\nonumber \\
 & \,\,\,\,\,\,\,\,+\frac{1}{2}\sum_{j<k}\left[i\Delta\varphi(\bar{\theta}_{j,k}^{(0)})\left(\delta\bar{u}_{j}+\delta\bar{u}_{k}\right)\right]F_{N}(\{\bar{\theta}^{(0)}\}).\label{eq:FFmuterm}
\end{align}
We will calculate this expression also as the residue of the $F$-term,
which - as we will see - gives the contribution of a virtual process
involving the operator.

For the $\langle\{m\}\vert\mathcal{O}\vert\{n\}\rangle_{L}$ form
factor, an analogous formula to \eqref{eq:mutermrecollection} can
be derived, where in the numerator another term like \eqref{eq:FFmuterm}
appears containing all the $\mu$-terms of the particles in state
$\langle\{m\}\vert$. 

\section{$F$-term correction and its residue}

In the paper \cite{fieldtheoretical} the one-particle form factor
with the $F$-term correction $\langle0\vert\mathcal{O}\vert\bar{\theta}^{(1)}\rangle_{L}^{(1)}$
was calculated by approximating the Fourier-transform $\Gamma(\omega,q)$
of the finite volume (Euclidean) two-point function $\langle\mathcal{O}(t,x)\mathcal{O}(0,0)\rangle_{L}$
in the mirror model, and extracting the residue of the pole appearing
on the $\omega$ plane at the one-particle energy $E_{n}(L)$. This
residue  is just $\left|\langle0\vert\mathcal{O}\vert n\rangle_{L}\right|^{2}$
as can be easily seen from the Fourier-space version of the spectral
representation \eqref{eq:FVcorr}.\footnote{
\begin{align*}
\Gamma(\omega,q) & =\int_{-\infty}^{\infty}dt\frac{1}{L}\int_{-\frac{L}{2}}^{\frac{L}{2}}dxe^{i(\omega t+qx)}\langle\mathcal{O}(t,x)\mathcal{O}(0,0)\rangle_{L}=\\
 & =\sum_{N}\sum_{\{n\}}\left|\langle0\vert\mathcal{O}\vert\{n\}\rangle\right|^{2}\left\{ \frac{\delta_{q-p_{\{n\}}(L)}}{E_{\{n\}}(L)+i\omega}+\frac{\delta_{q+p_{\{n\}}(L)}}{E_{\{n\}}(L)-i\omega}\right\} 
\end{align*}
} In the mirror model the trace which evaluates the two-point function
in the $L$-channel can be approximated by an expansion over the lowest
multi(mirror-)particle states.Whilst calculating matrix elements of
the operator among these, an infinite volume form factor like $F_{1,2}(\theta_{1}\vert\theta_{2},\theta_{3})$
appeared. The Dirac-delta function in the crossing formula \eqref{eq:crossing}
for this form factor needed a regularization since its square appeared
in calculations; the same regularization was also used in \cite{preprint}
to generalize the $\Ordo(e^{-mL})$ Lüscher-correction to arbitrary
non-diagonal finite volume form factors.

\subsection{The formal derivation of the $F$-term}

In the general case\footnote{In this subsection we work in the shG model, where $\mu$-terms are
absent.} the $F$-term correction $\delta^{(F)}F_{N+M}$ to the infinite volume
form factor is again separated from the corrections to the normalization
of the states like in \eqref{eq:mutermrecollection}:
\[
\langle\{m\}\vert\mathcal{O}\vert\{n\}\rangle_{L}^{(1)}=\frac{F_{N+M}(\{\bar{\vartheta}^{(1)}+i\pi\},\{\bar{\theta}^{(1)}\})+\delta^{(F)}F_{N+M}(\{\bar{\vartheta}^{(0)}+i\pi\},\{\bar{\theta}^{(0)}\})}{\sqrt{\prod_{i<j}S(\bar{\vartheta}_{j,i}^{(1)})\rho_{M}^{(1)}(\{\bar{\vartheta}^{(1)}\})\prod_{i<j}S(\bar{\theta}_{i,j}^{(1)})\rho_{N}^{(1)}(\{\bar{\theta}^{(1)}\})}}.
\]
\begin{figure}[h]
\begin{centering}
\includegraphics[width=3cm]{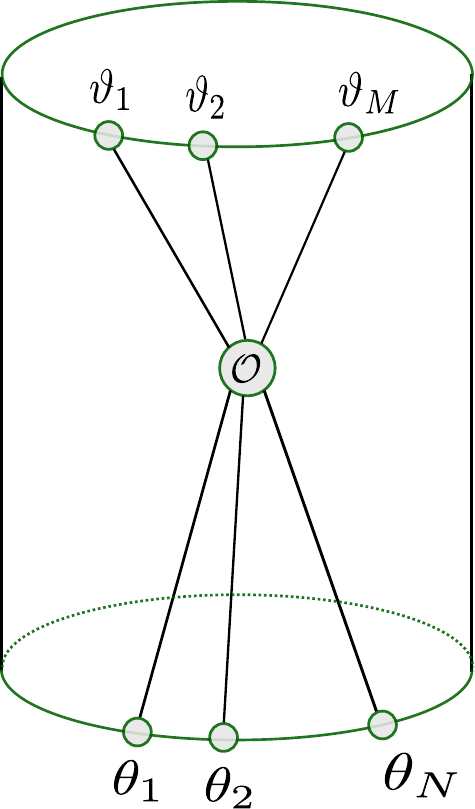}~~~~~~~~~~~~~~~~~\includegraphics[width=5cm]{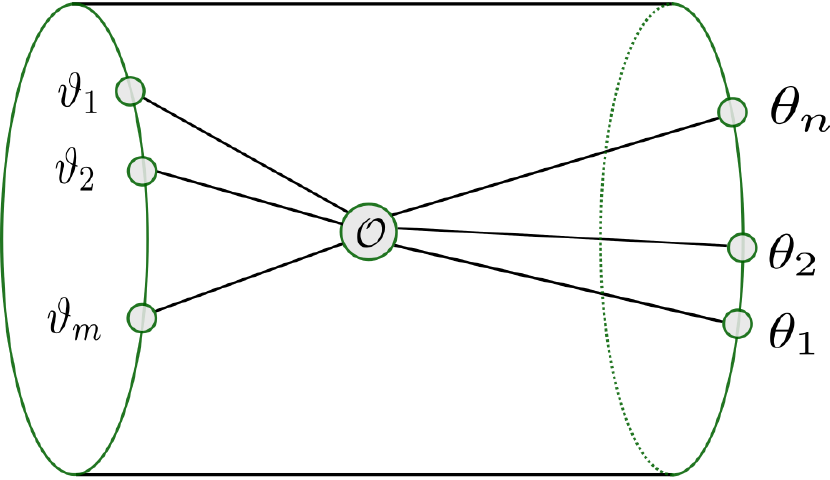}
\par\end{centering}
\caption{Left: Illustration to finite volume form factor $\langle\vartheta_{M},\ldots,\vartheta_{1}\vert\mathcal{O}\vert\theta_{1},\ldots,\theta_{N}\rangle_{L}$.
Right: The same matrix element in the ``thermal'' channel, as a
trace $\mbox{Tr}(e^{-LH}\mathcal{O}_{N,M})$. Source: \cite{preprint}}

\label{fig:cylinders}
\end{figure}

The idea for extending the one-particle result for this general case
is based on calculating an appropriately normalized $L$-channel trace
(see the right hand side of Fig. \ref{fig:cylinders}): 
\begin{equation}
F_{N+M}(\{\vartheta+i\pi\},\{\theta\})+\delta^{(F)}F_{N+M}(\{\vartheta+i\pi\},\{\theta\})\;\leftrightarrow\;\frac{\mbox{Tr}(e^{-LH}\mathcal{O}_{N,M})}{\sqrt{\mbox{Tr}_{N}(e^{-LH})\mbox{Tr}_{M}(e^{-LH})}}\label{eq:normtrace}
\end{equation}
where the norm contains the partition functions for excited states
$Z_{N}(L)=\Tr_{N}(e^{-LH})$ introduced in subsection \ref{subsec:L=0000FCscher's-F-term-corrections}.
However now the in- and outgoing states can differ in particle numbers
and momenta, and so one cannot make the boundary conditions periodic
for the time direction and compactify it, like in the case of scattering
states. This is why the size $R$ is not indicated in the Hamiltonian,
we now work on an infinite cylinder. The square root in the denominator
makes the partition functions correspond to the left- and right halves
of the figure on the right of \ref{fig:cylinders} accordingly.

 $\mathcal{O}_{N,M}$ is an operator in the mirror model which is
also not a local one (like the defect operator in \ref{subsec:L=0000FCscher's-F-term-corrections})
and corresponds to the original $\mathcal{O}$ sandwiched between
the $M$ and $N$-particle excited states. In the end, intuitively,
the formula calculates the ``thermal'' expectation value of this
operator at inverse temperature $L$.

After inserting complete systems the trace for the operator is 
\begin{align*}
\mbox{Tr}(e^{-LH}\mathcal{O}_{N,M}) & =\sum_{\nu,\mu}\langle\nu\vert\mathcal{O}_{N,M}\vert\mu\rangle\langle\mu\vert\nu\rangle e^{-E_{\nu}L}=\\
 & =F_{N+M}+\int\frac{du}{2\pi}\int\frac{dv}{2\pi}\langle v\vert\mathcal{O}_{N,M}\vert u\rangle\langle u\vert v\rangle e^{-mL\cosh v}+\ldots
\end{align*}
where the first term comes by taking $\vert\mu\rangle$ and $\vert\nu\rangle$
to be the vacuum, and one-particle states with rapidities $u$ and
$v$ for the second. 

\begin{figure}[h]
\begin{centering}
\includegraphics[width=5cm]{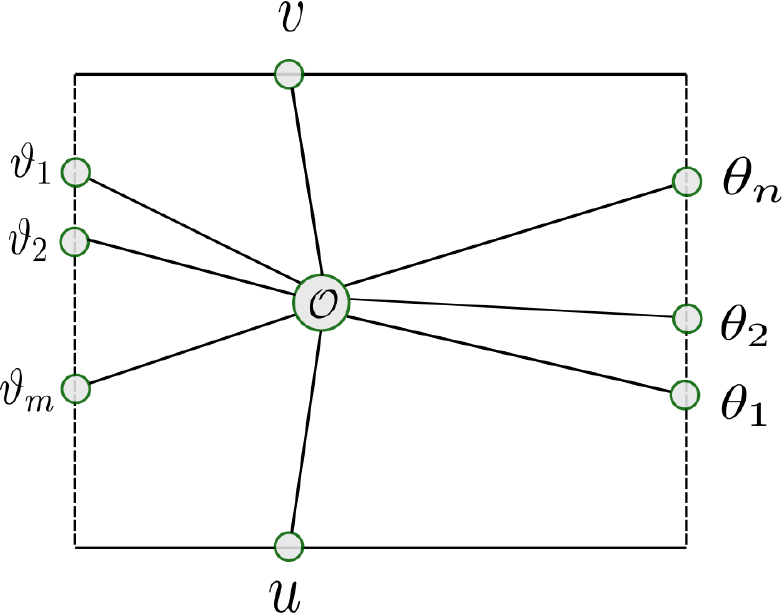}
\par\end{centering}
\caption{The matrix element $\langle v\vert\mathcal{O}_{N,M}\vert u\rangle$
of the non-local operator in the $L$-channel with an ingoing mirror
particle with $u$ and an outgoing one with $v$. The directions of
the particle lines can be intuitively identified with the imaginary
parts of the arguments in \ref{eq:Onm}. Source: \cite{preprint} }

\label{fig:Onm}
\end{figure}

With the same insertions the partition function becomes 
\[
\mbox{Tr}_{N}(e^{-LH})=1+\int\frac{du}{2\pi}\int\frac{dv}{2\pi}\langle u\vert v\rangle\prod_{j}S(\frac{i\pi}{2}+\theta_{j}-v)\langle v\vert u\rangle e^{-mL\cosh v}+\ldots\;.
\]
In the matrix element one should cross the outgoing (mirror) particle
with $v$:
\begin{align}
\langle v\vert\mathcal{O}_{N,M}\vert u\rangle & =F_{N+M+2}(v+i\pi-i\epsilon,\{\vartheta+\frac{i\pi}{2}\},u,\{\theta-\frac{i\pi}{2}\})+\nonumber \\
 & +2\pi\delta(v-u)\prod_{j}S(\frac{i\pi}{2}+\vartheta_{j}-u)F_{N+M}(\{\vartheta+\frac{i\pi}{2}\},\{\theta-\frac{i\pi}{2}\}),\label{eq:Onm}
\end{align}
where for the rapidities of the in- and outgoing real particles the
$\vartheta_{j}+\frac{i\pi}{2}$ and $\theta_{j}-\frac{i\pi}{2}$ choices
are illustrated on figure \ref{fig:Onm}.

Since the states are normalized in infinite volume as $\langle u\vert v\rangle=2\pi\delta(u-v)$,
each of the integrals contains the square of the delta function. The
regularization 
\[
2\pi\delta(u-v)=\frac{i}{u-v+i\epsilon}-\frac{i}{u-v-i\epsilon}
\]
(where a finite $\epsilon>0$ corresponds to a finite $R$ volume,
and so $\epsilon\to0$ implies $R\to\infty$) was shown to be correct
in \cite{fieldtheoretical} for the one-particle form factor. By using
this everywhere, we evaluate the integral in $v$ first. There will
be simple and double poles at $v=u\pm i\epsilon$. One can shift the
contour for example above $i\epsilon$, and pick up the residues only
for $u+i\epsilon$. However the shifted contour contributes zero in
the $\epsilon\to0$ limit, since it completely misses $v=u$ in the
$\delta$-function (since it lies above both of the poles), and so
the result of the $v$ integral is just the mentioned residue.

In the end the result becomes \cite[app. B]{preprint}
\begin{align}
 & \delta^{(F)}F_{N+M}(\{\vartheta+i\pi\},\{\theta\})=\nonumber \\
 & =\int\frac{dv}{2\pi}\{F_{M+N+2}^{c}(v+i\pi,\{\vartheta+\frac{i\pi}{2}\},v,\{\theta-\frac{i\pi}{2}\})-\nonumber \\
 & -\partial_{v}\frac{i}{2}[\prod_{j=1}^{M}S((\vartheta_{j}+\frac{i\pi}{2})-v)-\label{eq:Fcreg}\\
 & \quad\quad\;-\prod_{k=1}^{N}S(v-(\theta_{j}-\frac{i\pi}{2}))]F_{M+N}(\{\vartheta+\frac{i\pi}{2}\},\{\theta-\frac{i\pi}{2}\})\}e^{-mL\cosh v}\nonumber 
\end{align}
where $F_{M+N+2}^{c}(\ldots)\equiv F_{M+N+2}^{\left[+1\right]}(\overset{\bullet}{v}+i\pi,\{\vartheta+\frac{i\pi}{2}\},\overset{\circ}{v},\{\theta-\frac{i\pi}{2}\})$
is the so-called connected form factor with a subtracted kinematical
pole. The integrand can be rewritten with the symmetrical version
of the subtraction as discussed in App. \ref{chap:Finite-parts-of},
since the second term is just half of the kinematical residue (see
r.h.s in \eqref{eq:KIN}) for this case, differentiated:
\begin{align}
 & \delta^{(F)}F_{N+M}(\{\vartheta+i\pi\},\{\theta\})=\label{eq:FregFterm}\\
 & =\int\frac{dv}{2\pi}F_{M+N+2}^{r}(v+i\pi,\{\vartheta+\frac{i\pi}{2}\},v,\{\theta-\frac{i\pi}{2}\})e^{-mL\cosh v},\nonumber 
\end{align}
where we call $F_{M+N+2}^{r}$ the regulated form factor after \cite{fieldtheoretical}:
\begin{align}
 & F_{M+N+2}^{r}(v+i\pi,\{\vartheta+\frac{i\pi}{2}\},v,\{\theta-\frac{i\pi}{2}\})=\label{eq:Fr}\\
 & =\lim_{\epsilon\to0}\{F_{M+N+2}(v+\frac{\epsilon}{2}+i\pi,\{\vartheta+\frac{i\pi}{2}\},v-\frac{\epsilon}{2},\{\theta-\frac{i\pi}{2}\})-\nonumber \\
 & \;-\frac{i}{\epsilon}[\prod_{j=1}^{M}S((\vartheta_{j}+\frac{i\pi}{2})-v)-\nonumber \\
 & \;\quad-\prod_{k=1}^{N}S(v-(\theta_{j}-\frac{i\pi}{2}))]F_{M+N}(\{\vartheta+\frac{i\pi}{2}\},\{\theta-\frac{i\pi}{2}\})\}\nonumber 
\end{align}
i.e. $F_{M+N+2}^{r}(\ldots)\equiv F_{M+N+2}^{\left[0\right]}(\overset{\bullet}{v}+i\pi,\{\vartheta+\frac{i\pi}{2}\},\overset{\bullet}{v},\{\theta-\frac{i\pi}{2}\})$
according to App. \ref{chap:Finite-parts-of}.

\subsection{Pole structure of the regulated form factor and the residues of the
$F$-term\label{subsec:Pole-structure-of}}

The only remaining correspondence between the F- and $\mu$-term corrections
to show is that the appropriate contour in
\begin{equation}
\delta^{(F)}F_{N}(\{\theta\})=\int\frac{dv}{2\pi}F_{N+2}^{r}(v+i\pi,v,\{\theta-\frac{i\pi}{2}\})e^{-mL\cosh v}\label{eq:FtermofFF}
\end{equation}
gives $\delta^{(\mu)}F_{N}(\{\theta\})$ - at least when one takes
the result at the solutions of the quantization condition. Although
the symmetric definition of $F^{r}$ is ``natural'', it makes the
calculation of the residues at $v=\theta_{k}\mp i(u-\frac{\pi}{2})=\theta_{k}\pm\frac{i\pi}{6}$
subtle, therefore we use the other types of subtractions instead (see
App. \ref{chap:Finite-parts-of}):
\begin{align}
F_{N+2}^{r}(v+i\pi,v,\{\theta-\frac{i\pi}{2}\}) & =F_{N+2}^{\left[\pm1\right]}(\overset{{\bullet\atop \circ}}{v}+i\pi,\overset{{\circ\atop \bullet}}{v},\{\theta-\frac{i\pi}{2}\})\mp\label{eq:FrtoFc}\\
 & \quad\mp\frac{i}{2}\partial_{v}\left(1-\prod_{j}S(v-\theta_{j}+\frac{i\pi}{2})\right)F_{N}(\{\theta\}).\nonumber 
\end{align}
The distribution of poles between the two terms on the right hand
side will be particularly simple this way:
\begin{itemize}
\item at $v=\theta_{k}-i(u-\frac{\pi}{2})=\theta_{k}+\frac{i\pi}{6}$, the
$F_{N+2}^{\left[+1\right]}$ has a simple pole
\item the same is true for $v=\theta_{k}+i(u-\frac{\pi}{2})=\theta_{k}-\frac{i\pi}{6}$
and $F_{N+2}^{\left[-1\right]}$
\item the derivative term contains a double pole for both cases.
\end{itemize}
As shown in App. \ref{sec:Poles-in-the}, the regulated form factor
then has a singularity structure (for the pole below the real axis):
\begin{align}
 & F_{N+2}^{r}(v+i\pi,v,\{\theta-\frac{i\pi}{2}\})\vert_{v=\theta_{k+}-\frac{i\pi}{2}}=\nonumber \\
 & =-\frac{\Gamma^{2}}{2\delta^{2}}\left[\prod_{j\neq k}S(\hat{\theta}_{k+,j})\right]F_{N}(\{\theta\})-\frac{i\Gamma^{2}}{\delta}\left[\prod_{j<k}S(\hat{\theta}_{k+,j})\right]^{'}\prod_{j>k}S(\hat{\theta}_{k+,j})F_{N}(\{\theta\})-\nonumber \\
 & -\frac{\Gamma}{\delta}\prod_{j\neq k}S(\hat{\theta}_{k+,j})F_{N,k+}^{b}(\{\theta\})+\mathcal{O}(\delta^{0})\label{eq:polestructure}
\end{align}
where $\theta_{k\pm}=\hat{\theta}_{k\pm}+i\delta=\theta_{k}\pm iu+i\delta$
and $[\ldots]'$ means that we differentiating the S-matrices in their
arguments inside the brackets and use the product rule ; whereas $F_{N,k\pm}^{b}(\{\theta\})\equiv F_{N+1}^{[\pm1]}(\ldots,\overset{{\bullet\atop \circ}}{\theta}_{k}+iu,\overset{{\circ\atop \bullet}}{\theta}_{k}-iu,\ldots)$
is similar to $F_{N,k}^{b}(\{\theta\})$ appearing in \eqref{eq:F2N}
(see later).

With reordering the terms, and encapsulating the singularities\footnote{Notice that in the first term there are first order poles also in
addition to the second order one coming from $S'(\theta_{k+,k})$.} into $S'(\theta_{k+,k})\propto\delta^{-2}$ and $S(\theta_{k+,k})\propto\delta^{-1}$:
\begin{align}
 & F_{N+2}^{r}(v+i\pi,v,\{\theta-\frac{i\pi}{2}\})\vert_{v=\theta_{k+}-\frac{i\pi}{2}}=\label{eq:rewrittenpolestructure}\\
 & =\overbrace{-\frac{i}{2}S'(\theta_{k+,k})\prod_{j\neq k}S(\theta_{k+,j})}^{\Rightarrow\frac{i}{2}\partial_{k}\delta u_{k+}}F_{N}(\{\theta\})+\nonumber \\
 & +\overbrace{\prod_{j}S(\theta_{k+,j})}^{\Rightarrow\delta u_{k+}}\Gamma^{-1}F_{N,k+}^{b}(\{\theta\})+\nonumber \\
 & +\overbrace{\frac{i}{2}\left(\left[\prod_{j<k}S(\theta_{k+,j})\right]^{'}\prod_{j\geq k}S(\theta_{k+,j})-\prod_{j\leq k}S(\theta_{k+,j})\left[\prod_{j>k}S(\theta_{k+,j})\right]^{'}\right)}^{\Rightarrow-\frac{1}{2}\sum_{j<k}i\partial_{j}\delta u_{k+}+\frac{1}{2}\sum_{j>k}i\partial_{j}\delta u_{k+}}F_{N}(\{\theta\}).\nonumber 
\end{align}
Here the expressions shown above the terms are their contributions
in the residues of the $F$-term integrand \eqref{eq:FtermofFF} -
see App. \ref{sec:Correspondence-to-the}. One can calculate the residues
for the complex conjugate pole in the same way, and by using the rule
\eqref{eq:residuerule} we get:
\begin{align*}
 & \delta^{(\mu)}F_{N}(\{\theta\})=\\
 & =\frac{1}{2}\sum_{k,\pm}\pm i\text{Res}_{v\to\theta_{k}\mp\frac{i\pi}{6}}\left\{ F_{N+2}^{r}(v+i\pi,v,\{\theta_{j}-\frac{i\pi}{2}\})e^{-mL\cosh v}\right\} =\\
 & =\sum_{k,\pm}\bigl\{\Gamma^{-1}F_{N,k\pm}^{b}(\{\theta\})\delta u_{k\pm}-\frac{i}{2}F_{N}(\{\theta\})(\mp\partial_{k}+\sum_{j<k}\partial_{j}-\sum_{j>k}\partial_{j})\delta u_{k\pm}\bigr\},
\end{align*}
where the sum is over the pair of poles, and with the relations $\partial_{j}\delta u_{k\pm}=\pm i\partial_{j}Q_{k\pm}^{(0)}\delta u_{j\pm}$
derived in App. \ref{chap:Derivation-of-the} the above formula can
be rewritten, and taken at the solutions $\theta_{j}=\bar{\theta}_{j}^{(0)}$,
which gives the same formula \eqref{eq:FFmuterm} as the bound state
method for the $\mu$-term (see App. \ref{sec:Correspondence-to-the}):
\begin{align*}
\delta^{(\mu)}F_{N}(\{\bar{\theta}^{(0)}\}) & =\sum_{k}\left\{ \frac{2}{\Gamma}F_{N,k}^{b}(\{\bar{\theta}^{(0)}\})-\frac{1}{2}\partial_{k}Q_{k}^{(0)}(\{\bar{\theta}^{(0)}\})F_{N}(\{\bar{\theta}^{(0)}\})\right\} \delta\bar{u}_{k}\\
 & \,\,\,\,\,\,\,\,+\frac{1}{2}\sum_{j<k}\left[i\Delta\varphi(\bar{\theta}_{j,k}^{(0)})\left(\delta\bar{u}_{j}+\delta\bar{u}_{k}\right)\right]F_{N}(\{\bar{\theta}^{(0)}\}).
\end{align*}

\lhead[\chaptername~\thechapter]{\rightmark}

\rhead[\leftmark]{}

\lfoot[\thepage]{}

\cfoot{}

\rfoot[]{\thepage}

\chapter{Conclusion\label{chap:Conclusion}}

In the paper \cite{preprint} a formula was proposed for the $\Ordo(e^{-mL})$
exponential volume corrections of finite volume non-diagonal form
factors.

\begin{figure}[h]
\begin{centering}
\includegraphics{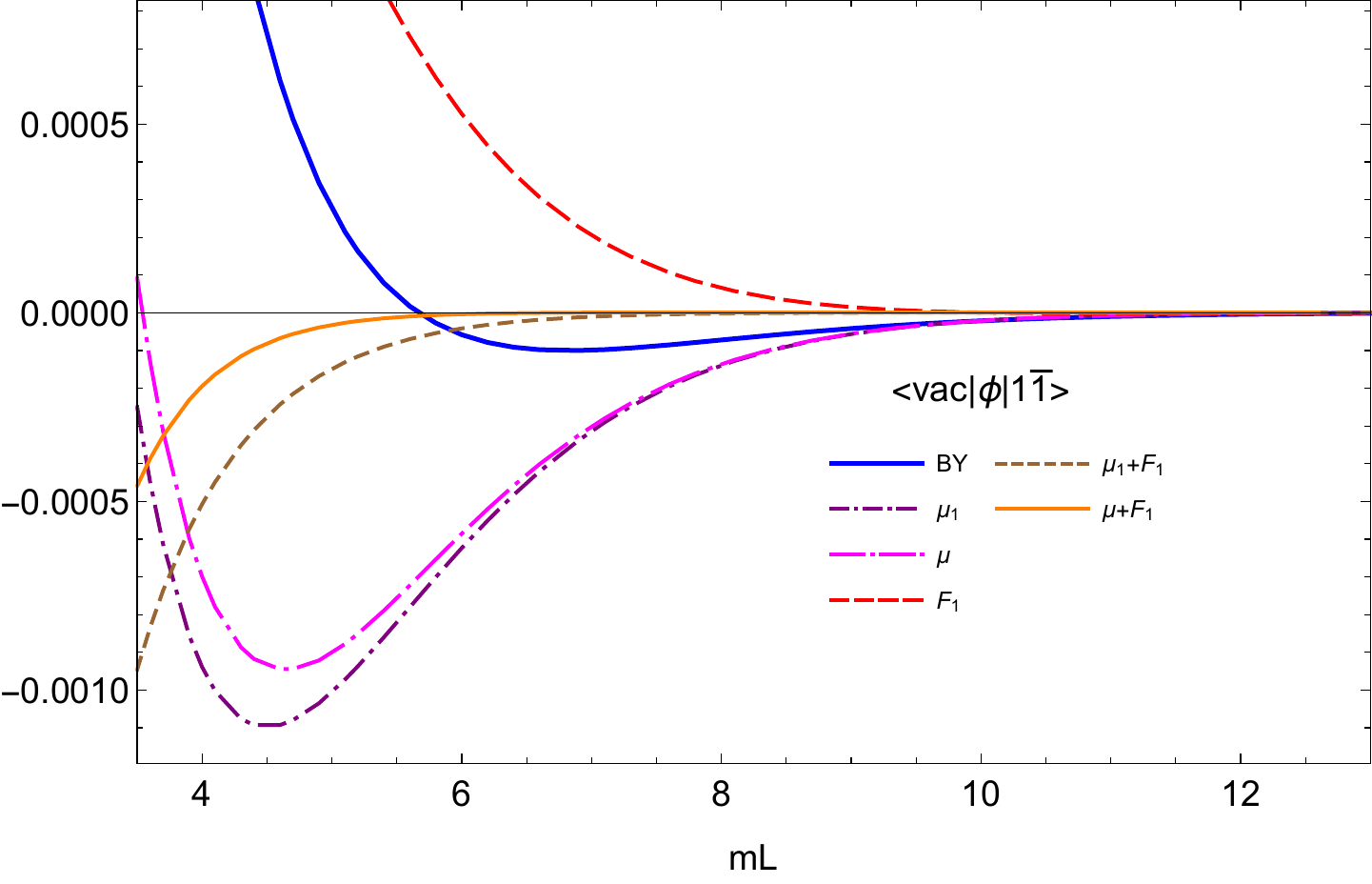}
\par\end{centering}
\caption{Comparison of the same kind of volume corrections to the form factor
$\langle0\vert\Phi\vert\{1,-1\}\rangle_{L}$ as explained below the
figure \ref{fig:LYenexp} for the energy  (here $\Phi$ is the deformation
field of the scaling Lee-Yang model, whose form factor was ``measured'').
The figure shows the difference between (the absolute values of) the
analytic formula and the TCSA result. Source: \cite{preprint}}

\label{fig:LYFVFF01}
\end{figure}

The numerical checks - coming from the TCSA method - carried out in
the LY model (see \eqref{fig:LYFVFF01}) showed that by adding this
$F$-term to the already available corrections one could indeed approximate
the exact result better.

\begin{figure}[h]
\begin{centering}
\includegraphics[height=6cm]{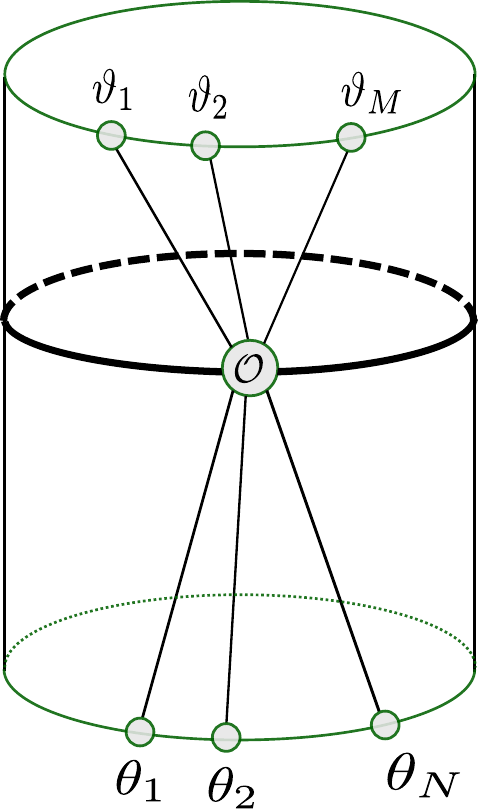}
\par\end{centering}
\caption{Interpretation of the formula \eqref{eq:FregFterm}. A member of a
virtual particle pair wraps the cylinder and both of them reaches
the operator. Source: \cite{preprint}}

\label{fig:LuscherF}
\end{figure}

The formula itself is an integral over the momentum of a mirror particle
pair, and the integrand has a form factor \eqref{eq:Fr}, which contains
this pair along with the physical particles from the original finite
volume form factor (see Fig. \ref{fig:LuscherF}). The kinematical
pole appearing due to the pair needs to be regulated in a - mathematically
``natural'' - symmetrical way.

The universal contour shown in Fig. \eqref{fig:contour} for the integrals
- which gave both the $\mu$-terms and the $F$-term correction in
case of the energy - applies for the newly developed form factor formula
in the same way. This fact was checked by extending a previously available
derivation on Lüscher's $\mu$-term corrections to the finite volume
non-diagonal form factor, in which a physical particle appears as
a bound state of two with complex rapidities.

This latter method offers not only the leading exponential correction,
but - as noted in \cite{pozsgaymu} for the energy - it also seems
to sum up all the higher $\mu$- term corrections. Moreover, the relation
between $F$- and $\mu$-terms  may be utilized to ``reengineer''
the integral terms  for higher exponential orders from this expansion
in powers of $e^{-\mu L}$, in which case the calculations in this
work could come handy. 

Since the derivations used the generic structure of the form factor
axioms, they could also serve as a starting point for analyzing leading
volume corrections corresponding to the lightest particles in general
diagonally scattering models with minimal modifications.

As stated in \cite{preprint}, in an unpublished work of Zoltán Bajnok
it was shown that the $F$-term also coincides with the appropriate
expansion of the existing exact formula for the finite volume diagonal
form factor \cite{pozsgayexactI} in the diagonal limit \cite{bajnokchao}.
This relation may help in the future to check the higher-order exponential
corrections in an exact way for the non-diagonal case.

\cleardoublepage{}

\lhead[]{Acknowledgments}

\rhead[Acknowledgments]{}

\chapter*{Acknowledgments}

\addcontentsline{toc}{chapter}{Acknowledgments} 

Here, I would like to thank Zoltán Bajnok: he introduced me to this
beautiful topic while encouraging me along the path to understand
it with a lot of helpful discussions, and he always showed great patience
towards me when having difficulties.

I would also like to thank Áron Bodor and Márton Lájer   for the fruitful
times we worked and understood issues together.

I am grateful to my family and to my friends for that they are supporting
me in studying physics and helping me through setbacks and hard times.

\appendix

\lhead[\chaptername~\thechapter]{\rightmark}

\rhead[\leftmark]{}

\lfoot[\thepage]{}

\cfoot{}

\rfoot[]{\thepage}

\chapter{Finite parts of form factors\label{chap:Finite-parts-of}}

For a later use we define different ways of subtracting kinematical
and dynamical poles from FF-s.

A singularity appears between two arguments if their difference is
a specific imaginary value ($\pi$ in case of kinematical and the
fusion angle in case of the dynamical pole). The residue of this singularity
will depend on the base point of the limit, and it is fixed by \eqref{eq:KIN}
and \eqref{eq:DYN} to be the second argument - there $\theta$. In
a general case (we can use permutation axiom \eqref{eq:PERM} to move
the emphasized arguments into the first and second position):
\[
\Res_{\theta'=\theta}F(\ldots,\theta'+ia,\ldots,\theta-ib,\ldots)=iR(\theta).
\]
However, if we would like to subtract the pole we can vary this base
point between $\theta'$ and $\theta$, because it only changes the
finite part:
\begin{align*}
 & F^{\left[\alpha\right]}(\ldots,\theta+ia,\ldots,\theta-ib,\ldots)\equiv\\
 & \equiv\lim_{\epsilon\to0}\{F(\ldots,\theta+ia+\epsilon\left(\frac{1+\alpha}{2}\right),\ldots,\theta-ib-\epsilon\left(\frac{1-\alpha}{2}\right),\ldots)-\frac{i}{\epsilon}R(\theta)\}.
\end{align*}
The specific cases $\alpha=\pm1,0$ will be useful. The difference
between them is the derivative of the residue:
\begin{equation}
F^{\left[\pm1\right]}(\ldots,\theta+ia,\ldots,\theta-ib,\ldots)=F^{\left[0\right]}(\ldots,\theta+ia,\ldots,\theta-ib,\ldots)\pm\frac{i}{2}\partial_{\theta}R(\theta).\label{eq:alpha}
\end{equation}
Let us introduce a notation to indicate which pair of arguments the
singularity was subtracted from, and at which value of $\alpha$
\[
\begin{cases}
F^{\left[+1\right]}(\ldots,\overset{\bullet}{\theta}+ia,\ldots,\overset{\circ}{\theta}-ib,\ldots) & \alpha=+1\\
F^{\left[\,0\,\right]}(\ldots,\overset{\bullet}{\theta}+ia,\ldots,\overset{\bullet}{\theta}-ib,\ldots) & \alpha=0\\
F^{\left[-1\right]}(\ldots,\overset{\circ}{\theta}+ia,\ldots,\overset{\bullet}{\theta}-ib,\ldots) & \alpha=-1
\end{cases}.
\]
The particular cases which we encounter in calculations are:
\begin{align*}
 & F_{N+1}^{\left[\pm1\right]}(\theta_{1},\ldots,\overset{{\bullet\atop \circ}}{\theta}_{k}+iu,\overset{{\circ\atop \bullet}}{\theta}_{k}-iu,\ldots,\theta_{N})=\\
 & =F_{N+1}^{\left[0\right]}(\theta_{1},\ldots,\overset{\bullet}{\theta}_{k}+iu,\overset{\bullet}{\theta}_{k}-iu,\ldots,\theta_{N})\pm\frac{i}{2}\partial_{\theta_{k}}F_{N}(\theta_{1},\ldots,\theta_{N}),
\end{align*}

for dynamical poles, and 

\begin{align*}
 & F_{M+N+2}^{[\pm1]}(\overset{{\bullet\atop \circ}}{v}+i\pi,\{\vartheta+\frac{i\pi}{2}\},\overset{{\circ\atop \bullet}}{v},\{\theta-\frac{i\pi}{2}\})=\\
 & =F_{M+N+2}^{[0]}(\overset{\bullet}{v}+i\pi,\{\vartheta+\frac{i\pi}{2}\},\overset{\bullet}{v},\{\theta-\frac{i\pi}{2}\})\pm\\
 & \pm\partial_{v}\frac{i}{2}[\prod_{j=1}^{M}S((\vartheta_{j}+\frac{i\pi}{2})-v)-\\
 & -\prod_{k=1}^{N}S(v-(\theta_{j}-\frac{i\pi}{2}))]F_{M+N}(\{\vartheta+\frac{i\pi}{2}\},\{\theta-\frac{i\pi}{2}\})
\end{align*}

for kinematical ones.

\chapter{Derivation of the $\mu$-term for elementary form factors\label{chap:Derivation-of-the}}

\section{Useful relations based on bound state quantization\label{sec:Useful-relations-based}}

From the definition \eqref{eq:dupm} of $\delta u_{j\pm}$ by using
\eqref{eq:LYSmxpoles} and \eqref{eq:LYbootstrap}:

\begin{align*}
 & e^{\pm iQ_{k\pm}^{(0)}(\{\theta_{j}\pm i(u+\delta u_{j\pm})\}_{j=1}^{N})}=1=\\
 & =\overbrace{S(2i(u+\delta u_{k\pm}))}^{\frac{\Gamma^{2}}{2\delta u_{k\pm}}}e^{\pm imL\sinh(\theta_{k}\pm iu)}\times\\
 & \times\prod_{j\neq k}[\overbrace{S(\theta_{k,j}-iu\pm iu)S(\theta_{k,j}+iu\pm iu)}^{S(\theta_{k,j}\pm iu)}]^{\pm1}+\Ordo(e^{-\mu L}),
\end{align*}
where we could use the bootstrap equations only because the $\Ordo(e^{-\mu L})$
corrections to the fusing angle in $\theta_{j}\pm i(u+\delta u_{j\pm})$
became negligible\footnote{For simplicity we follow only the leading order in the notation and
usually write equalities in the sense that the two sides differs in
an $\Ordo(e^{-2\mu L})$ term. } - since we just determined the leading order here. The resulting
functional form of the $\delta u_{j\pm}$-s is
\[
\delta u_{k\pm}(\{\theta\})=\frac{\Gamma^{2}}{2}e^{\pm imL\sinh(\theta_{k}\pm iu)}\prod_{j:j\neq k}S(\theta_{k,j}\pm iu)^{\pm1},
\]
where instead of following the corrections of $\theta_{j}$-s, we
use the generic variables. As mentioned in \eqref{sec:-term-correction},
these two necessarily coincide at the solution of the BY equations
and become real:
\begin{equation}
\delta u_{k+}(\{\bar{\theta}^{(0)}\})=\delta u_{k-}(\{\bar{\theta}^{(0)}\})=\delta\bar{u}_{k}.\label{eq:onshell}
\end{equation}
Now we establish some useful relations among derivatives. We denote
derivatives as $\partial_{j\pm}\equiv\partial_{\theta_{j\pm}}$, and
introduce the indices $\rho,\sigma=\pm$. The matrix $\partial_{j\rho}Q_{k\sigma}^{(0)}(\{\theta_{\pm}\})$
is symmetric for $j\rho\leftrightarrow k\sigma$ since its off-diagonal
elements are $\partial_{j\rho}\delta(\theta_{k\sigma,j\rho})=-\varphi(\theta_{k\sigma,j\rho})$,
and the function $\varphi(\theta)=\partial_{\theta}\delta(\theta)$
is symmetric, i.e. $\varphi(\theta)=\varphi(-\theta)$ by unitarity
of $\delta$.

Taking the functions \eqref{eq:Q} and \eqref{eq:antiQ} at the $(0)$-th
iteration, i.e. with $Q_{k\pm}^{(0)}$, one gets $Q_{k}^{(0)}(\{\theta_{\pm}\})$
and $\tilde{Q}_{k}^{(0)}(\{\theta_{\pm}\})$.\footnote{The function $Q_{k}^{(0)}(\{\theta_{\pm}\})=Q_{k+}^{(0)}(\{\theta_{\pm}\})+Q_{k-}^{(0)}(\{\theta_{\pm}\})$
is not to be confused with $Q_{k}^{(0)}(\{\theta\})$, they only coincide
for $\theta_{j\pm}=\theta_{j}\pm iu$, which is not generally true
(!)} One can also similarly introduce the variables and derivatives 
\begin{align*}
\theta_{j\pm} & =\theta_{j}\pm\tilde{\theta}_{j}\;\begin{cases}
\partial_{j}=\partial_{j+}+\partial_{j-}\\
\tilde{\partial}_{j}=\partial_{j+}-\partial_{j-}
\end{cases}
\end{align*}
The symmetry of the above mentioned Jacobian presents itself as $\partial_{i}Q_{j}^{(0)}(\{\theta_{\pm}\})$
and $\tilde{\partial}_{i}\tilde{Q}_{j}^{(0)}(\{\theta_{\pm}\})$ being
symmetric in $i\leftrightarrow j$, and $\tilde{\partial}_{i}Q_{j}^{(0)}(\{\theta_{\pm}\})=\partial_{j}\tilde{Q}_{i}^{(0)}(\{\theta_{\pm}\})$.
Also $\partial_{i}Q_{j\pm}^{(0)}(\{\theta_{\pm}\})=\partial_{j\pm}Q_{i}^{(0)}(\{\theta_{\pm}\}),\;\tilde{\partial}_{i}Q_{j\pm}^{(0)}(\{\theta_{\pm}\})=\partial_{j\pm}\tilde{Q}_{i}^{(0)}(\{\theta_{\pm}\})$
is true.

One sees that the only term in $Q_{k\pm}^{(0)}(\{\theta_{\pm}\})$
(see \eqref{eq:mutermqc}) as a function of variables $\theta_{j},\tilde{\theta}_{j}$
which does not depend on $\theta_{j}$ is the one where the pole comes
from: $-i\ln S(\pm\tilde{\theta}_{k})$. This way we can write
\begin{align}
\partial_{j}\delta u_{k\pm}(\{\theta\}) & =\partial_{j}\overbrace{\frac{\Gamma^{2}}{2}\frac{e^{\pm iQ_{k\pm}^{(0)}(\{\theta_{\pm}\})}}{S(\tilde{\theta}_{k})}}^{\delta u_{k\pm}}=\label{eq:duderivative}\\
 & =\pm i\partial_{j}Q_{k\pm}^{(0)}(\{\theta\pm iu\})\delta u_{k\pm}(\{\theta\}).\nonumber 
\end{align}
Let us take $Q_{k}^{(0)}(\{\theta_{\pm}\})$ at $\theta_{j\pm}=\theta_{j}\pm i(u+\delta u_{j\pm})$,
and expand it in $\delta u_{j\pm}$-s:
\begin{align*}
 & Q_{k}^{(0)}(\{\theta_{j}\pm i(u+\delta u_{j\pm})\}_{j=1}^{N})=\\
 & =\overbrace{Q_{k}^{(0)}(\{\theta\pm iu\})}^{Q_{k}^{(0)}(\{\theta\})}+\sum_{j}\{i\delta u_{j+}\overbrace{\partial_{j+}Q_{k}^{(0)}}^{\partial_{k}Q_{j+}^{(0)}}(\{\theta\pm iu\})-i\delta u_{j-}\overbrace{\partial_{j-}Q_{k}^{(0)}}^{\partial_{k}Q_{j-}^{(0)}}(\{\theta\pm iu\})\}
\end{align*}
where the r.h.s. by \eqref{eq:duderivative} is the same as that of
\eqref{eq:rewrittenqc}. At the solution we get the same expansion
either starting directly from 
\begin{align*}
 & Q_{k}^{(0)}(\{\bar{\theta}_{j}^{(\mu)}\pm i(u+\delta\bar{u}_{j})\}_{j=1}^{N})\equiv Q_{k}^{(\mu)}(\{\bar{\theta}^{(\mu)}\})=\\
 & =Q_{k}^{(0)}(\{\bar{\theta}^{(\mu)}\})+i\sum_{j}\delta\bar{u}_{j}\partial_{k}\tilde{Q}_{j}^{(0)}(\{\bar{\theta}^{(0)}\pm iu\}))=2\pi n_{k},
\end{align*}
or taking \eqref{eq:rewrittenqc} at $\{\bar{\theta}^{(\mu)}\}$.
In the first case the equations $\tilde{Q}_{k}^{(0)}(\{\bar{\theta}^{(\mu)}\pm i\bar{u}^{(\mu)})\})=2\pi(n_{k+}-n_{k-})$
are also necessary; in the second case the $\delta u_{j\pm}(\{\theta\})$-s
take care of this extra $N$ number of constraints in $Q_{k}^{(\mu)}(\{\theta\})$.

We should also show that the functions $\delta u_{k\pm}$ are the
residual contributions of the integral $\delta\Phi$ as stated in
\eqref{eq:Res}. Substituting $\epsilon=\theta-(\theta_{k}\pm\frac{i\pi}{6})$
and using the crossing symmetry of $S$ to eliminate appearing $i\pi$-s
for the $\theta_{k}-\frac{i\pi}{6}$ case:
\begin{align*}
 & \Res_{\theta=(\theta_{k}\mp iu)\pm i\frac{\pi}{2}=\theta_{k}\pm\frac{i\pi}{6}}\prod_{j}S(\theta_{j}+i\frac{\pi}{2}-\theta)e^{-mL\cosh\theta}=\\
 & =\overbrace{\lim_{\epsilon=0}\epsilon S(iu\mp\epsilon)}^{\mp i\Gamma^{2}}\prod_{j:j\neq k}\overbrace{S(\pm\theta_{j,k}+iu)}^{S(\theta_{k,j}\mp iu)^{\mp1}}e^{\mp mL\cosh(\theta_{k}\mp iu)}=\mp2i\delta u_{k\mp}(\{\theta\})
\end{align*}

\section{Density of states}

First we show the validity of expansion \eqref{eq:densities}. We
rewrite the Jacobian determinant of the exact quantization conditions
$Q_{k\pm}(\{\theta_{\pm}\})$ in terms of $Q_{k}(\{\theta\},\{\tilde{\theta}\})$
and $\tilde{Q}_{k}(\{\theta\},\{\tilde{\theta}\})$ :
\begin{align}
\rho_{2N}(\{\theta_{\pm}\}) & =\det\frac{\partial\left(Q_{1+},Q_{1-},\ldots,Q_{N+},Q_{N-}\right)}{\partial\left(\theta_{1+},\theta_{1-},\ldots,\theta_{N+},\theta_{N-}\right)}=\frac{1}{4^{N}}\begin{vmatrix}\left[\tilde{\partial}\tilde{Q}\right] & \left[\tilde{\partial}Q\right]\\
\left[\partial\tilde{Q}\right] & \left[\partial Q\right]
\end{vmatrix}=\nonumber \\
 & =\frac{1}{4^{N}}\det\left[\tilde{\partial}\tilde{Q}\right]\det\left(\underbrace{\left[\partial Q\right]-\left[\partial\tilde{Q}\right]\left[\tilde{\partial}\tilde{Q}\right]^{-1}\left[\tilde{\partial}Q\right]}_{\equiv M}\right)\label{eq:rho2N}
\end{align}

where we subtracted and added simultaneously the members in each pair
of rows with $\partial_{i\pm}$ and in columns with $Q_{j\pm}$, such
that the blocks $\left[\tilde{\partial}\tilde{Q}\right]_{ij}=\tilde{\partial}_{i}\tilde{Q}_{j}(\{\theta_{\pm}\})$,
$\left[\tilde{\partial}Q\right]_{ij}=\left[\partial\tilde{Q}\right]_{ji}=\tilde{\partial}_{i}Q_{j}(\{\theta_{\pm}\})$
and $\left[\partial Q\right]_{ij}=\partial_{i}Q_{j}(\{\theta_{\pm}\})$
show up. A $1/2$ factor comes from each step of this reorganization
resulting in $1/4^{N}$. Next we used that for block matrices 
\[
\det\begin{pmatrix}A & B\\
C & D
\end{pmatrix}=\det A\det\left(D-CA^{-1}B\right),
\]
if $A$ is invertible \cite{blockdet}, which is true for $\left[\tilde{\partial}\tilde{Q}\right]$
in the general case. From the explicit form \eqref{eq:mutermqc} of
$Q_{j\pm}^{(0)}$-s one can get the identity
\[
\tilde{\partial}_{j}\tilde{Q}_{j}^{(0)}(\{\theta\},\{\tilde{\theta}\})=4\varphi(2\tilde{\theta}_{j})+\partial_{j}Q_{j}^{(0)}(\{\theta\},\{\tilde{\theta}\})
\]
and substitute $\tilde{\theta}_{j}=i(u+\delta\bar{u}_{j})$ in it,
which shows that - since $\varphi(2i(u+\delta\bar{u}_{k}))\simeq(2\delta\bar{u}_{k})^{-1}$
has a pole coming from the singularity of the S-matrix - the singular
behaviour of $\rho_{2N}^{(0)}$ comes from the determinant of $[\tilde{\partial}\tilde{Q}^{(0)}]$. 

This gets compensated by the prefactor in \eqref{eq:densities}, which
simplifies if we rewrite it with the help of $\varphi$: 
\[
\left(\frac{2\delta\bar{u}_{k}}{\Gamma}\right)^{2}S(2i(u+\delta\bar{u}_{k}))=2\delta\bar{u}_{k}\left(1+2\Gamma^{-2}\delta\bar{u}_{k}S_{0}\right)+\Ordo(\delta\bar{u}_{k}^{3})=\varphi(2i(u+\delta\bar{u}_{k}))^{-1}+\Ordo(\delta\bar{u}_{k}^{3}).
\]
To see this we divide each line of the matrix $[\tilde{\partial}\tilde{Q}^{(0)}]$
in the determinant with the appropriate $4\varphi(2i\bar{u}_{k})\simeq2\delta\bar{u}_{k}^{-1},\;\bar{u}_{k}^{(\mu)}=u+\delta\bar{u}_{k}$
from the product:
\begin{align}
 & \left(\frac{2\delta\bar{u}_{k}}{\Gamma}\right)^{2}S(2i\bar{u}_{k})\frac{1}{4^{N}}\det[\tilde{\partial}\tilde{Q}^{(0)}]_{\theta_{j\pm}=\bar{\theta}_{j\pm}^{(\mu)}}=\nonumber \\
 & =\prod_{k}\left(\frac{1}{4\varphi(2i\bar{u}_{k}^{(\mu)})}\right)\det\begin{bmatrix}4\varphi(2i\bar{u}_{1}^{(\mu)})+\partial_{1}Q_{1}^{(0)} & \cdots & \tilde{\partial}_{1}\tilde{Q}_{N}^{(0)}\\
\vdots & \ddots & \vdots\\
\tilde{\partial}_{N}\tilde{Q}_{1}^{(0)} & \cdots & 4\varphi(2i\bar{u}_{N}^{(\mu)})+\partial_{N}Q_{N}^{(0)}
\end{bmatrix}\nonumber \\
 & =\det\begin{bmatrix}1+\frac{1}{2}\partial_{1}Q_{1}^{(0)}\delta\bar{u}_{1} & \cdots & \frac{1}{2}\tilde{\partial}_{1}\tilde{Q}_{N}^{(0)}\delta\bar{u}_{1}\\
\vdots & \ddots & \vdots\\
\frac{1}{2}\tilde{\partial}_{N}\tilde{Q}_{1}^{(0)}\delta\bar{u}_{N} & \cdots & 1+\frac{1}{2}\partial_{N}Q_{N}^{(0)}\delta\bar{u}_{N}
\end{bmatrix}\simeq\underline{1+\sum_{k}\frac{1}{2}\partial_{k}Q_{k}^{(0)}(\{\bar{\theta}^{(0)}\})\delta\bar{u}_{k}}\label{eq:onehalfcorr}
\end{align}
where in the last equality we used $\det(I+X)\simeq1+\Tr(X)$. The
matrix under the second determinant in the last line of \eqref{eq:rho2N}
simplifies as
\begin{align}
M_{ij} & \equiv\left(\partial_{i}Q_{j}^{(0)}-\partial_{i}\tilde{Q}_{k}^{(0)}[\tilde{\partial}\tilde{Q}^{(0)}]_{kl}^{-1}\tilde{\partial}_{l}Q_{j}^{(0)}\right)_{\{\theta_{\pm}\}=\{\bar{\theta}_{\pm}^{(\mu)}\}}\label{eq:determinant}\\
 & \simeq\underline{\partial_{i}Q_{j}^{(0)}(\{\bar{\theta}_{\pm}^{(\mu)}\})-\sum_{k}\frac{1}{2}\partial_{i}\tilde{Q}_{k}^{(0)}\tilde{\partial}_{k}Q_{j}^{(0)}\delta\bar{u}_{k}},\nonumber 
\end{align}
since at the leading order $[\tilde{\partial}\tilde{Q}^{(0)}]_{\theta_{j\pm}=\bar{\theta}_{j\pm}^{(\mu)}}^{-1}\simeq\diag(\delta\bar{u}_{1}/2,\ldots,\delta\bar{u}_{N}/2)$.

Now we approach from the other direction, and manipulate the r.h.s.
of \eqref{eq:densities} where $\rho^{(\mu)}(\{\theta\})=\det J^{(\mu)}(\{\theta\})$:
\begin{align*}
J_{ij}^{\left(\mu\right)}(\{\theta\}) & =\partial_{i}Q_{j}^{\left(0\right)}(\{\theta\})+\partial_{i}\partial_{j}\sum_{k}\left(\delta u_{k+}(\{\theta\})+\delta u_{k-}(\{\theta\})\right)=\\
 & =\partial_{i}Q_{j}^{\left(0\right)}+i\partial_{i}\left(\partial_{j}Q_{k+}^{(0)}\delta u_{k+}-\partial_{j}Q_{k-}^{(0)}\delta u_{k-}\right)=\\
 & =\partial_{i}Q_{j}^{\left(0\right)}+\sum_{k}\{i\left(\partial_{i}\partial_{j}Q_{k+}^{(0)}\delta u_{k+}-\partial_{i}\partial_{j}Q_{k-}^{(0)}\delta u_{k-}\right)-\\
 & -\left(\partial_{i}Q_{k+}^{(0)}\partial_{j}Q_{k+}^{(0)}\delta u_{k+}+\partial_{i}Q_{k-}^{(0)}\partial_{j}Q_{k-}^{(0)}\delta u_{k-}\right)\}.
\end{align*}
Taking this at the solution gives:
\begin{align}
 & J_{ij}^{\left(\mu\right)}(\{\bar{\theta}^{(\mu)}\})=\nonumber \\
 & =\partial_{i}Q_{j}^{\left(0\right)}(\{\bar{\theta}^{(\mu)}\})+\sum_{k}\left[i\partial_{i}\partial_{j}\tilde{Q}_{k}^{(0)}-\left(\partial_{i}Q_{k+}^{(0)}\partial_{j}Q_{k+}^{(0)}+\partial_{i}Q_{k-}^{(0)}\partial_{j}Q_{k-}^{(0)}\right)\right]\delta\bar{u}_{k}=\nonumber \\
 & =\underline{\partial_{i}Q_{j}^{\left(0\right)}(\{\bar{\theta}^{(\mu)}\})+\sum_{k}\left[i\partial_{i}\partial_{j}\tilde{Q}_{k}^{(0)}-\frac{1}{2}\left(\partial_{i}\tilde{Q}_{k}^{(0)}\partial_{j}\tilde{Q}_{k}^{(0)}+\partial_{i}Q_{k}^{(0)}\partial_{j}Q_{k}^{(0)}\right)\right]\delta\bar{u}_{k}},\label{eq:Jij}
\end{align}
where in the last line we extended the expression as
\begin{align*}
 & \partial_{i}Q_{k+}\partial_{j}Q_{k+}+\partial_{i}Q_{k-}\partial_{j}Q_{k-}=\\
 & =\left(\partial_{i}Q_{k+}-\partial_{i}Q_{k-}\right)\left(\partial_{j}Q_{k+}-\partial_{j}Q_{k-}\right)+\left(\partial_{i}Q_{k+}+\partial_{i}Q_{k-}\right)\left(\partial_{j}Q_{k+}+\partial_{j}Q_{k-}\right)=\\
 & =\partial_{i}\tilde{Q}_{k}\partial_{j}\tilde{Q}_{k}+\partial_{i}Q_{k}\partial_{j}Q_{k}.
\end{align*}
We get back to \eqref{eq:determinant}, and expand the first term:
\begin{align*}
\partial_{i}Q_{j}^{(0)}(\{\bar{\theta}_{\pm}^{(\mu)}\}) & =\partial_{i}Q_{j}^{(0)}(\{\bar{\theta}^{(\mu)}\})+\sum_{k}i(\partial_{k+}\partial_{i}Q_{j}^{(0)}\delta\bar{u}_{k}-\partial_{k-}\partial_{i}Q_{j}^{(0)}\delta\bar{u}_{k})=\\
 & =\partial_{i}Q_{j}^{(0)}(\{\bar{\theta}^{(\mu)}\})+\sum_{k}i\partial_{i}\partial_{j}\tilde{Q}_{k}^{(0)}\delta\bar{u}_{k},
\end{align*}
the last line comes by using $\tilde{\partial}_{k}Q_{j}^{(0)}=\partial_{j}\tilde{Q}_{k}^{(0)}$.
With this, \eqref{eq:determinant} looks just like the beginning of
\eqref{eq:Jij}, and so their relation is:
\[
M_{ij}=J_{ij}^{\left(\mu\right)}(\{\bar{\theta}^{(\mu)}\})+\overbrace{\frac{1}{2}\sum_{k}\partial_{i}Q_{k}^{(0)}\partial_{j}Q_{k}^{(0)}\delta\bar{u}_{k}}^{\equiv\frac{1}{2}\delta Q_{ij}}.
\]
Taking the determinant of $M$ and factoring out $J^{(\mu)}$ we have:
\begin{align*}
\det M & =\det\left(J^{\left(\mu\right)}+\delta Q\right)=\det J^{(\mu)}\det(I+J^{(\mu)-1}\delta Q)\simeq\\
 & \simeq\det J^{(\mu)}\det(I+\frac{1}{2}[\partial Q^{(0)}]^{-1}\delta Q)\simeq\underline{\det J^{(\mu)}\left(1+\frac{1}{2}\Tr([\partial Q^{(0)}]^{-1}\delta Q)\right)}
\end{align*}
since $\delta Q$ is $\Ordo(e^{-\mu L})$ and this way we can drop
such corrections in $J^{(\mu)-1}$ too. The term simplifies as
\[
\frac{1}{2}\Tr\left([\partial Q^{(0)}]^{-1}\delta Q\right)=\frac{1}{2}\sum_{j}\sum_{i}\sum_{k}[\partial Q^{(0)}]_{ji}^{-1}\partial_{i}Q_{k}^{(0)}\partial_{j}Q_{k}^{(0)}\delta\bar{u}_{k}=\frac{1}{2}\sum_{k}\partial_{k}Q_{k}^{(0)}\delta\bar{u}_{k},
\]
which appears in \eqref{eq:onehalfcorr} already once. Assembling
together the l.h.s of \eqref{eq:densities} (taking everything at
$\{\bar{\theta}^{(\mu)}\}$) we have
\[
\left(\frac{2\delta\bar{u}_{k}}{\Gamma}\right)^{2}S(2i\bar{u}_{k})\frac{1}{4^{N}}\det[\tilde{\partial}\tilde{Q}^{(0)}]\det M\simeq\rho^{(\mu)}(\{\bar{\theta}^{(\mu)}\})\left(1+\frac{1}{2}\sum_{k}\partial_{k}Q_{k}^{(0)}\delta\bar{u}_{k}\right)^{2},
\]
 which indeed gives the r.h.s of \eqref{eq:densities} after expanding
the square.

\section{Phase correction}

By substituting $\theta_{k\pm}=\theta_{k}\pm i(u+\delta u_{k\pm}(\{\theta\}))$
and using the bootstrap for the product of S-matrices in \eqref{eq:mutermtemplate}
\[
\prod_{\rho,\sigma=\pm}S(\theta_{i\rho,j\sigma})=S(\theta_{i,j})\prod_{\rho,\sigma=\pm}\left(1\overbrace{-\varphi(\theta_{i\rho,j\sigma})}^{+\partial_{j\sigma}Q_{i\rho}^{(0)}}(\rho\delta u_{i\rho}-\sigma\delta u_{j\sigma})\right).
\]
At the solutions $\{\bar{\theta}^{(\mu)}\}$ this becomes
\begin{align}
\prod_{\rho,\sigma=\pm}S(\bar{\theta}_{i\rho,j\sigma}^{(\mu)}) & =S(\bar{\theta}_{i,j}^{(\mu)})\left(1+\sum_{\rho,\sigma=\pm}\partial_{j\sigma}Q_{i\rho}^{(0)}(\rho\delta\bar{u}_{i}-\sigma\delta\bar{u}_{j})\right)=\label{eq:phasecorr}\\
 & =S(\bar{\theta}_{i,j}^{(\mu)})\left(1+\partial_{j}\tilde{Q}_{i}^{(0)}\delta\bar{u}_{i}-\tilde{\partial}_{j}Q_{i}^{(0)}\delta\bar{u}_{j}\right)\nonumber 
\end{align}
and also
\[
\prod_{\rho,\sigma=\pm}S(\bar{\theta}_{i\rho,j\sigma}^{(\mu)})=S(\bar{\theta}_{i,j}^{(\mu)})\{1-[\overbrace{\varphi(\bar{\theta}_{i,j}^{(0)}+2iu)-\varphi(\bar{\theta}_{i,j}^{(0)}-2iu)}^{i\Delta\varphi(\bar{\theta}_{i,j}^{(0)})}](\delta\bar{u}_{i}+\delta\bar{u}_{j})\},
\]
since the terms like $\varphi(\bar{\theta}_{i,j}^{(0)})(\delta\bar{u}_{i}-\delta\bar{u}_{j})$
drop out. The $\Delta\varphi(\theta)$ function is real analytic and
so this correction is purely imaginary in contrast to the other $\mu$-terms
which are real.

\section{Form factor}

In the numerator of \eqref{eq:mutermtemplate} we can start to successively
expand dynamical singularities since they are non-overlapping. They
appear among arguments in disjoint pairs, and if we choose each of
the two argument from different pairs they will not be close to pole
position: as discussed in \eqref{sec:Form-factors-in}, non-diagonal
finite volume form factors cannot have coinciding rapidities in the
``fermionic'' case which we deal with.

Substituting $\theta_{j\pm}=\theta_{j}\pm i(u+\delta_{j})$ in $F_{4}(\theta_{1+},\theta_{1-},\theta_{2+},\theta_{2-})$,
where $\delta$-s are small parameters:
\begin{align*}
F_{4}(\{\theta_{\pm}\}) & =\frac{\Gamma}{2\delta_{1}}F_{3}(\theta_{1},\theta_{2+},\theta_{2-})+F_{4}^{\left[0\right]}(\overset{\bullet}{\theta}_{1}+iu,\overset{\bullet}{\theta}_{1}-iu,\theta_{2+},\theta_{2-})+\Ordo(\delta_{1})=\\
 & =\frac{\Gamma}{2\delta_{1}}\left[\frac{\Gamma}{2\delta_{2}}F_{2}(\theta_{1},\theta_{2})+F_{3}^{\left[0\right]}(\theta_{1},\overset{\bullet}{\theta}_{2}+iu,\overset{\bullet}{\theta}_{2}-iu)+\Ordo(\delta_{2})\right]+\\
 & +\frac{\Gamma}{2\delta_{2}}F_{3}^{\left[0\right]}(\overset{\bullet}{\theta}_{1}+iu,\overset{\bullet}{\theta}_{1}-iu,\theta_{2})+\Ordo(1)+\Ordo(\delta_{1}),
\end{align*}
where the subtraction of both poles gave an $\Ordo(1)$ part. By multiplying
the l.h.s. with $\frac{2\delta_{1}}{\Gamma}\frac{2\delta_{2}}{\Gamma}$
we have
\[
\frac{2\delta_{1}}{\Gamma}\frac{2\delta_{2}}{\Gamma}F_{4}(\{\theta_{\pm}\})=F_{2}(\theta_{1},\theta_{2})+\sum_{k=1}^{2}\frac{2\delta_{k}}{\Gamma}F_{2,k}^{b}(\theta_{1},\theta_{2})+\Ordo(\delta^{2}),
\]
where $F_{N,k}^{b}$ is defined in \eqref{sec:-term-correction} and
$\Ordo(\delta^{2})$ means that the remaining part is quadratic in
$\delta_{1},\delta_{2}$. The generalization to \eqref{eq:F2N} is
straightforward.

\chapter{Residues of the $F$-term for elementary form factors}

\section{Poles of the regulated form factor\label{sec:Poles-in-the}}

Now we calculate the residue of the pole in $F_{N+2}^{\left[-1\right]}(\overset{\circ}{v}+i\pi,\overset{\bullet}{v},\{\theta-\frac{i\pi}{2}\})$
at $v=\theta_{k}-\frac{i\pi}{6}$ whose relevance was discussed in
\eqref{subsec:Pole-structure-of}. To achieve this we investigate
the singularities of $F_{N+2}(v+i\pi,v-\epsilon,\{\theta-\frac{i\pi}{2}\})$
at $v=\theta_{k}+iu-\frac{i\pi}{2}+i\delta$, in terms of the infinitesimals
$\epsilon$ and $\delta$. The $\mathcal{O}(\delta^{0})$ terms are
not important to us, since we need the residue of the integrand in
\eqref{eq:FtermofFF}.

We move the arguments of the subtraction close to $\theta_{k}-\frac{i\pi}{2}$
by permutation and periodicity:
\begin{align}
 & F_{N+2}(v+i\pi,v-\epsilon,\ldots,\theta_{k}-\frac{i\pi}{2},\ldots)=\nonumber \\
 & =\prod_{j<k}S(v-\epsilon-\theta_{j}+\frac{i\pi}{2})F_{N+2}(v+i\pi,\ldots,v-\epsilon,\theta_{k}-\frac{i\pi}{2},\ldots)=\\
 & =\prod_{j<k}S(v-\theta_{j}+\frac{i\pi}{2}-\epsilon)\prod_{j>k}S(\theta_{j}-v+\frac{i\pi}{2})F_{N+2}(\ldots,v-\epsilon,\theta_{k}-\frac{i\pi}{2},v-i\pi,\ldots)\label{eq:sandwich}
\end{align}
It is also allowed to shift all of the arguments of the FF by $\frac{i\pi}{2}$
(see \eqref{eq:LOR} for scalar operators). If we look at the emphasized
arguments $v+\frac{i\pi}{2}-\epsilon,\theta_{k},v-\frac{i\pi}{2}$
at the pole:
\begin{align*}
\theta_{k}+iu\underbracket{+i\delta-\epsilon,\quad\overbracket{\theta_{k},\quad\theta_{k}+iu+}^{\mathrm{dyn.}}}_{\mathrm{kin.}}i\delta-i\pi
\end{align*}
and we think of the FF as a function of the differences of its arguments
(which we can do for scalar operators), we see that there is a new
dynamical pole between the second and the third argument in $\delta$
(since $\pi-u=2u$), and the original kinematical one is present between
the first and the third in $\epsilon$. Therefore the $\delta$ and
$\epsilon$ singularities are independent and the structure of $F_{N+2}$
in general looks like ($\theta_{k\pm}\equiv\theta_{k}\pm iu+i\delta$):
\[
F_{N+2}(\ldots,\theta_{k+}\underbracket{-\epsilon,\;\overbracket{\theta_{k},\;\theta_{k+}}^{\delta}}_{\epsilon}-i\pi\ldots)=\frac{A}{\delta\epsilon}+\frac{B}{\delta}+\frac{C}{\epsilon}+\mathcal{O}(1)
\]
If we use the dynamical axiom, we can get the terms proportional to
$\delta^{-1}$:
\begin{align}
 & F_{N+2}(\ldots,\theta_{k+}-\epsilon,\theta_{k},\theta_{k}-2iu+i\delta,\ldots)=\label{eq:AB}\\
 & =\frac{i\Gamma}{(-i\delta)}F_{N+1}(\ldots,\theta_{k+}-\epsilon,\theta_{k}-iu+i\delta,\ldots)+\mathcal{O}(\delta^{0})\nonumber \\
 & =\frac{1}{\delta}\left(\frac{A}{\epsilon}+B\right)+\mathcal{O}(\delta^{0}).\nonumber 
\end{align}
Note that we chose the base point (in the sense of App. \ref{chap:Finite-parts-of})
so that the newly formed dynamical singularity in $F_{N+1}$ becomes
a $\epsilon^{-1}$ pole. This way we can guarantee that our expansion
has indeed the same form as the right hand side of \eqref{eq:AB}.

There is only one step left - expanding the new dynamical pole in
$\epsilon$ - to get the coefficients $A$ and $B$:
\[
F_{N+1}(\ldots,\theta_{k+}-\epsilon,\theta_{k-},\ldots)=\frac{i\Gamma}{(-\epsilon)}F_{N}(\ldots,\theta_{k}+i\delta,\ldots)+F_{N+1}^{\left[+1\right]}(\ldots,\overset{\bullet}{\theta}_{k+},\overset{\circ}{\theta}_{k-},\ldots)+\mathcal{O}(\epsilon),
\]
where $\delta$-s in the arguments of $F_{N}$ and $F_{N+1}^{\left[+1\right]}$
can be neglected due to the $\delta^{-1}$ factor in \eqref{eq:AB}.
We can conclude that
\begin{align*}
A & =i\Gamma^{2}F_{N}(\{\theta\})\\
B & =-\Gamma F_{N+1}^{\left[+1\right]}(\ldots,\overset{\bullet}{\theta}_{k}+iu,\overset{\circ}{\theta}_{k}-iu,\ldots)\equiv-\Gamma F_{N,k+}^{b}(\{\theta\}).
\end{align*}
If we write the prefactor of S-matrices in \eqref{eq:sandwich} at
$v=\theta_{k+}-\frac{i\pi}{2}$ as ($\hat{\theta}_{k+,j}\equiv\theta_{k,j}+iu$):
\begin{align*}
 & \prod_{j<k}S(\theta_{k+,j}-\epsilon)\prod_{j>k}S(\theta_{j,k+}+i\pi)=\\
 & =\overbrace{\prod_{j\neq k}S(\hat{\theta}_{k+,j})}^{D}-\epsilon\overbrace{\left[\prod_{j<k}S(\hat{\theta}_{k+,j})\right]^{'}\prod_{j>k}S(\hat{\theta}_{k+,j})}^{E}+\mathcal{O}(\epsilon^{2})+\mathcal{O}(\delta)=\\
 & =D-\epsilon E+\mathcal{O}(\epsilon^{2})+\mathcal{O}(\delta),
\end{align*}
by putting things together we see that the subtracted FF indeed contributes
with a $\delta^{-1}$ pole (we should keep in mind that the explicit
$\epsilon^{-1}$ pole is subtracted by the definition of $F_{N+2}^{\left[-1\right]}$):
\begin{align*}
 & F_{N+2}^{\left[-1\right]}(\overset{\circ}{v}+i\pi,\overset{\bullet}{v},\{\theta-\frac{i\pi}{2}\})\vert_{v=\theta_{k+}-\frac{i\pi}{2}}=\\
 & =\left(D-\epsilon E+\mathcal{O}(\epsilon^{2})+\mathcal{O}(\delta)\right)\left(\frac{A}{\delta\epsilon}+\frac{B}{\delta}+\frac{C}{\epsilon}+\mathcal{O}(1)\right)-\frac{D}{\epsilon}\left(\frac{A}{\delta}+C\right)=\\
 & =\frac{-EA+DB}{\delta}+\mathcal{O}(\delta^{0})=\\
 & =\frac{-i\Gamma^{2}}{\delta}\left[\prod_{j<k}S(\hat{\theta}_{k+,j})\right]^{'}\prod_{j>k}S(\hat{\theta}_{k+,j})F_{N}(\{\theta\})-\frac{\Gamma}{\delta}\prod_{j\neq k}S(\hat{\theta}_{k+,j})F_{N,k+}^{b}(\{\theta\})+\mathcal{O}(\delta^{0}).
\end{align*}
The derivative term in \eqref{eq:FrtoFc} at the same position has
a particular structure where the simple pole in $S(\theta_{k+,k})=S(i(u+\delta))$
produces a single $\delta^{-2}$ term in the end. If we have two analytic
functions $f,g$ where $f$ has a singularity at $z_{0}$,
\begin{align*}
f(z) & =\frac{f_{-1}}{z-z_{0}}+\mathcal{O}(1)\\
g(z) & =g_{0}+g_{1}(z-z_{0})+\mathcal{O}((z-z_{0})^{2})
\end{align*}
their Leibniz-rule will look like
\begin{align*}
\partial_{z}[f(z)g(z)] & =\left(-\frac{f_{-1}}{(z-z_{0})^{2}}+\mathcal{O}(1)\right)\left(g_{0}+g_{1}(z-z_{0})+\mathcal{O}((z-z_{0})^{2})\right)+\\
 & +\left(\frac{f_{-1}}{z-z_{0}}+\mathcal{O}(1)\right)\left(g_{1}+\mathcal{O}(z-z_{0})\right)=\\
 & =-\frac{f_{-1}g_{0}}{(z-z_{0})^{2}}+\mathcal{O}(1)=f'(z)g(z_{0})+\mathcal{O}(1).
\end{align*}
So the contribution of this term is
\begin{align*}
 & \mp\frac{i}{2}\partial_{v}\left(1-\prod_{j}S(v-\theta_{j}+\frac{i\pi}{2})\right)\vert_{v=\theta_{k+}-\frac{i\pi}{2}}F_{N}(\{\theta\})=\\
 & =\frac{i}{2}\left[\prod_{j}S(\theta_{k+,j})\right]^{'}F_{N}(\{\theta\})=\\
 & =\frac{i}{2}S'(\theta_{k+,k})\left[\prod_{j\neq k}S(\hat{\theta}_{k+,j})\right]F_{N}(\{\theta\})+\mathcal{O}(\delta^{0}).
\end{align*}
Altogether, the regulated form factor is then
\begin{align}
 & F_{N+2}^{r}(v+i\pi,v,\{\theta-\frac{i\pi}{2}\})\vert_{v=\theta_{k+}-\frac{i\pi}{2}}=\nonumber \\
 & =-\frac{\Gamma^{2}}{2\delta^{2}}\left[\prod_{j\neq k}S(\hat{\theta}_{k+,j})\right]F_{N}(\{\theta\})-\frac{i\Gamma^{2}}{\delta}\left[\prod_{j<k}S(\hat{\theta}_{k+,j})\right]^{'}\prod_{j>k}S(\hat{\theta}_{k+,j})F_{N}(\{\theta\})-\nonumber \\
 & -\frac{\Gamma}{\delta}\prod_{j\neq k}S(\hat{\theta}_{k+,j})F_{N,k+}^{b}(\{\theta\})+\mathcal{O}(\delta^{0}).\label{eq:polestructureappendix}
\end{align}
The treatment of the other pole at $v=\theta_{k-}+\frac{i\pi}{2}$
is very similar, but one should start with $F_{N+2}^{\left[+1\right]}(\overset{\bullet}{v}+i\pi,\overset{\circ}{v},\{\theta-\frac{i\pi}{2}\})$
as mentioned before.

\section{Correspondence to the $\mu$-terms\label{sec:Correspondence-to-the}}

The S-matrix and its derivative at the $t$-channel pole from \eqref{eq:LYSmxpoles}:
\begin{align*}
S(\theta) & =-\frac{i\Gamma^{2}}{\theta-iu}+\Ordo(1) & S'(\theta) & =\frac{i\Gamma^{2}}{\left(\theta-iu\right)^{2}}+\Ordo(1).
\end{align*}
Since we are allowed to modify the formula \eqref{eq:polestructureappendix}
with $\Ordo(\delta^{0})$ terms, the first and second order poles
in it can be replaced by
\begin{align*}
S(\theta_{k+,k}) & =-\frac{\Gamma^{2}}{\delta}+\mathcal{O}(\delta^{0}) & S'(\theta_{k+,k}) & =-\frac{i\Gamma^{2}}{\delta^{2}}+\mathcal{O}(\delta^{0}),
\end{align*}
which looks like
\begin{align*}
 & F_{N+2}^{r}(v+i\pi,v,\{\theta-\frac{i\pi}{2}\})\vert_{v=\theta_{k+}-\frac{i\pi}{2}}=\\
 & =-\frac{i}{2}S'(\theta_{k+,k})\left[\prod_{j\neq k}S(\hat{\theta}_{k+,j})\right]F_{N}(\{\theta\})+\\
 & +iS(\theta_{k+,k})\left[\prod_{j<k}S(\hat{\theta}_{k+,j})\right]^{'}\prod_{j>k}S(\hat{\theta}_{k+,j})F_{N}(\{\theta\})+\\
 & +S(\theta_{k+,k})\prod_{j\neq k}S(\hat{\theta}_{k+,j})\Gamma^{-1}F_{N,k+}^{b}(\{\theta\})+\mathcal{O}(\delta^{0}).
\end{align*}
By replacing $\hat{\theta}_{k+,j}\to\theta_{k+,j}=\hat{\theta}_{k+,j}+i\delta$
in the first term of the above formula extra $\delta^{-1}$ poles
appear, which we compensate with
\[
-\frac{i}{2}S(\theta_{k+,k})\left[\prod_{j\neq k}S(\theta_{k+,j})\right]^{'}F_{N}(\{\theta\}),
\]
and this term combines with the second to give \eqref{eq:rewrittenpolestructure}.
If we then take 
\[
\frac{i}{2}\Res_{v=\hat{\theta}_{k+}-\frac{i\pi}{2}=\theta_{k}-\frac{i\pi}{6}}\left\{ F_{N+2}^{r}(v+i\pi,v,\{\theta_{j}-\frac{i\pi}{2}\})e^{-mL\cosh v}\right\} ,
\]
terms like 
\begin{align*}
-\frac{1}{2}\Res_{\delta=0}\left\{ \prod_{j}S(\theta_{k+,j})e^{imL\sinh\theta_{k+}}\right\}  & =\delta u_{k+}(\{\theta\})\\
-\frac{1}{2}\Res_{\delta=0}\left\{ iS'(\theta_{k+,l})\prod_{j\neq l}S(\theta_{k+,j})e^{imL\sinh\theta_{k+}}\right\}  & =-i\partial_{l}\delta u_{k+}(\{\theta\})
\end{align*}
will appear in it and we get 
\begin{equation}
\Gamma^{-1}F_{N,k\pm}^{b}(\{\theta\})\delta u_{k\pm}(\{\theta\})-\frac{i}{2}F_{N}(\{\theta\})(\mp\partial_{k}+\sum_{j<k}\partial_{j}-\sum_{j>k}\partial_{j})\delta u_{k\pm}(\{\theta\})\label{eq:residuepm}
\end{equation}
if we repeat the same steps for the residue at $v=\hat{\theta}_{k-}+\frac{i\pi}{2}=\theta_{k}+\frac{i\pi}{6}$
of the integral. Adding the first term for $+$ and $-$, and taking
every function at $\theta_{j}=\bar{\theta}_{j}^{(0)},\;\forall j$
gives 
\[
2\Gamma^{-1}F_{N,k}^{b}(\{\bar{\theta}^{(0)}\})\delta\bar{u}_{k}
\]
since $F_{N,k\pm}^{b}(\{\theta\})=F_{N,k}^{b}(\{\theta\})\pm i\Gamma\partial_{k}F_{N}(\{\theta\})$
and $\delta\bar{u}_{k}=\delta u_{k+}(\{\bar{\theta}^{(0)}\})=\delta u_{k-}(\{\bar{\theta}^{(0)}\})$.

Then the second term in \eqref{eq:residuepm} by using $\partial_{j}\delta u_{k\pm}=\pm i\partial_{j}Q_{k\pm}^{(0)}(\{\theta_{\pm}\})\delta u_{k\pm}$
derived in App. \ref{sec:Useful-relations-based} looks 
\[
\frac{1}{2}F_{N}(\{\theta\})\left[-\partial_{j}Q_{k\pm}^{(0)}(\{\theta_{\pm}\})\pm\left(\sum_{j<k}-\sum_{j>k}\right)\partial_{j}Q_{k\pm}^{(0)}(\{\theta_{\pm}\})\right]\delta u_{k\pm}(\{\theta\})
\]
and taking the required sum at the $(0)$-th order solution gives
\[
\frac{1}{2}F_{N}(\{\bar{\theta}^{(0)}\})\left[-\partial_{k}Q_{k}^{(0)}(\{\bar{\theta}^{(0)}\})+\left(\sum_{j<k}-\sum_{j>k}\right)\partial_{j}\tilde{Q}_{k}^{(0)}(\{\bar{\theta}^{(0)}\})\right]\delta\bar{u}_{k}.
\]
Next we sum up the contributions of the particles:
\begin{align}
\delta^{(\mu)}F_{N}(\{\bar{\theta}^{(0)}\}) & =\sum_{k}\left\{ 2\Gamma^{-1}F_{N,k}^{b}(\{\bar{\theta}^{(0)}\})\delta\bar{u}_{k}-\frac{1}{2}\partial_{k}Q_{k}^{(0)}(\{\bar{\theta}^{(0)}\})F_{N}(\{\bar{\theta}^{(0)}\})\right\} \delta\bar{u}_{k}+\nonumber \\
 & +\frac{1}{2}\sum_{k}\left(\sum_{j:j<k}-\sum_{j:j>k}\right)\partial_{j}\tilde{Q}_{k}^{(0)}(\{\bar{\theta}^{(0)}\})\delta\bar{u}_{k},\label{eq:rawmuterm}
\end{align}
and reorder the very last term:
\[
-\sum_{j,k:j>k}\overbrace{\partial_{j}\tilde{Q}_{k}^{(0)}}^{\tilde{\partial}_{k}Q_{j}^{(0)}}\delta\bar{u}_{k}\overset{j\leftrightarrow k}{=}-\sum_{k,j:k>j}\tilde{\partial}_{j}Q_{k}^{(0)}\delta\bar{u}_{j},
\]
after which the last line of \eqref{eq:rawmuterm} looks as:
\[
\left(\sum_{j,k:j<k}-\sum_{j,k:j>k}\right)\partial_{j}\tilde{Q}_{k}^{(0)}\delta\bar{u}_{k}=\sum_{j,k:j<k}\left(\partial_{j}\tilde{Q}_{k}^{(0)}\delta\bar{u}_{k}-\tilde{\partial}_{j}Q_{k}^{(0)}\delta\bar{u}_{j}\right)
\]
which is the same as \eqref{eq:phasecorr}, and one can rewrite it
in terms of $\Delta\varphi(\bar{\theta}_{j,k}^{(0)})$ to finally
get \eqref{eq:FFmuterm}.

\cleardoublepage{}

\lhead[]{\rightmark}

\rhead[\leftmark]{}

\bibliographystyle{utphys}
\bibliography{Refs}

\clearpage{}

\lhead[]{Nomenclature}

\rhead[Nomenclature]{}

\printnomenclature[2.5cm]{}
\end{document}